\documentclass[prd,preprint,superscriptaddress,amsmath,amssymb,nofootinbib]{revtex4}
\usepackage{graphicx}
\usepackage{dcolumn}
\usepackage{bm}
\usepackage{amssymb}
\usepackage{amsmath}
\usepackage{epsfig}    
\usepackage{color}
\usepackage{slashed}
\usepackage{hhline}
\usepackage{tikz}
\usetikzlibrary{decorations.pathmorphing}

\def\be{\begin{equation}}
\def\ee{\end{equation}}
\newcommand{\bea}{\begin{eqnarray}}
\newcommand{\eea}{\end{eqnarray}}
\newcommand{\nn}{\nonumber}

\begin{document}

\title{Zee models with a non-invertible $Z_M$ symmetry}

\author{Huiji Jin} 
\email{jinhuiji@stu.scu.edu.cn}
\affiliation{College of Physics, Sichuan University, Chengdu 610065, China}

\author{Takaaki Nomura}
\email{nomura@scu.edu.cn}
\affiliation{College of Physics, Sichuan University, Chengdu 610065, China}

\author{Hiroshi Okada}
\email{hiroshi3okada@htu.edu.cn}
\affiliation{Department of Physics, Henan Normal University, Xinxiang 453007, China}

\date{\today}

\begin{abstract}
We investigate Zee models by incorporating a non-invertible $Z_M$ symmetry. The models are systematically classified based on their symmetry assignments, which dictate structures for the Yukawa couplings and the neutrino mass matrix. By evaluating the consistency of these mass structures with current experimental data, we identify the viable model candidates. 
Focusing on a representative benchmark model based on the non-invertible $Z_7$ symmetry, we perform a detailed numerical analysis.
Our results yield characteristic predictions for neutrino observables and charged lepton flavor violating processes within the allowed parameter space.
 \end{abstract}
\maketitle

\section{Introduction}

The origin of neutrino masses and mixings remains one of the compelling questions in particle physics, 
pointing directly toward 
the physics beyond the standard model (BSM). 
In fact, numerous BSM frameworks have been proposed to generate neutrino masses, ranging from tree-level mechanisms to various loop-level contributions~\cite{Cai:2017jrq}.
Within these frameworks,
a key challenge is to achieve a predictive structure of neutrino mass matrix which naturally accommodates the observed neutrino oscillation data.

Among the various radiative mechanisms,
the Zee model stands out as a highly attractive candidate, where the neutrino masses are induced at the one-loop level~\cite{Zee:1980ai}.
By extending the scalar sector with a second Higgs doublet and a charged scalar field, the model provides a rich phenomenological landscape.
While the general Zee model can fit current neutrino data, its predictability is often limited by the large number of free parameters in the Yukawa sector~\cite{Herrero-Garcia:2017xdu}.
To enhance predictability, various symmetries have been adopted;
for instance, the original Zee model with $Z_2$ symmetry has already been excluded by neutrino experimental data~\cite{Koide:2001xy,Frampton:2001eu,He:2003ih}.
Subsequent studies have explored flavor symmetries, including 
 global $U(1)$ and discrete symmetries~\cite{Fukuyama:2010ff,Kanemura:2015maa,Nomura:2019dhw,Matsui:2021khj} as well as recent developments in modular flavor symmetries~\cite{Feruglio:2017spp,Nomura:2021pld, Nomura:2024nwh}.
A promising new direction is the application of non-invertible symmetries, which have recently garnered significant attention in the formal and phenomenological literature.

Unlike traditional group-based symmetries, non-invertible symmetries provide selection rules that do not follow a group structure; for comprehensive reviews and applications, see~\cite{Gomes:2023ahz,Schafer-Nameki:2023jdn,Bhardwaj:2023kri,Shao:2023gho} and~\cite{Kobayashi:2024yqq,Kobayashi:2024cvp,Delgado:2024pcv,Funakoshi:2024uvy,Kobayashi:2025znw,Liang:2025dkm,Kobayashi:2025ldi,Kobayashi:2025cwx,Kobayashi:2025lar,Nomura:2025sod,Dong:2025jra,Suzuki:2025oov,Cordova:2022fhg,Nomura:2025yoa,Chen:2025awz,Okada:2025kfm,Kobayashi:2025thd,Suzuki:2025bxg,Jangid:2025krp,Jangid:2025thp,Nomura:2025tvz,Suzuki:2025kxz,Okada:2025adm,Nakai:2025thw,Okada:2026gxl,Kang:2026osw,Kashav:2026jjg,Okada:2026iob,Okada:2026bpp,Xu:2026nwh,Okada:2026pek,Nomura:2026hcu}, respectively.
These selection rules from non-invertible symmetries can provide interaction structures which are not realized by invertible(group) symmetries. 
Furthermore, a particularly interesting feature is that such symmetries can be broken by radiative corrections even if they remain exact at the tree level; see~\cite{Heckman:2024obe,Kaidi:2024wio,Suzuki:2025oov,Kobayashi:2025cwx}.
This scheme offers a novel conceptual framework for constructing BSM physics models and leads to unique predictions.

In this work, we investigate the application of non-invertible $Z_M$ symmetry, denoted by $Z_M^{NI}$, within the Zee model framework.
Remarkably $Z_M^{NI}$ symmetry leads to a diverse set of non-invertible selection rules, which can be motivated by string theory~\cite{Kobayashi:2024yqq}.
In constructing our model, we consider a softly-broken $Z^{NI}_M$ symmetry. 
We systematically classify the models in terms of transformation properties of the fields under $Z_M^{NI}$ and identify viable candidates by analyzing 
the resulting neutrino mass structures. 
After classification, we perform a detailed numerical analysis of a representative benchmark model,
presenting predictions for neutrino sector and lepton flavor violations in charged lepton decays to illustrate the efficacy of  our framework.

This paper is organized as follows. In sec.~II, we discuss our framework in constructing models showing some formulas and classify models under $Z_M^{NI}$. In sec.~III, we investigate a benchmark model and show predicted neutrino observables and lepton flavor violations. We summarize the work in sec.~IV.

\section{Models}

In this section, we discuss the construction of Zee models with non-invertible $Z_M$ symmetry.
The field contents of the models are the same as the original Zee model where we introduce a second Higgs doublet $\Phi$ and $SU(2)$ singlet charged scalar field $S^+$ with hyper charge 1. 
Then we apply softly broken non-invertible $Z_M$ symmetry, denoted by $Z_M^{NI}$ hereafter, to the SM leptons and new scalar fields.

Under $Z_M^{NI}$ symmetry, we assign a class $[g^k]$ for fields where $g$ is the generator of the original $Z_M$ discrete symmetry and $k$ is an integer running $0$ to $M-1$~\footnote{ 
A class is defined using the automorphism such that 
\begin{equation}
[g^k] = \{ h g^k h^{-1} \ | \ h = e, r  \} \quad (k = 0, 1, \cdots, M-1), \nn
\end{equation}
where $\{e,r\}$ is associated with gauged $Z_2$ symmetry on orbifold satisfying $e g^{-1} e = g$ and  $r g r^{-1} = g^{-1}$. The detailed explanations can be referred to refs.~\cite{Kobayashi:2024yqq,Kobayashi:2024cvp,Kobayashi:2025znw} }.
Then the products of two classes are given by
\begin{equation}
[g^k] [g^{k'}] = [g^{k+k'}] + [g^{M-k+k'}]. \label{eq:products}
\end{equation}
If we assign $[g^k]$ and $[g^{k'}]$ to fields $\phi$ and $\phi'$ the term $\phi \phi'$ is invariant when $[g^{k+k'}]$ or $[g^{M-k+k'}]$ coincides with trivial class $[g^0]$.
This rule can be easily extended for interaction terms including three and four fields.
In this work, we assign classes of $Z_M^{NI}$ for three generations of SM leptons as $\{[g^{k_{\ell 1}}], [g^{k_{\ell 2} }], [g^{k_{\ell 3}}] \}$ to distinguish them. We also assign classes $[g^{k_\Phi}]$ and $[g^{k_S}]$ for $\Phi$ and $S^+$ respectively.
The field contents and charge assignments are summarized in the Table~\ref{tab:1}.
We first show general formulas of the models and then discuss lepton flavor structure which is determined by the assignment of $Z_M^{NI}$.

\begin{table}[t!]
\begin{tabular}{|c||c|c|c|c|c|}\hline\hline  
& ~$L_L$~ & ~$ \ell_R$~& ~$\Phi_1$~& ~$\Phi_2$~ & ~$S^+$  \\\hline\hline 
$SU(2)_L$   & $\bm{2}$  & $\bm{1}$  & $\bm{2}$ & $\bm{2}$ & $\bm{1}$  \\\hline 
$U(1)_Y$    & $-\frac12$  & $-1$  & $\frac12$ & $\frac12$  & $1$    \\\hline
$Z_M^{NI}$   & $ \{[g^{k_{\ell 1}}], [g^{k_{\ell 2}}], [g^{k_{\ell 3}} ] \}$  & $ \{[g^{k_{\ell 1}}], [g^{k_{\ell 2}}], [g^{k_{\ell 3}} ] \}$ & $[g^{k_\Phi}]$  & $[g^0]$ & $[g^{k_S}]$        \\\hline
\end{tabular}
\caption{Charge assignments of the leptons and scalar fields
under $SU(2)_L\otimes U(1)_Y \otimes Z_M^{NI}$.}\label{tab:1}
\end{table}

The Yukawa interactions in lepton sector are written by
\begin{equation}
\mathcal{L}_{Y_\ell} = y^\ell \overline{L_L} \ell_R \Phi_2 + y^\Phi \overline{L_L} \ell_R \Phi_1 + f \overline{L^c_L} (i \sigma_2) L_L S^+ + h.c. \, ,
\end{equation}
where $\sigma_2$ is the second Pauli matrix.
Here we omitted the flavor indices and the structures of the Yukawa couplings depend on the assignment of $Z^{NI}_M$ classes to fields as discussed below.
The scalar potential is given by
\begin{align}
V = &  \mu^2_1 |\Phi_1|^2 + \mu_{2}^2 |\Phi_2|^2 - \mu_{12}^2 (\Phi_1^\dagger \Phi_2 + h.c.) + \mu (\Phi_2^T i \sigma_2 \Phi_1 S^- + h.c. ) \nn \\
& + \frac12 \lambda_1 |\Phi_1|^4 + \frac12 \lambda_2 |\Phi_2|^4 + \lambda_3 |\Phi_1|^2 |\Phi_2|^2 + \lambda_4 |\Phi_1^\dagger \Phi_2|^2 + \frac{\lambda_5}{2} \{ (\Phi_1^\dagger \Phi_2)^2 + h.c. \}. 
\end{align}
In our framework, we consider $Z^{NI}_M$ is softly broken. Then the term $\Phi_2 i \sigma_2 \Phi_1 S^-$ is always allowed in any assignment of $Z^{NI}_M$ classes to $\Phi_1$ and $S^\pm$.
The assumption of soft breaking is a reasonable choice since terms forbidden by $Z_M^{NI}$ symmetry at tree level often appear through radiative correction and mass dimensional couplings would not be suppressed.

Note that the scalar potential is the same as the original Zee model and we do not analyze it in details. In the following, we just summarize the mass eigenstates in scalar sector to calculate a neutrino mass matrix.

\subsection{Mass eigenstates}

Here we show mass eigenstates in the models and derive interactions in the mass basis.
Firstly we write Higgs doublet fields in terms of the Higgs basis defined by
\begin{equation}
\begin{pmatrix} \Phi_1 \\ \Phi_2 \end{pmatrix} = 
\begin{pmatrix} c_\beta & - s_\beta \\ s_\beta & c_\beta \end{pmatrix}
\begin{pmatrix} H' \\ \Phi \end{pmatrix},
\end{equation}
where $s_\beta = \sin \beta$ and $c_\beta = \cos \beta$ with $\tan \beta = \langle \Phi_2 \rangle/\langle \Phi_1 \rangle$.
The doublet fieleds in the Higgs basis can be written as
\begin{equation}
H' = \begin{pmatrix} G^+ \\ \frac{1}{\sqrt{2}} (h'_1 + v + i G^0) \end{pmatrix}, \quad
\Phi = \begin{pmatrix} H^+ \\ \frac{1}{\sqrt{2}} (h'_2+ i A), \end{pmatrix},
\end{equation}
where $G^+$ and $G^0$ are Namubu-Goldstone bosons absorbed by massive SM gauge bosons $W^+$ and $Z$, and $v\simeq 246$ GeV is the Higgs vacuum expectation value (VEV).
The mass eigenstates for CP-even Higgs bosons are defined as 
\begin{equation}
\begin{pmatrix} h'_1 \\ h'_2 \end{pmatrix} =
\begin{pmatrix} c_{\alpha -\beta} & -s_{\alpha - \beta} \\ s_{\alpha - \beta} & c_{\alpha - \beta} \end{pmatrix}
\begin{pmatrix} H \\ h \end{pmatrix}
\end{equation}
where the mixing angle $\alpha$ is determined by the parameters in the Higgs potential.
Here $h$ is identified as the SM Higgs boson while $H$ corresponds to heavy neutral CP-even one.
We also write mass eigenstates of charged scalar bosons such that
\begin{equation}
\begin{pmatrix} H^\pm \\ S^\pm \end{pmatrix} =
\begin{pmatrix} c_{\chi} & -s_{\chi} \\ s_{\chi} & c_{\chi} \end{pmatrix}
\begin{pmatrix} H_1^\pm \\ H_2^\pm \end{pmatrix},
\end{equation}

The lepton Yukawa interactions in the scalar mass basis, with alignment limit ($s_{\alpha -\beta} = 1$), are given by
\begin{align}
\mathcal{L}_{Y_\ell} = & \frac{v}{\sqrt{2}} (c_\beta y^\Phi + s_\beta y^\ell) \overline{\ell_L} \ell_R
+ \frac{1}{\sqrt{2}} (c_\beta y^\Phi + s_\beta y^\ell) h \overline{\ell_L} \ell_R + \frac{1}{\sqrt{2}} (s_\beta y^\Phi - c_\beta y^\ell) H \overline{\ell_L} \ell_R
 \nn \\
& +\frac{i}{\sqrt{2}} (-s_\beta y^\Phi + c_\beta y^\ell) A \overline{\ell_L} \ell_R
+c_\chi (-s_\beta  y^\Phi + c_\beta y^\ell) H_1^+ \overline{\nu_L} \ell_R +s_\chi (s_\beta  y^\Phi - c_\beta y^\ell) H_2^+ \overline{\nu_L} \ell_R \nn \\
& + f (\overline{\nu^C_L} \ell_L - \overline{\ell^C_L} \nu_L) (s_\chi H_1^+ + c_\chi H_2^+) + h.c. \label{eq:int-mass-basis1}
\end{align}
The first term of RHS in Eq.~\eqref{eq:int-mass-basis1} is a mass term of charged leptons.
Thus the charged lepton mass matrix can be written by
\begin{equation}
M_\ell = \frac{c_\beta v}{\sqrt{2}} (y^\Phi + t_\beta y^\ell).
\end{equation}
The mass matrix is diagonalized via mixing matrices $V_R$ and $V_L$ as $m_\ell \equiv$ diag.$(m_e,m_\mu,m_\tau)= V_L^\dag M_\ell V_R$, which are associated with transformation $\ell_{L(R)} \to V_{L(R)} \ell^m_{L(R)}$ with $\ell^m_{L(R)}$ being mass eigenstates.
Therefore, the mass matrix satisfies $V_L^\dag M_\ell M_\ell^\dagger V_L ={\rm diag.}(|m_e|^2,|m_\mu|^2,|m_\tau|^2)$.

\subsection{Neutrino mass}


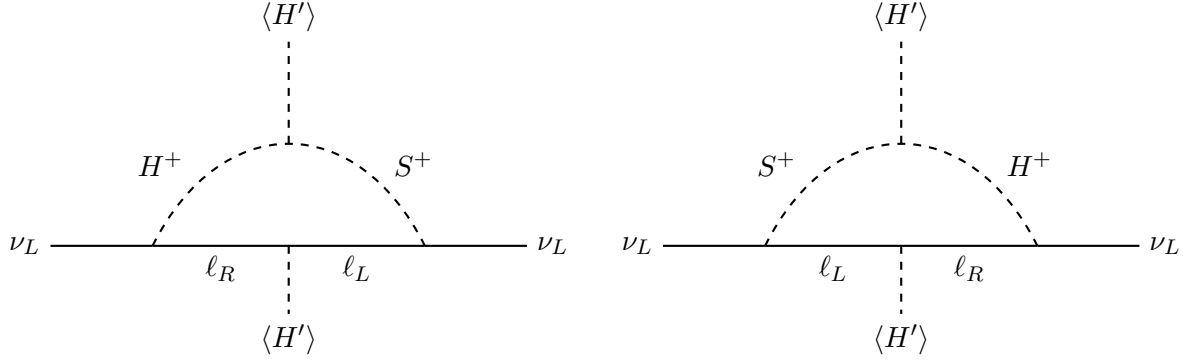
\begin{figure}[t]
\begin{tikzpicture}[scale=0.9]


\draw[thick] (-3.5,0) -- (-2,0);
\draw[thick] (2,0) -- (3.5,0);

\draw[thick] (-2,0) -- (2,0);

\draw[dashed,thick]
(-2,0) .. controls (-1,2) and (1,2) .. (2,0);

\draw[dashed,thick] (0,1.5) -- (0,3);
\draw[dashed,thick] (0,0) -- (0,-1);

\node[left] at (-3.5,0) {$\nu_L$};
\node[right] at (3.5,0) {$\nu_L$};

\node[below] at (-1,0) {$\ell_R$};
\node[below] at (1,0) {$\ell_L$};

\node[left] at (-1.4,1.2) {$H^+$};
\node[right] at (1.4,1.2) {$S^+$};

\node[above] at (0,3) {$\langle H' \rangle$};
\node[below] at (0,-1) {$\langle H' \rangle$};


\begin{scope}[xshift=9cm]

\draw[thick] (-3.5,0) -- (-2,0);
\draw[thick] (2,0) -- (3.5,0);

\draw[thick] (-2,0) -- (2,0);

\draw[dashed,thick]
(-2,0) .. controls (-1,2) and (1,2) .. (2,0);

\draw[dashed,thick] (0,1.5) -- (0,3);
\draw[dashed,thick] (0,0) -- (0,-1);

\node[left] at (-3.5,0) {$\nu_L$};
\node[right] at (3.5,0) {$\nu_L$};

\node[below] at (-1,0) {$\ell_L$};
\node[below] at (1,0) {$\ell_R$};

\node[left] at (-1.4,1.2) {$S^+$};
\node[right] at (1.4,1.2) {$H^+$};

\node[above] at (0,3) {$\langle H' \rangle$};
\node[below] at (0,-1) {$\langle H' \rangle$};

\end{scope}

\end{tikzpicture}
\caption{One-loop diagrams inducing neutrino masses. \label{fig:diagram}}
\end{figure}


The neutrino mass matrix is generated by one-loop diagram as in the original Zee model. 
Neutrino masses are induced by diagrams in Fig.~\ref{fig:diagram}.
The relevant Yukawa interactions for neutrino mass generation are written by 
\begin{equation}
\mathcal{L}_{Y_\ell} \supset (s_\beta c_\chi H_1^+ +  s_\beta s_\chi H_2^+)Y_{ij} \overline{\nu_{L_i}} \ell^m_{R_j} 
+ (s_\chi H_1^+ + c_\chi H_2^+) F_{ij} \left(\overline{\nu^C_{L_i}} \ell^m_{L_j} - \overline{\ell^{m C}_{L_i}} \nu_{L_j} \right) + h.c. , \label{eq:Y-charged}
\end{equation}
where the relevant Yukawa couplings are given by 
\begin{equation}
Y \equiv (- y^\Phi + \cot \beta y^\ell) V_R, \quad F \equiv f V_L.
\label{eq:YF}
\end{equation}

Calculating one-loop diagrams, the neutrino mass matrix is approximately given by 
\begin{equation}
(m_\nu)_{ij} \simeq \frac{\sin 2\chi}{16 \pi^2} \ln \left( \frac{m^2_{H^+_2}}{m^2_{H^+_1}} \right) (F m_\ell Y^\dagger)_{ij} + (i \leftrightarrow j),
\end{equation}
where charged lepton masses in the log factor are ignored.

\subsection{Lepton flavor violation}

Here we formulate charged lepton flavor violations (CLFVs).
Firstly, we write Yukawa interactions relevant for CLFV in mass basis.
The relevant Yukawa interactions with neutral scalar bosons are 
\begin{equation}
\mathcal{L}_{Y_\ell} \supset (Y_h h + Y_H H - i Y_H A) \overline{\ell^m_L} \ell^m_R + h.c., 
\end{equation}
where the couplings are given by
\begin{align}
Y_h & \equiv \frac{1}{\sqrt{2}} V^\dagger_L (c_\beta y^\Phi + s_\beta y^\ell) V_R, \nn \\
Y_H & \equiv \frac{1}{\sqrt{2}} V^\dagger_L (s_\beta y^\Phi - c_\beta y^\ell) V_R.
\end{align}
Note that Yukawa coupling matrix for the $H$ and $A$ is the same. Also $Y_h$ should be diagonal and does not contribute to CLFV process.
The relevant Yukawa interactions associated with charged scalar bosons are already given in Eq.~\eqref{eq:Y-charged} that are used to generate neutrino masses.

\begin{figure}[t]
\begin{tikzpicture}[scale=0.9]


\draw[thick] (-3.5,0) -- (-2,0);
\draw[thick] (2,0) -- (3.5,0);

\draw[thick] (-2,0) -- (2,0);

\draw[dashed,thick]
(-2,0) .. controls (-1,2) and (1,2) .. (2,0);

\draw[thick,decorate, decoration={snake}] (0,1.5) -- (1.5,3);

\node[left] at (-3.5,0) {$\ell_i$};
\node[right] at (3.5,0) {$\ell_j$};

\node[below] at (0,0) {$\nu$};

\node[left] at (-1.4,1.2) {$H_{1,2}^+$};

\node[above] at (0,2.5) {$\gamma$};


\begin{scope}[xshift=9cm]

\draw[thick] (-3.5,0) -- (-2,0);
\draw[thick] (2,0) -- (3.5,0);

\draw[thick] (-2,0) -- (2,0);

\draw[dashed,thick]
(-2,0) .. controls (-1,2) and (1,2) .. (2,0);

\draw[thick,decorate, decoration={snake}] (0,0) -- (1,-1);

\node[left] at (-3.5,0) {$\ell_i$};
\node[right] at (3.5,0) {$\ell_j$};

\node[below] at (-1,0) {$\ell$};
\node[below] at (1,0) {$\ell$};

\node[left] at (-1.4,1.2) {$H/A$};

\node[below] at (0,-1) {$\gamma$};

\end{scope}

\end{tikzpicture}
\caption{One-loop diagrams inducing CLFV process $\ell_i \to \ell_j \gamma$. \label{fig:diagram2}}
\end{figure}
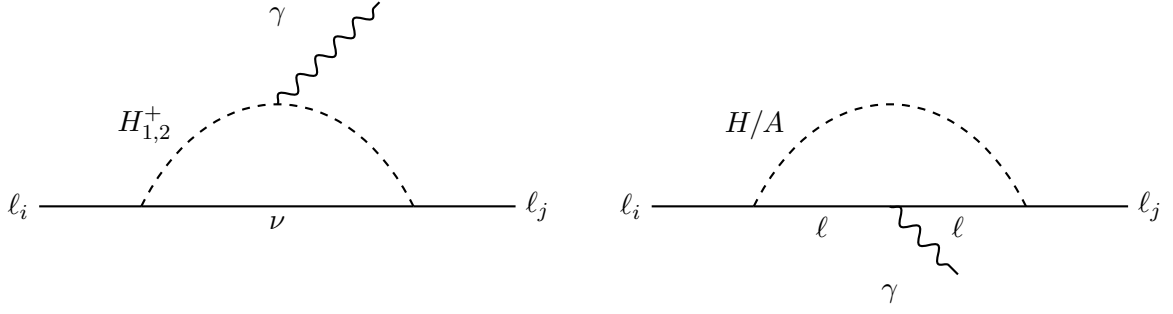

The CLFV decay processes, $\ell_i \to \ell_j \gamma$, are induced at one-loop level via diagrams in Fig.~\ref{fig:diagram2}.
Estimating the diagrams, we can write the branching ratio (BR) of the CLFV decay as follows
\begin{equation}
BR(\ell_i \to \ell_j \gamma) = \frac{48 \pi^3 \alpha_{em} C_{ij}}{G_F^2} \left( |A_L|^2 + |A_R|^2 \right), \label{eq:BRCLFV}
\end{equation}
where $G_F$ is the Fermi constant and $\alpha_{em}$ is the electromagnetic fine structure constant.
Here $A_{L(R)}$ can be obtained by summing up all diagrams where the detailed forms are summarized in the Appendix~\ref{sec:appendix-CLFV}; the same amplitude also induce $\mu \to e$ conversion process~\cite{Kuno:1999jp,Kitano:2002mt,Davidson:2018kud,COMET:2009qeh} where we omit to discuss the process since it is highly suppressed in our benchmark model analyzed below. 
Furthermore, three body CLFV decays $\ell^\mp_i \to \ell^\mp_j \ell^\mp_{k} \ell^\pm_{l}$ are induced at tree level by exchanging heavy neutral bosons via Yukawa coupling $Y_H$.
Here we focus on $\mu \to eee$ and $\tau \to \mu \mu \mu$ CLFV processes which provides clear signal at the experiments.
The BRs are then given by~\cite{Herrero-Garcia:2017xdu} 
\begin{align}
& BR(\mu \to eee) \simeq \frac{1}{64 G^2_F m_H^4} \left[ |(Y_H)^*_{11} (Y_H)_{12}|^2  + |(Y_H)_{11} (Y_H)^*_{21}|^2 \right] BR(\mu \to e \nu \bar \nu), \nn \\
& BR(\tau \to \mu \mu \mu) \simeq \frac{1}{64 G^2_F m_H^4} \left[ |(Y_H)^*_{22} (Y_H)_{23}|^2  + |(Y_H)_{22} (Y_H)^*_{32}|^2 \right] BR(\tau \to \mu \nu \bar \nu),
\end{align}
where we ignored charged lepton masses and took $m_A = m_H$.
The current upper bounds are given by~\cite{MEGII:2025gzr, MEGII:2023ltw, BaBar:2009hkt, Belle:2021ysv,SINDRUM:1987nra,Hayasaka:2010np}
\begin{align}
& { BR} (\mu \to e \gamma) < 1.5 \times 10^{-13} \, , \quad { BR} (\tau \to e \gamma) < 3.3 \times 10^{-8} \, , \quad { BR} (\tau \to \mu \gamma) < 4.2 \times 10^{-8} \, \nn \\
& BR(\mu \to eee) < 1.0 \times 10^{-12}, \quad BR(\tau \to \mu \mu \mu) < 2.1 \times 10^{-8}.
\end{align}
We impose these constraints in numerical analysis.

\subsection{Exploring viable models}

In this subsection, we explore viable models.
The models are classified by the assignments of $Z_M$ classes to the fields in Table~\ref{tab:1}.
The structures of Yukawa couplings $f$ and $y^\Phi$ are then determined by the choice of the assignment of classes.
Note that the Yukawa coupling matrix $y^\ell$ is diagonal, since we assign the same classes for $L_L$ and $\ell_R$, and $\Phi_2$ has no charge under $Z_M^{NI}$.

For illustration, we investigate a simple case under $Z_4^{NI}$.
In the first example, three generations of SM leptons are assigned to have $\{[g^0], [g^1], [g^2] \}$ classes, and both $\Phi_1$ and $ S^+$ belong to the class $[g^1]$.
The structure of Yukawa couplings are 
\begin{equation}
f : \begin{pmatrix} 0 & \times & 0 \\ \times  & 0 & \times \\ 0 & \times & 0 \end{pmatrix}, \quad
y^\Phi: \begin{pmatrix} 0 & \times & 0 \\ \times  & 0 & \times \\ 0 & \times & 0 \end{pmatrix}, \quad (\text{example 1})
\end{equation}
where "$\times$" indicates a non-zero component.
For example,  $\overline{(L_{L})_1} (\ell_R)_1 \Phi_1$ is not allowed since the products of classes $[g^0][g^0][g^1]$ does not contain trivial class $[g^0]$. The other components can be also understood in the same manner.

The structure of neutrino mass is determined by the products of matrices $F m_\ell Y^\dagger = f M_\ell (-y^\Phi + \cot \beta y^\ell)^\dagger$ and its transpose.
Then structure of $m_\nu$ in this case is 
\begin{align}
 m_\nu : \begin{pmatrix} \times & \times & \times \\ \times  & \times & \times \\ \times & \times & \times \end{pmatrix}, \qquad
 m_\nu : \begin{pmatrix} \times & 0 & \times \\ 0  & \times & 0 \\ \times & 0 & \times \end{pmatrix} \quad (t_\beta \to \infty),
 \quad (\text{example 1})
\end{align}
where we show the structure for the limit of $t_\beta \to \infty$ as a reference.
Thus, we do not have specific structure for finite $t_\beta$ while two-zero structure appears in the limit $t_\beta \to \infty$; this two-zero structure cannot fit neutrino data.
As the second example, we set the same assignments for SM leptons, while assignment of class for $\Phi_1(S^+)$ is chosen to be $[g^1]([g^2])$.
In this example, structure of $f$ is the same but that of $y^\Phi$ is changed to be 
\begin{equation}
y^\Phi: \begin{pmatrix} 0 & 0 & \times \\ 0  & \times & 0 \\ \times & 0 & 0 \end{pmatrix}, \quad (\text{example 2}).
\end{equation}
Then the structure of neutrino mass becomes 
\begin{equation}
 m_\nu : \begin{pmatrix} 0 & \times & 0 \\ \times  & 0 & \times \\ 0 & \times & 0 \end{pmatrix}, \quad (\text{example 2})
\end{equation}
where it does not change in the $t_\beta \to \infty$ limit.
Thus, in this example, we can not fit the neutrino data due to the structure.

As in the examples, the assignments of classes to fields in the models determine the structure of the neutrino mass matrix.
Then we explore viable models for $Z_M$ with $M=4$ to $M=7$ where $M=4$ is the minimal choice to distinguish three generations by three classes. 
We also impose following conditions,
\begin{itemize}
\item Assignments of classes to $L_L$ and $\ell_R$ are the same so that $y^\ell$ matrix is always diagonal, 

\item The SM quarks and $\Phi_2$ belong to the trivial class $[g^0]$ so that quarks only couple to $\Phi_2$ without restricting structure of Yukawa couplings to realize quark masses and mixings.
\end{itemize}
Under these conditions, structures of neutrino mass matrix are examined for different assignments of $Z_M^{NI}$ classes to the fields for each $M$.
Then we list viable models providing a structure of neutrino mass which is possible to fit the neutrino data. 
The complete lists of viable models for each $M$ are shown in the appendix~\ref{sec:viable-models}.

\section{Numerical analysis of a benchmark model}

\begin{table}[t!]
\begin{tabular}{|c||c|c|c|c|c|}\hline\hline  
& ~$L_L$~ & ~$ \ell_R$~& ~$\Phi_1$~& ~$\Phi_2$~ & ~$S^+$  \\\hline\hline 
$SU(2)_L$   & $\bm{2}$  & $\bm{1}$  & $\bm{2}$ & $\bm{2}$ & $\bm{1}$  \\\hline 
$U(1)_Y$    & $-\frac12$  & $-1$  & $\frac12$ & $\frac12$  & $1$    \\\hline
$Z_7^{NI}$   & $ \{[g^{0}], [g^{1}], [g^{2} ] \}$  & $ \{[g^{0}], [g^{1}], [g^{2} ] \}$ & $[g^3]$  & $[g^0]$ & $[g^1]$        \\\hline
\end{tabular}
\caption{Charge assignments for our benchmark model
under $SU(2)_L\otimes U(1)_Y \otimes Z_7^{NI}$.}\label{tab:2}
\end{table}

In this section, we carry out numerical analysis for our benchmark model.
As a benchmark model, we choose $Z_7^{NI}$ model (1) in Table~\ref{tab:M4} where the assignment is given in Table~\ref{tab:2}.

\subsection{Description of the benchmark model}

In the model strucrues of Yukawa couplings are 
\begin{equation}
y^\ell : \begin{pmatrix} y^\ell_{11} & 0 & 0 \\ 0  & y^\ell_{22} & 0 \\ 0 & 0 & y^{\ell}_{33} \end{pmatrix}, \quad
y^\Phi : \begin{pmatrix} 0 & 0 & 0 \\ 0  & 0 & y^\Phi_{23} \\ 0 & y^\Phi_{32} & y^\Phi_{33} \end{pmatrix}, \quad
f : \begin{pmatrix} 0 & f_{12} & 0 \\ - f_{12}  & 0 & f_{23} \\ 0 & - f_{23} & 0 \end{pmatrix},
\end{equation}
where we explicitly write the components of the matrices.
Thus, charged lepton mass matrix becomes
\begin{align}
M_\ell = & \frac{v s_\beta}{\sqrt{2}} (\cot \beta y^\Phi + y^\ell) \nn \\
& = \frac{v s_\beta}{\sqrt{2}} \left[ \cot \beta \begin{pmatrix} 0 & 0 & 0 \\ 0  & 0 & y^\Phi_{23} \\ 0 & y^\Phi_{32} & y^\Phi_{33} \end{pmatrix} + \begin{pmatrix} y^\ell_{11} & 0 & 0 \\ 0  & y^\ell_{22} & 0 \\ 0 & 0 & y^{\ell}_{33} \end{pmatrix} \right].
\end{align}
In this case, the first generation of charged lepton does not mix with other generations, and the mixing matrix $V_{L(R)}$ has the structure of 
\begin{equation}
V_{L(R)} : \begin{pmatrix} 1 & 0 & 0 \\ 0  & \times & \times \\ 0 & \times & \times \end{pmatrix}. \label{eq:mnu}
\end{equation}
The electron mass is simply given by $m_e = v s_\beta y_{11}^\ell/\sqrt{2}$, which fixes the coupling $y^\ell_{11}$.
The mixing matrix $V_{L(R)}$ is calculated by numerically diagonalizing the mass matrix in the analysis below.
We also note that CLFV associated with electron is highly suppressed due to the structure of the Yukawa couplings. 

The neutrino mass is given by Eq.~\eqref{eq:mnu} and its structure is 
\begin{equation}
m_\nu : \begin{pmatrix} 0 & \times & \times \\ \times  & \times & \times \\ \times & \times & \times \end{pmatrix}, \quad 
m_\nu |_{t_\beta \to \infty} : \begin{pmatrix} 0 & 0 & \times \\ 0  & \times & \times \\ \times & \times & \times \end{pmatrix},
\end{equation}
where $m_\nu |_{t_\beta \to \infty}$ corresponds to the mass matrix under the $t_\beta \to \infty$ limit. 
An interesting point of the model is that we have one-zero texture with finite $t_\beta$ while two-zero texture is realized in the limit of $t_\beta \to \infty$ representing significant dependence of neutrino mass matrix to $t_\beta$ value.
We thus numerically explore the neutrino observables for different $t_\beta$ values below.

The mass eigenvalues for active neutrinos are obtained by diagonalizing the mass matrix as $D_\nu \equiv V_\nu^T m_\nu V_\nu$ where $D_\nu = \{D_{\nu_1}, D_{\nu_2}, D_{\nu_3} \}$.
The neutrino mixing are described by the standard parametrization in terms of the Pontecorvo-Maki-Nakagawa-Sakata (PMNS) mixing matrix {$U$}
defined by $U \equiv V_{e_L}^\dag V_\nu$,
and the Majorana phase is defined by $[1,e^{i\alpha_{21}/2},e^{i\alpha_{31}/2}]$~\cite{Okada:2019uoy}. 
Then, we estimate the Majorana phases using the formulas; 
\begin{align}
\cos \left( \frac{\alpha_{21}}{2} \right) = \frac{\text{Re}[U^*_{e1} U_{e2}] }{ c_{12} s_{12} c_{13}^2 }, \
\cos \left( \frac{\alpha_{31}}{2} - \delta_{CP} \right) = \frac{\text{Re}[U^*_{e1} U_{e3}] }{ c_{12} s_{13} c_{13}} , \\
 \sin \left( \frac{\alpha_{21}}{2} \right) = \frac{\text{Im}[U^*_{e1} U_{e2}] }{ c_{12} s_{12} c_{13}^2 }, \
\sin \left( \frac{\alpha_{31}}{2} - \delta_{CP} \right) =\frac{ \text{Im}[U^*_{e1} U_{e3}] }{ c_{12} s_{13} c_{13} },
\end{align}
where $c_{ij}(s_{ij})$ stands for $\cos \theta_{ij} (\sin \theta_{ij})$ with $\theta_{ij}$ being the mixing angle in the PMNS matrix.
In deriving phases, $\alpha_{21}/2$ and $\frac{\alpha_{31}}{2} - \delta_{CP}$
are subtracted from $\pi$, if $\cos(\alpha_{21}/2)$ and $\cos \left( \frac{\alpha_{31}}{2} - \delta_{CP} \right)$ are negative.
The Dirac CP-phase $\delta_{CP}$ is given by
\begin{align}
\cos\delta_{CP} &=\frac{-1}{2c_{12}s_{12}c_{23}s_{23}s_{13}} (|U_{31}|^2 - s_{12}^2 s_{23}^2-c_{12}^2c_{23}^2 s_{13}^2), \\
\sin\delta_{CP} &=\frac{1}{s_{23}c_{23}s_{12}c_{12}s_{13}c_{13}^2} {\rm Im}[U_{11}  U_{22} U_{12}^* U_{21}^* ]. \
\end{align}
For convenience, we rewrite the neutrino mass matrix as $m_\nu \equiv \kappa \tilde m_\nu$
where the factor $\kappa\equiv \sin 2 \chi$
is used to scale the neutrino mass. 
Therefore, $\kappa$ can be determined in terms of rescaled neutrino mass eigenvalues $\tilde D_\nu(\equiv D_\nu/\kappa)$ and atmospheric neutrino
mass-squared difference $\Delta m_{\rm atm}^2$ such that
\begin{align}
({\rm NH}):\  \kappa^2= \frac{|\Delta m_{\rm atm}^2|}{\tilde D_{\nu_3}^2-\tilde D_{\nu_1}^2},
\quad
({\rm IH}):\  \kappa^2= \frac{|\Delta m_{\rm atm}^2|}{\tilde D_{\nu_2}^2-\tilde D_{\nu_3}^2},
 \end{align}
where 
NH and IH stand for the normal
and inverted hierarchies, respectively. Subsequently, the solar neutrino mass-squared difference is obtained by the relation
\begin{align}
\Delta m_{\rm sol}^2= {\kappa^2}({\tilde D_{\nu_2}^2-\tilde D_{\nu_1}^2}).
 \end{align}
 
The effective mass for neutrino double beta decay, $m_{ee}$, is written by
\begin{equation}
m_{ee} =\kappa \left| \tilde D_{\nu_1} \cos^2 \theta_{12} \cos^2 \theta_{13} + \tilde D_{\nu_2} \sin^2 \theta_{12} \cos^2 \theta_{13} \, e^{i \alpha_{21}} + \tilde D_{\nu_3} \sin^2 \theta_{13} \, e^{i (\alpha_{31}-2 \delta_{\rm CP})} \right|.
\end{equation}
The strongest upper bounds are given by the KamLAND-Zen experiment~\cite{KamLAND-Zen:2024eml}. 
The upper bound on $m_{ee}$ at the 90\% CL is 
\begin{equation}
m_{ee} \leq (28 \ - \ 122) \ {\rm meV},
\end{equation}
where the range is due to the different method to estimate nuclear matrix elements.
Future experiments can provide stronger constraints such that $m_{ee} < (9-21)$ meV by LEGEND-1000~\cite{LEGEND:2021bnm} and $m_{ee} \leq (4.7-20.3)$ meV by mEXO~\cite{nEXO:2021ujk}.

In addition, we also consider direct search for neutrino mass where the model independent observable is $m^2_{\nu_e} \equiv \kappa^2\sum_i \tilde D^2_{\nu_i} |U_{ei}|^2$.
The current limit, at 90\% CL, by KATRIN~\cite{KATRIN:2024cdt} is 
\begin{equation}
m_{\nu_e} \leq 450 \ {\rm meV}.
\end{equation}

In numerical analysis, we also take into account the cosmological bounds on the sum of neutrino mass $\sum D_\nu$.
The upper bound is $\sum D_\nu \leq 120$ meV from Planck data with standard $\Lambda$CDM cosmological model~\cite{Planck:2018vyg}. 
More stringent constraint is given by combining baryon acoustic oscillation (BAO) data from DESI and CMB data leading the upper bound on the sum as $\sum D_{\nu}\le$ 72 meV~\cite{DESI:2024mwx}. 

\subsection{Numerical analysis}

We numerically analyze the lepton mass matrices for the benchmark model adopting the formulas in previous subsection.
Originally, we have 8 Yukawa couplings
\begin{equation}
\{y^\ell_{11}, y^\ell_{22}, y^\ell_{33}, y^\Phi_{23}, y^\Phi_{32}, y^\Phi_{33}, f_{12}, f_{23} \},
\end{equation}
where we choose $\{ y^\ell_{11},  y^\ell_{22},  y^\ell_{33} \}$ to be real parameters using phase redefinition for lepton fields. 
Then the couplings $\{ y^\ell_{11},  y^\ell_{22},  y^\ell_{33} \}$ are fixed to fit charged lepton masses. 
For estimating neutrino observables, we scan remaining free parameters as follows 
\begin{align}
|y^{\Phi}_{23,32,33}| \in [10^{-6}, 10^{-2}], \quad |f_{12,23}| \in [10^{-6}, 0.1],
\end{align}
where the complex phases of the couplings are scanned within $[-\pi, \pi]$.
Here we take $|y^\Phi_{23,32,33}|$ to be less than $10^{-2}$ since larger values of these couplings spoil to fit charged lepton masses for $t_\beta =1$.
In our analysis, we consider the cases of $t_\beta =1$, $100$ and $10000$ as reference values for illustration.
We then carry out chi-square analysis estimating the $\Delta \chi^2$ value of 
\begin{equation}
\Delta \chi^2 = \sum_{i}  \left( \frac{O_i^{\rm obs} - O_i^{\rm th}}{\delta O_i^{\rm exp}} \right)^2, \label{eq:chi-square}
\end{equation}
where $O_i^{\rm obs (th)}$ is the observed (theoretically) obtained value of the observables and $\delta O_i^{\rm exp}$ denotes the experimental error of $1\sigma$ level.
In the fitting, we take into account the neutrino observables $\{\sin^2 \theta_{12}, \sin^2 \theta_{13}, \sin^2 \theta_{23}, \Delta m^2_{\rm atm}, \Delta m^2_{sol} \}$ with global fit of Nufit 6.1~\cite{Esteban:2024eli}.
The other neutrino observables are estimated as output values such as CP-phases.

In the followings, we summarize the results of our numerical analysis showing predictions for neutrino observables and CLFV BRs. Firstly, we find no allowed parameters which can fit neutrino masses and mixing angles in the IH case.
We thus show our results in NH case for different $\tan \beta$ values below. 

\begin{figure}[t]\begin{center}
\includegraphics[width=50mm]{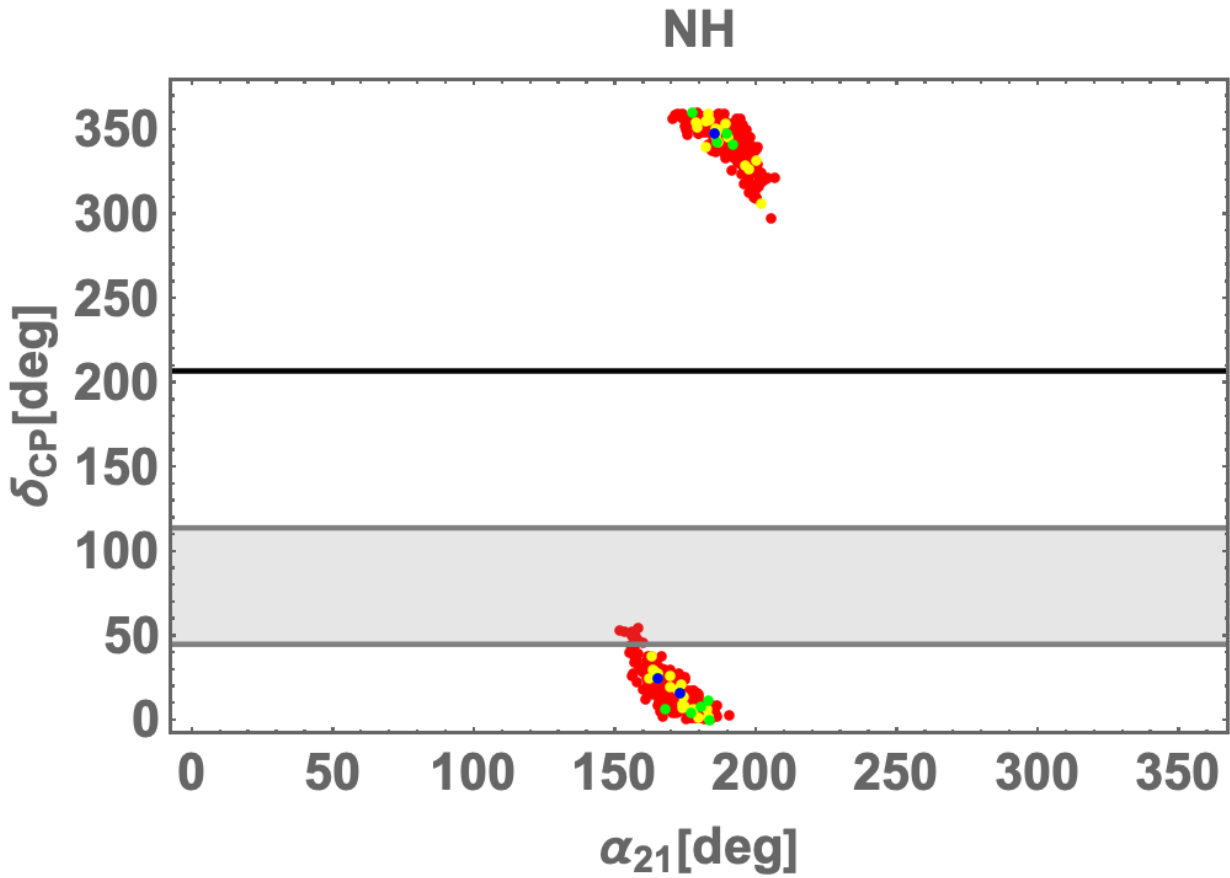} \quad
\includegraphics[width=50mm]{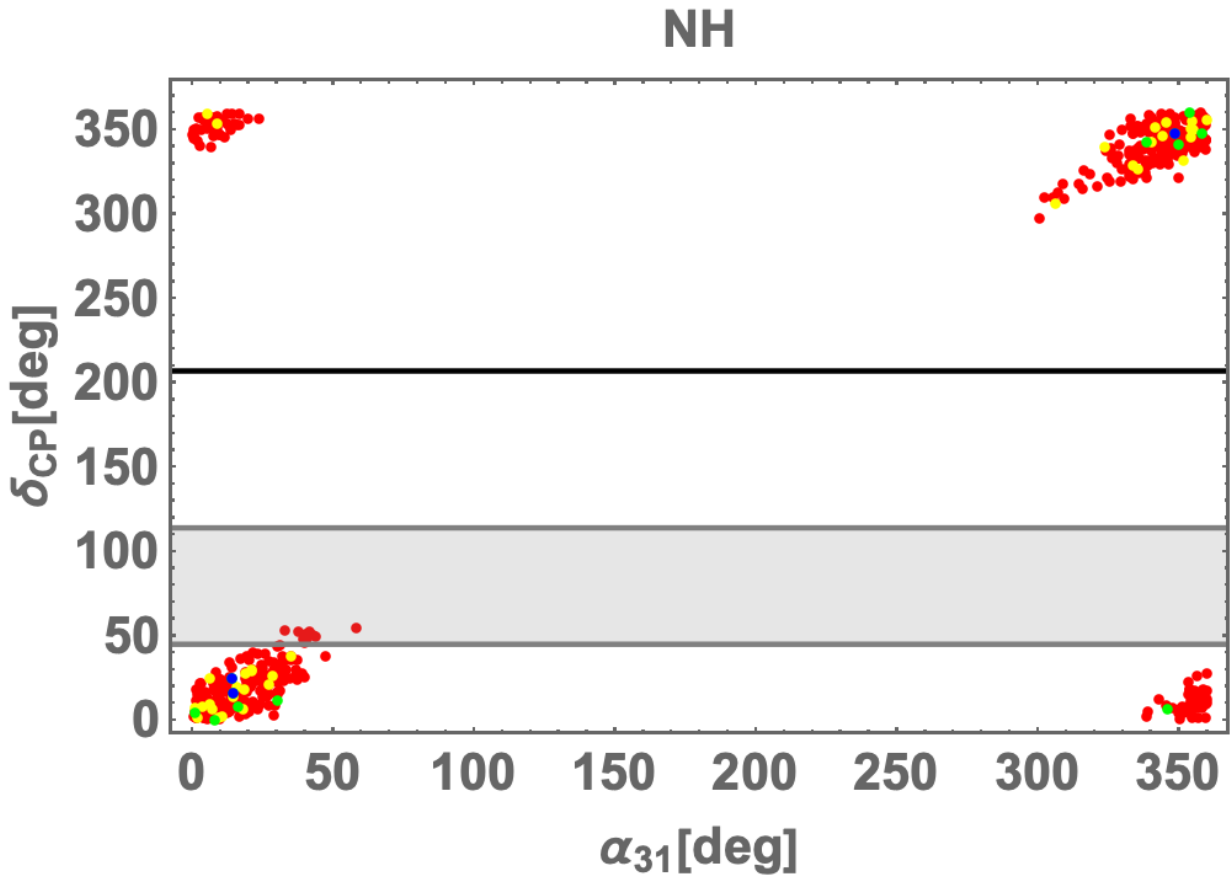} \quad
\includegraphics[width=50mm]{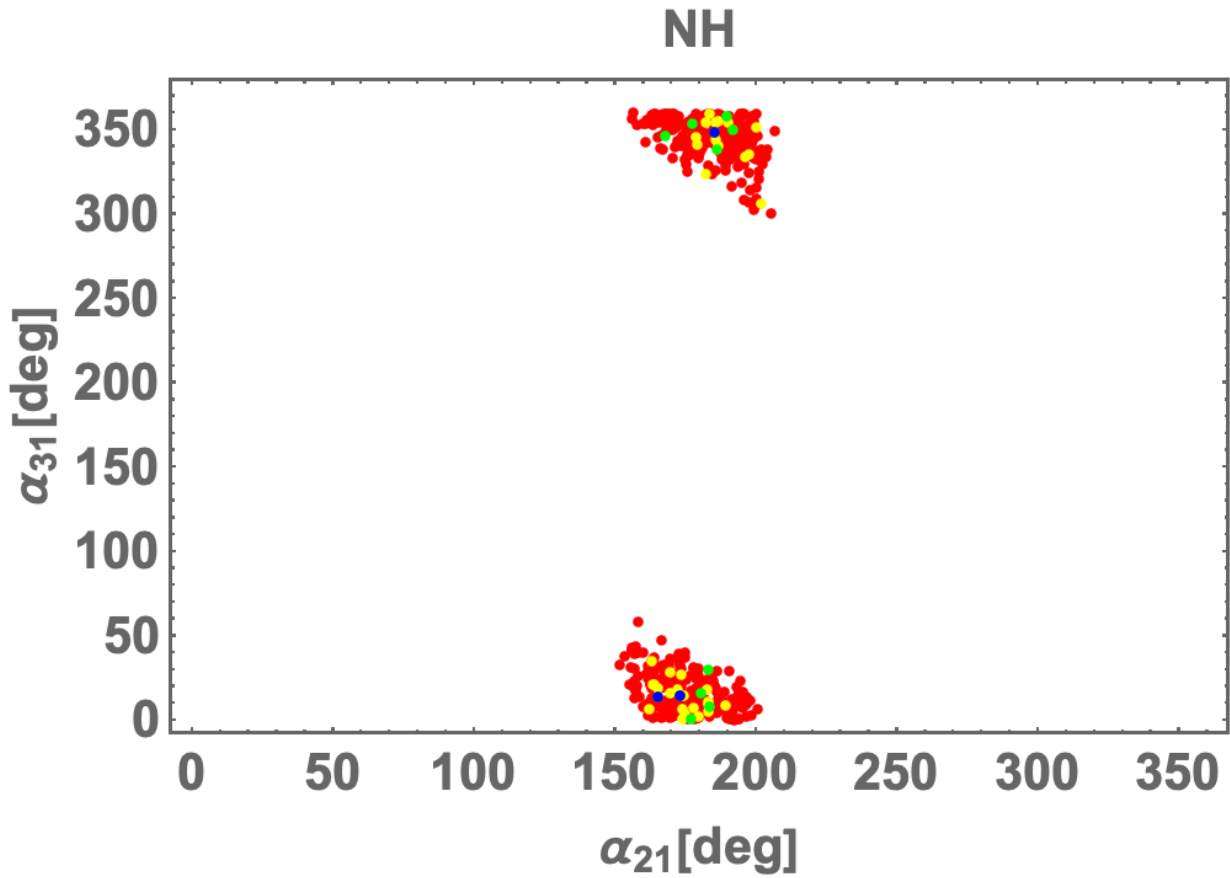}
\caption{Predicted values for neutrino observables from allowed parameter points on $\alpha_{21}$-$\delta_{\rm CP}$ (left), $\alpha_{31}$-$\delta_{\rm CP}$ (center) and $\alpha_{21}$-$\alpha_{31}$ (right) plane for $\tan \beta =1$. 
The color of points indicates an estimated $\chi^2$ value providing $0$-$1 \sigma$ (blue), $1$-$2 \sigma$ (green), $2$-$3 \sigma$ (yellow) and $3$-$5 \sigma$ (red) deviations; indications of color of points are common in the figures afterwards. In the left and center plots the gray region indicate the value of $\delta_{\rm CP}$ is deviated more than 3$\sigma$ in NuFit 6.1 while the horizontal black line represents the best fit value of $\delta_{\rm CP}$; they are common in the plots below showing the $\delta_{\rm CP}$.}
\label{fig:tb1A}
\end{center}\end{figure}

\begin{figure}[t]\begin{center}
\includegraphics[width=50mm]{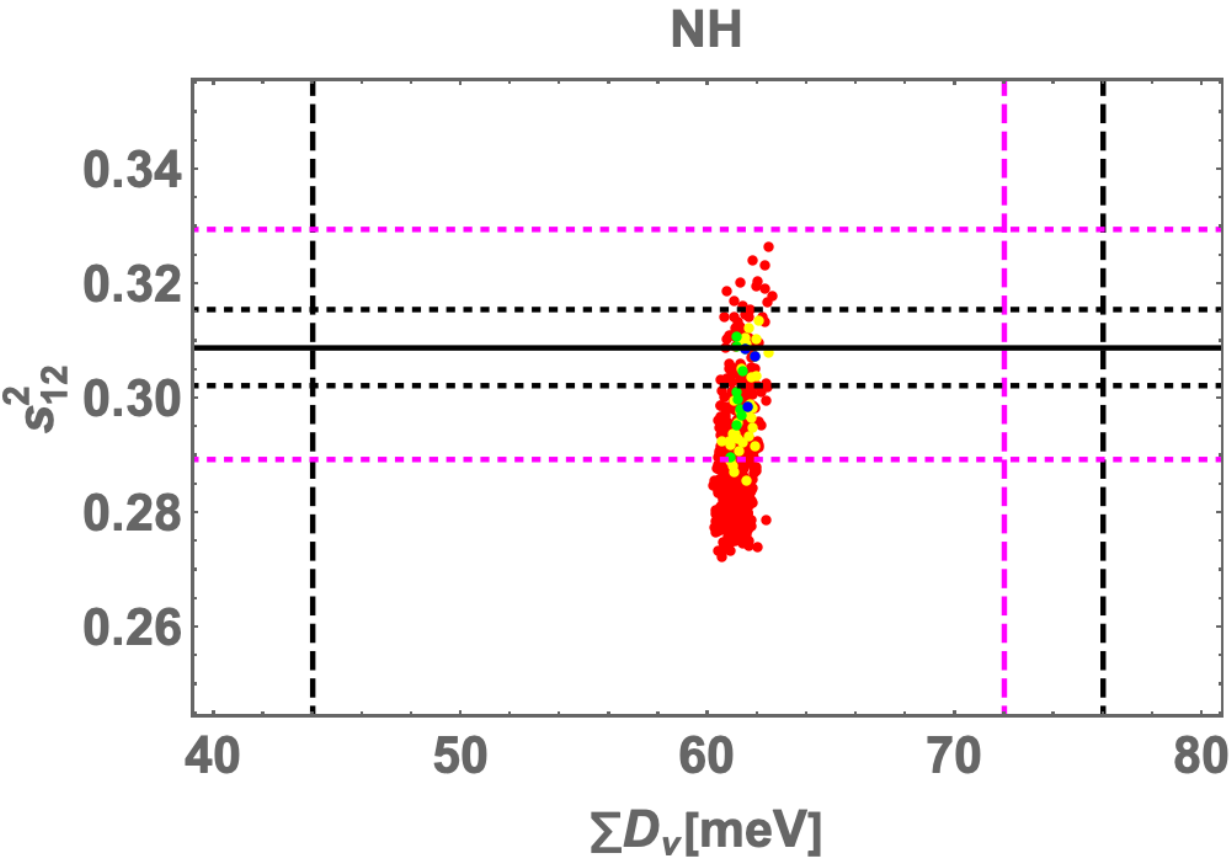} \quad
\includegraphics[width=50mm]{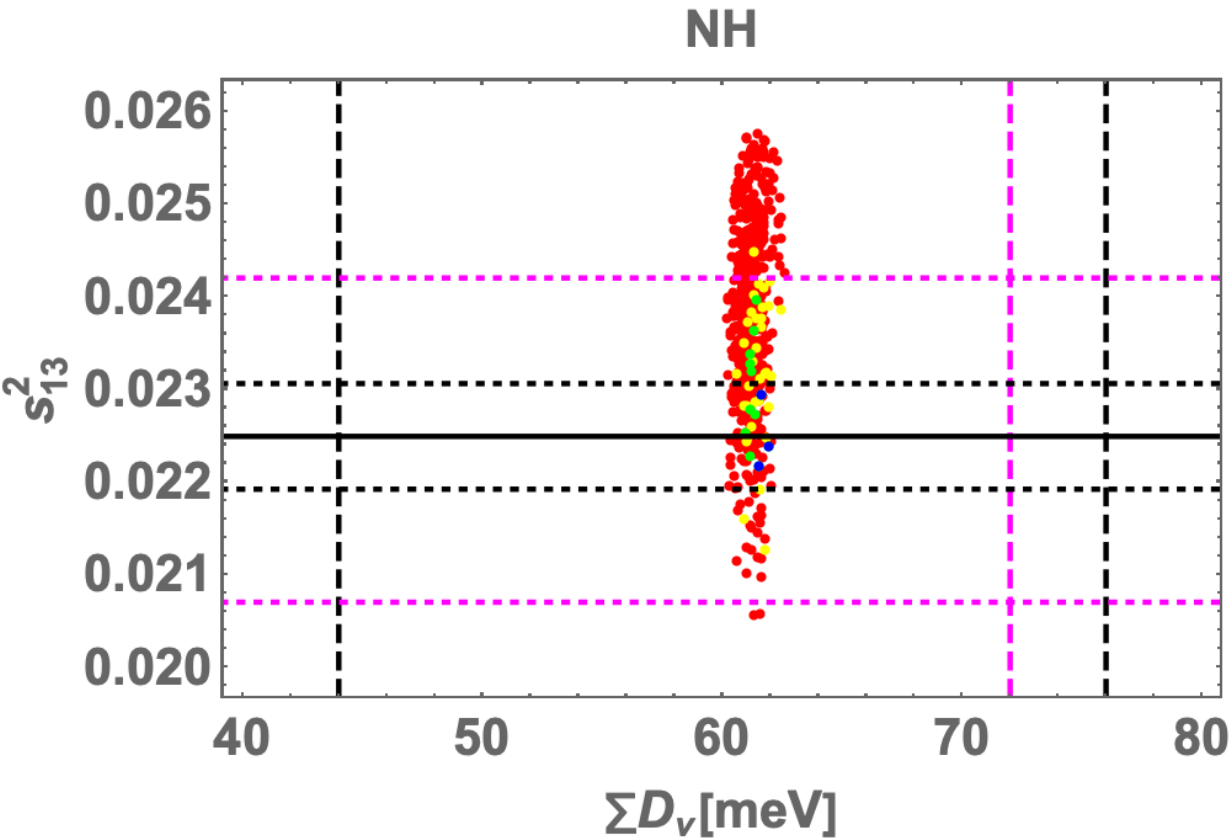} \quad
\includegraphics[width=50mm]{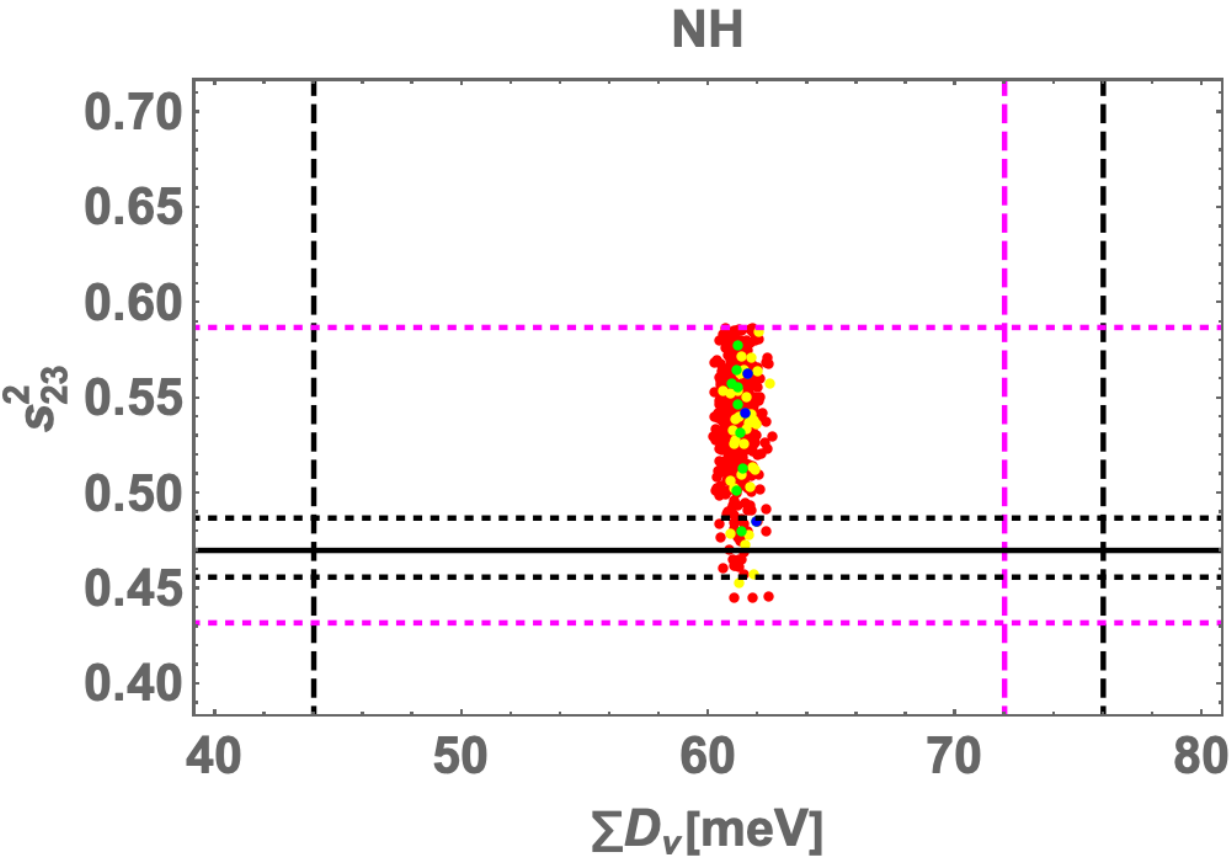}
\caption{Predicted values for neutrino observables from allowed parameter points on $\sum D_\nu$-$s^2_{12}$ (left), $\sum D_\nu$-$s^2_{13}$ (center) and $\sum D_\nu$-$s^2_{23}$ (right) plane for $\tan \beta =1$. The region between horizontal black(magenta) lines corresponding to the $1\sigma (3 \sigma)$ range. The vertical magenta dashed line indicate the upper limit of $\sum D_\nu$ by Planck+DESI data while the region between vertical black dashed lines can be tested by future CMB observations. The lines are common in the plots below showing mixing angles and $\sum m_\nu$.}
\label{fig:tb1B}
\end{center}\end{figure}

\begin{figure}[t]\begin{center}
\includegraphics[width=50mm]{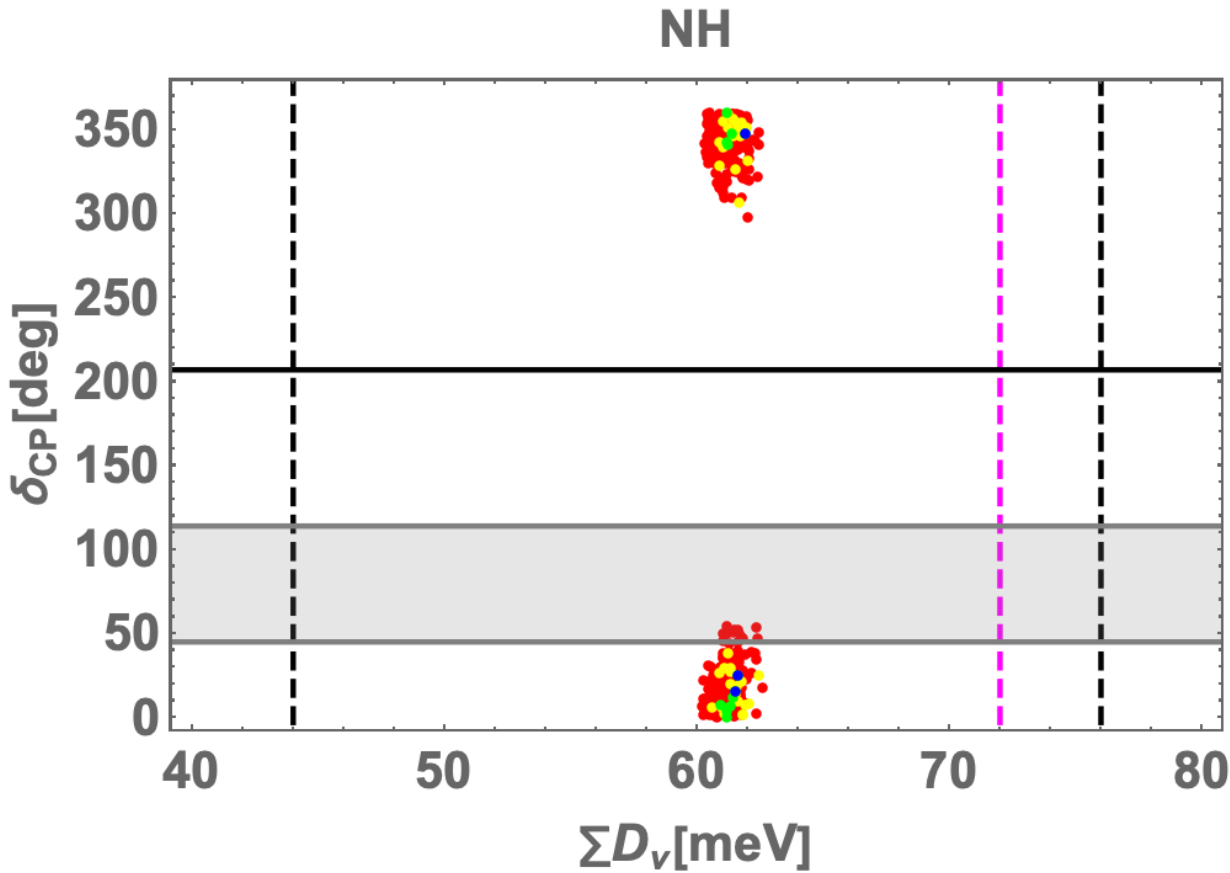} \quad
\includegraphics[width=50mm]{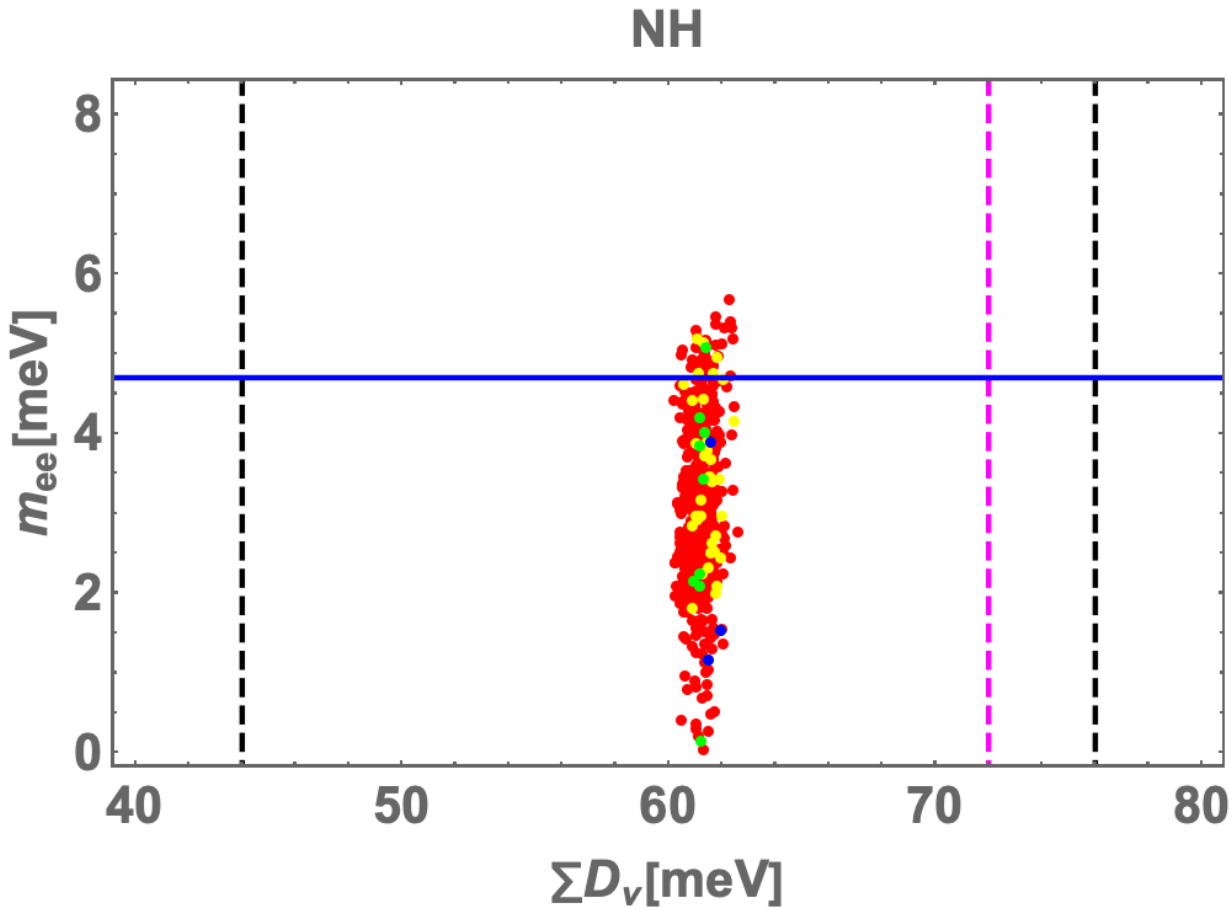} \quad
\includegraphics[width=50mm]{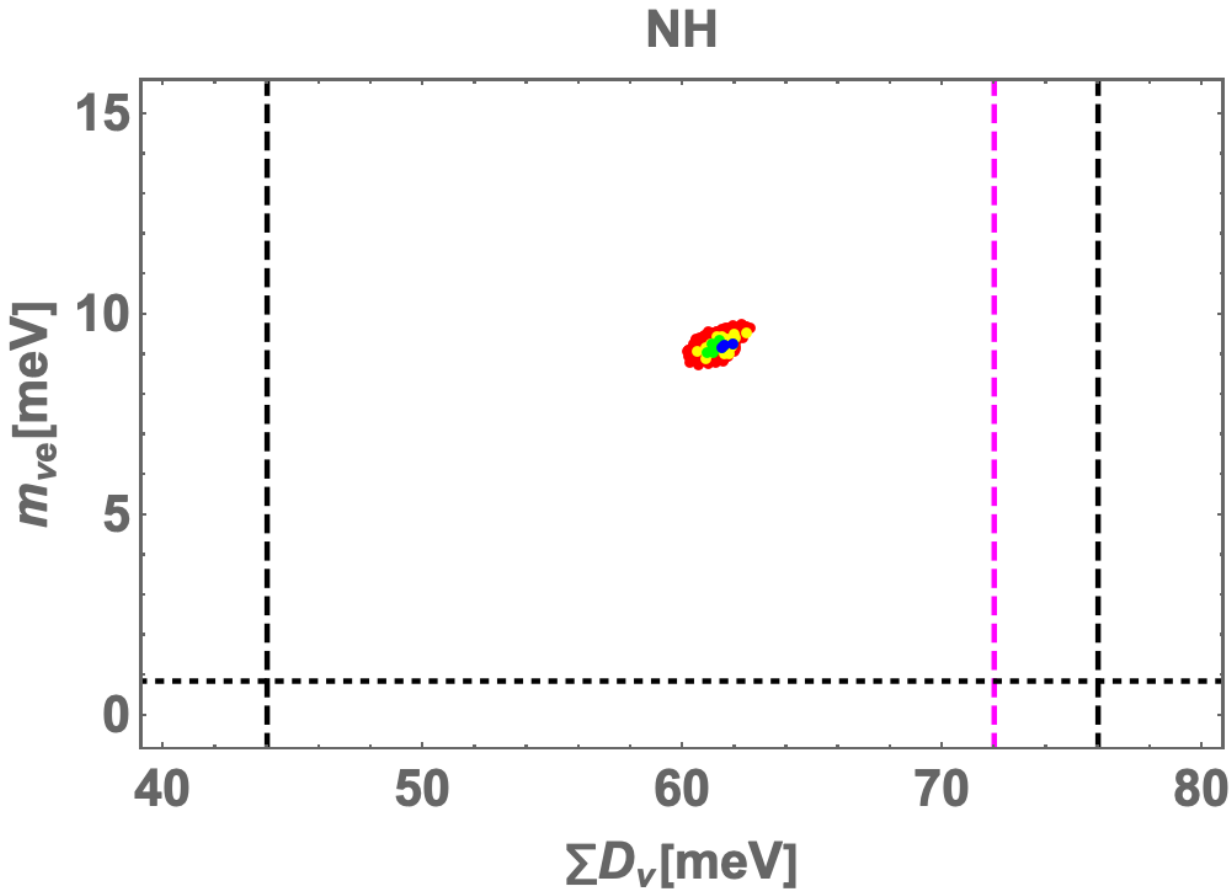}
\caption{Predicted values for neutrino observables from allowed parameter points on $\sum D_\nu$-$\delta_{\rm CP}$ (left), $\sum D_\nu$-$m_{ee}$ (center) and $\sum D_\nu$-$m_{\nu e}$ (right) plane for $\tan \beta =1$. The blue horizontal line in the center plot corresponds to the strongest future sensitivity by mEXO; it is common in the other plots below.  }
\label{fig:tb1C}
\end{center}\end{figure}

\begin{figure}[t]\begin{center}
\includegraphics[width=80mm]{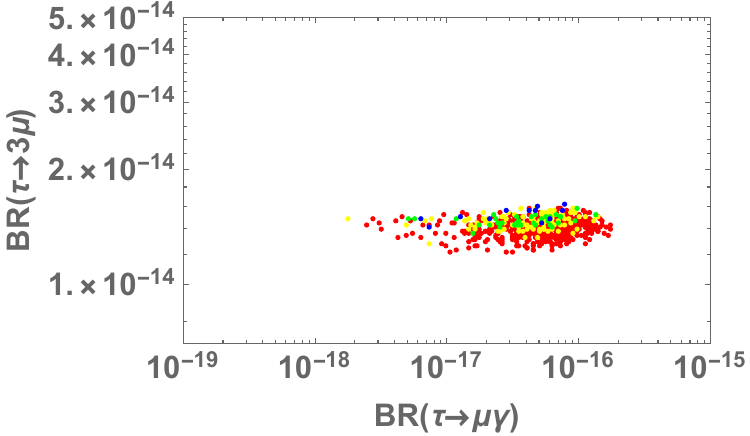}
\caption{The predicted BRs of CLFV processes $\tau \to \mu \gamma$ and $\tau \to 3 \mu$.}
\label{fig:tb1CLFV}
\end{center}\end{figure}

\noindent
(i){\bf Results of $t_\beta =1$ case} \\
Fig.~\ref{fig:tb1A} shows the predictions regarding Dirac and Majorana CP phases where each point corresponds to one allowed parameter sets and its color indicates an estimated $\chi^2$ value providing $0$-$1 \sigma$ (blue), $1$-$2 \sigma$ (green), $2$-$3 \sigma$ (yellow) and $3$-$5 \sigma$ (red) deviations; indications of color of points are common in the figures afterwards.
The left, center and right plots in Fig.~\ref{fig:tb1A} represent correlations on $\alpha_{21}$-$\delta_{\rm CP}$, $\alpha_{31}$-$\delta_{\rm CP}$ and $\alpha_{21}$-$\alpha_{31}$ plane, respectively. 
In the left and center plots the gray region indicate the value of $\delta_{\rm CP}$ is deviated more than 3$\sigma$ in NuFit 6.1 while the horizontal black line represents the best fit value of $\delta_{\rm CP}$.
We find that the value of $\delta_{CP}$ is preferred to be within 0-50 [deg] and 300-360 [deg] region.
On the other hand, the Majorana phase $\alpha_{21(31)}$ is predicted to be within around 150-205 [deg] (0-60 and 300-360 [deg]).

Fig.~\ref{fig:tb1B} shows our predictions on $\sum D_\nu$-$s^2_{12}$ (left), $\sum D_\nu$-$s^2_{13}$ (center) and $\sum D_\nu$-$s^2_{23}$ (right) planes.
The sum of neutrino masses is concentrated around 60-62 meV that is below the current cosmological limit.
For mixing angles, we find points in most of the region within $3 \sigma$ level. The region between horizontal black(magenta) lines corresponding to the $1\sigma (3 \sigma)$ range. The vertical magenta dashed line indicate the upper limit of $\sum D_\nu$ by Planck+DESI data while the region between vertical black dashed lines can be tested by future CMB observations~\cite{Abazajian:2019eic,Matsumura:2013aja,SajjadAthar:2021prg}.
The lines are common in the plots below showing mixing angles and $\sum m_\nu$.

Fig.~\ref{fig:tb1C} shows the predicted region on $\sum D_\nu$-$\delta_{\rm CP}$ (left), $\sum D_\nu$-$m_{ee}$ (center) and $\sum D_\nu$-$m_{\nu_e}$ (right) planes.
We find the range of $m_{ee}$ is $m_{ee} \lesssim 5.5$ meV.
The region for $m_{\nu_e}$ is localized around $m_{\nu_e} \sim 9$-$10$ meV. The blue horizontal line in the center plot corresponds to the strongest future sensitivity by mEXO; it is common in the other plots below.  .

Fig.~\ref{fig:tb1CLFV} shows our prediction regarding CLFV BRs where we show $BR(\tau \to \mu \gamma)$ and $BR(\tau \to 3 \mu)$ since the other BRs are forbidden or highly suppressed in the model due to the structure of the Yukawa matrices. We find that the value of $BR(\tau \to 3 \mu)$ is concentrated around $1.5 \times 10^{-14}$ while $BR(\tau \to \mu \gamma)$ has the range around $[2 \times 10^{-18}, 2 \times 10^{-16}]$. These values of BRs are sufficiently below the current bounds, and we do not find clear correlation for them.

\begin{figure}[t]\begin{center}
\includegraphics[width=50mm]{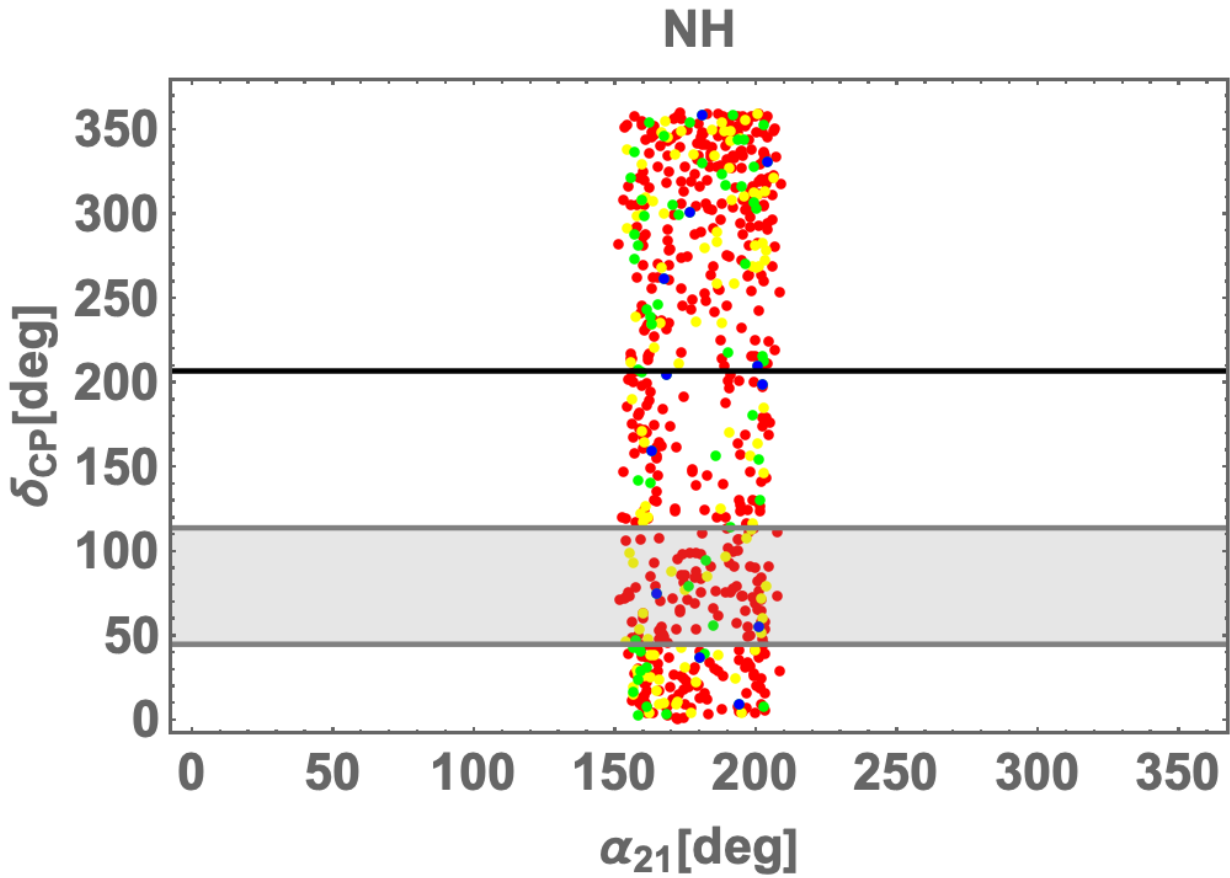} \quad
\includegraphics[width=50mm]{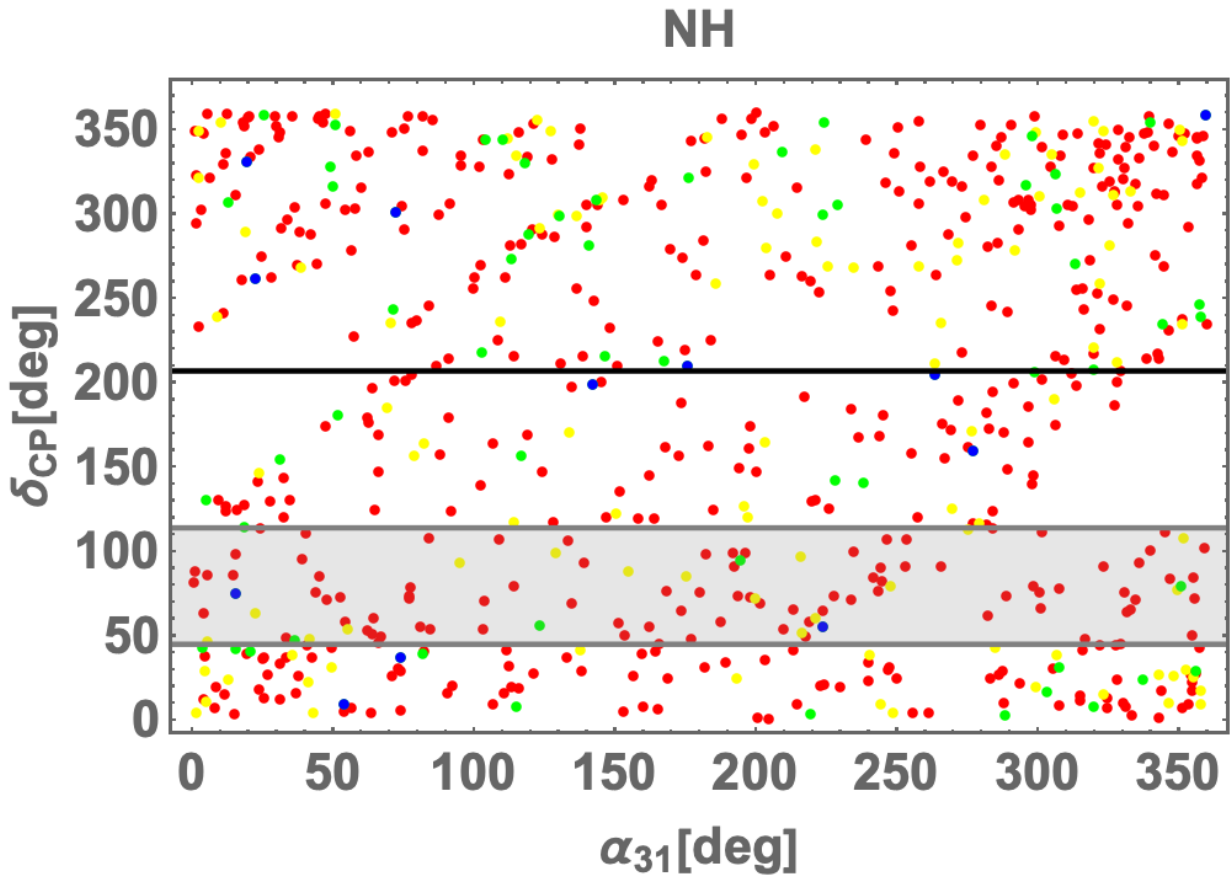} \quad
\includegraphics[width=50mm]{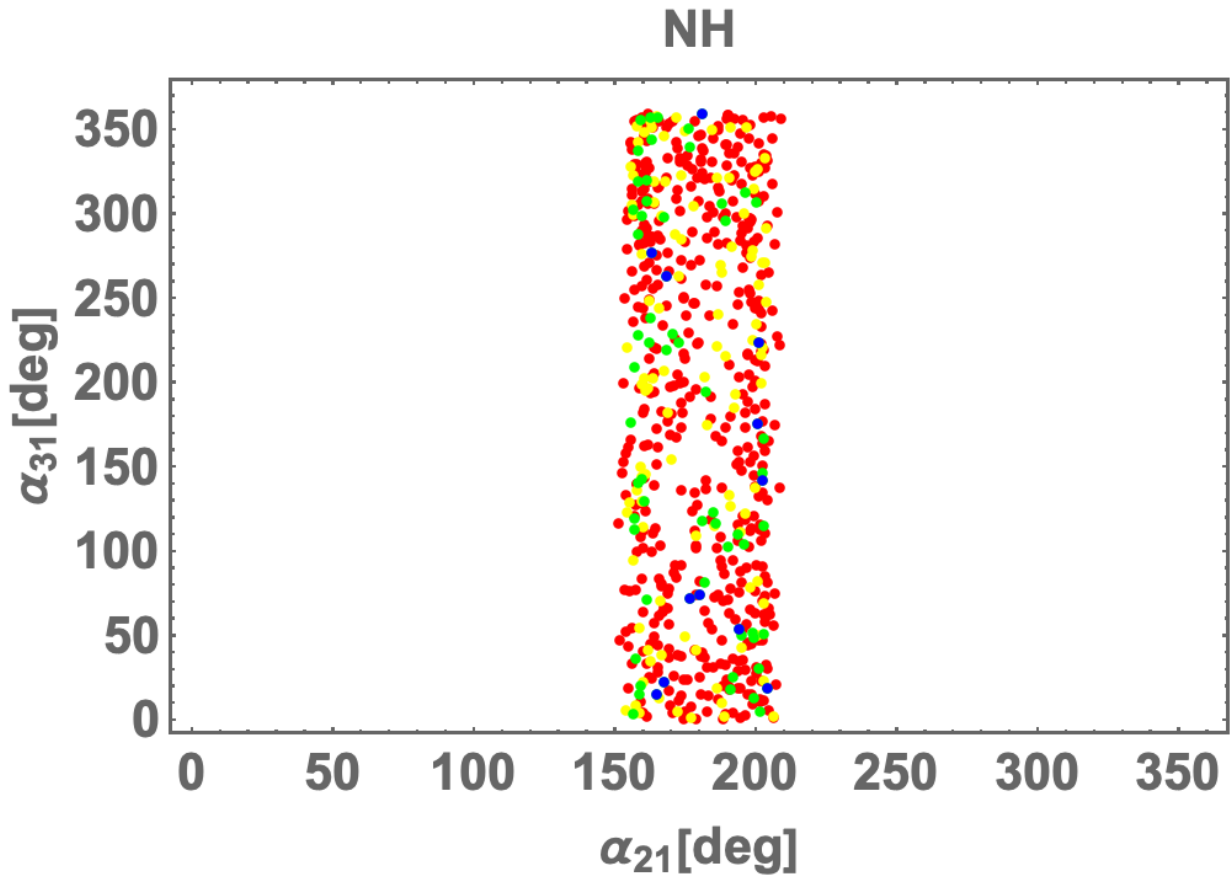}
\caption{Predicted values for neutrino observables from allowed parameter points on $\alpha_{21}$-$\delta_{\rm CP}$ (left), $\alpha_{31}$-$\delta_{\rm CP}$ (center) and $\alpha_{21}$-$\alpha_{31}$ (right) plane for $\tan \beta =100$.    }
\label{fig:tb100A}
\end{center}\end{figure}

\begin{figure}[t]\begin{center}
\includegraphics[width=50mm]{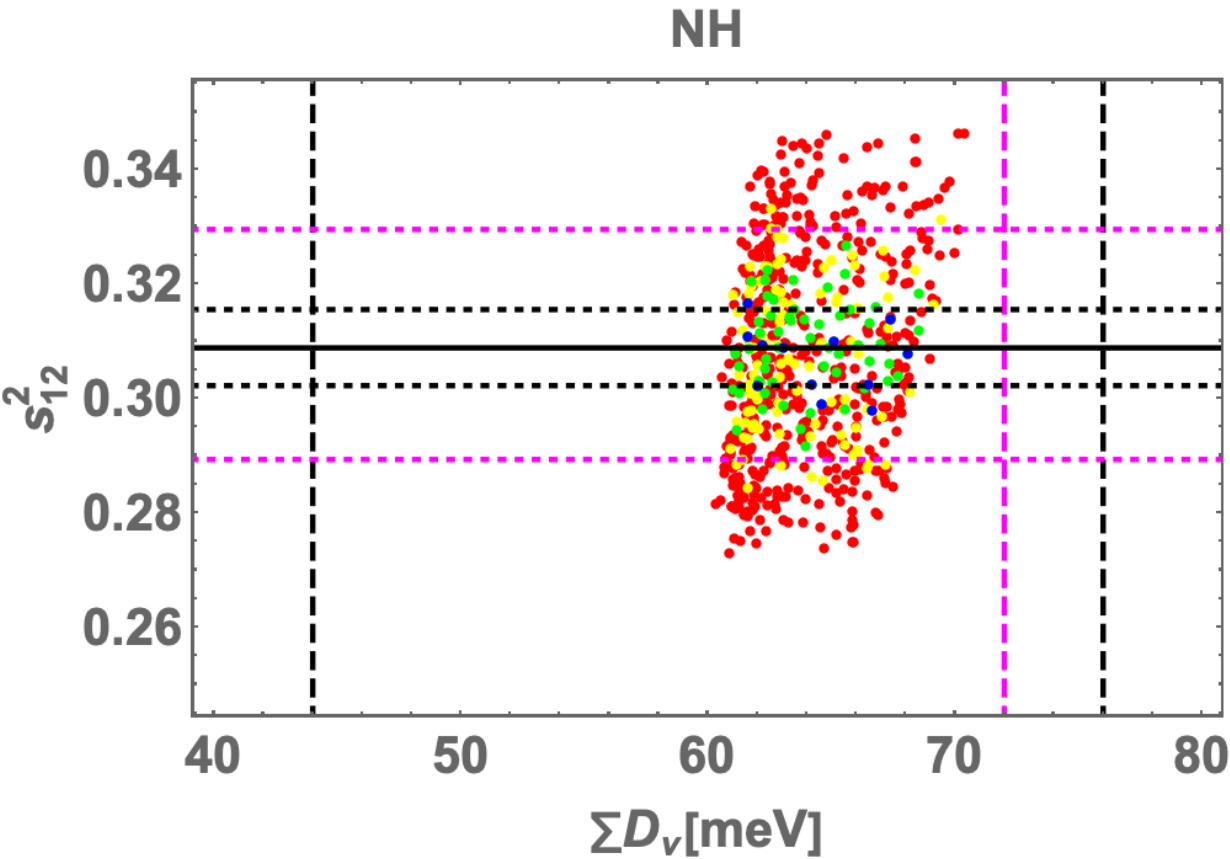} \quad
\includegraphics[width=50mm]{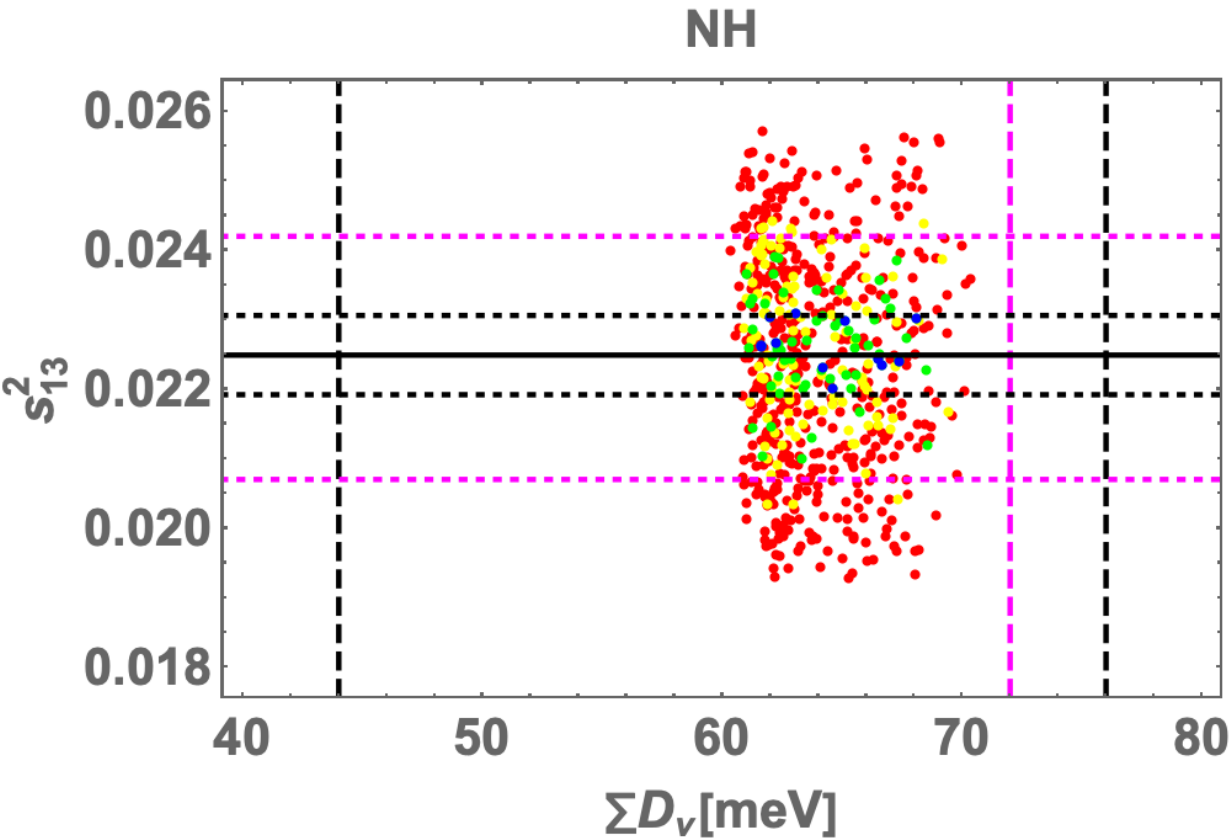} \quad
\includegraphics[width=50mm]{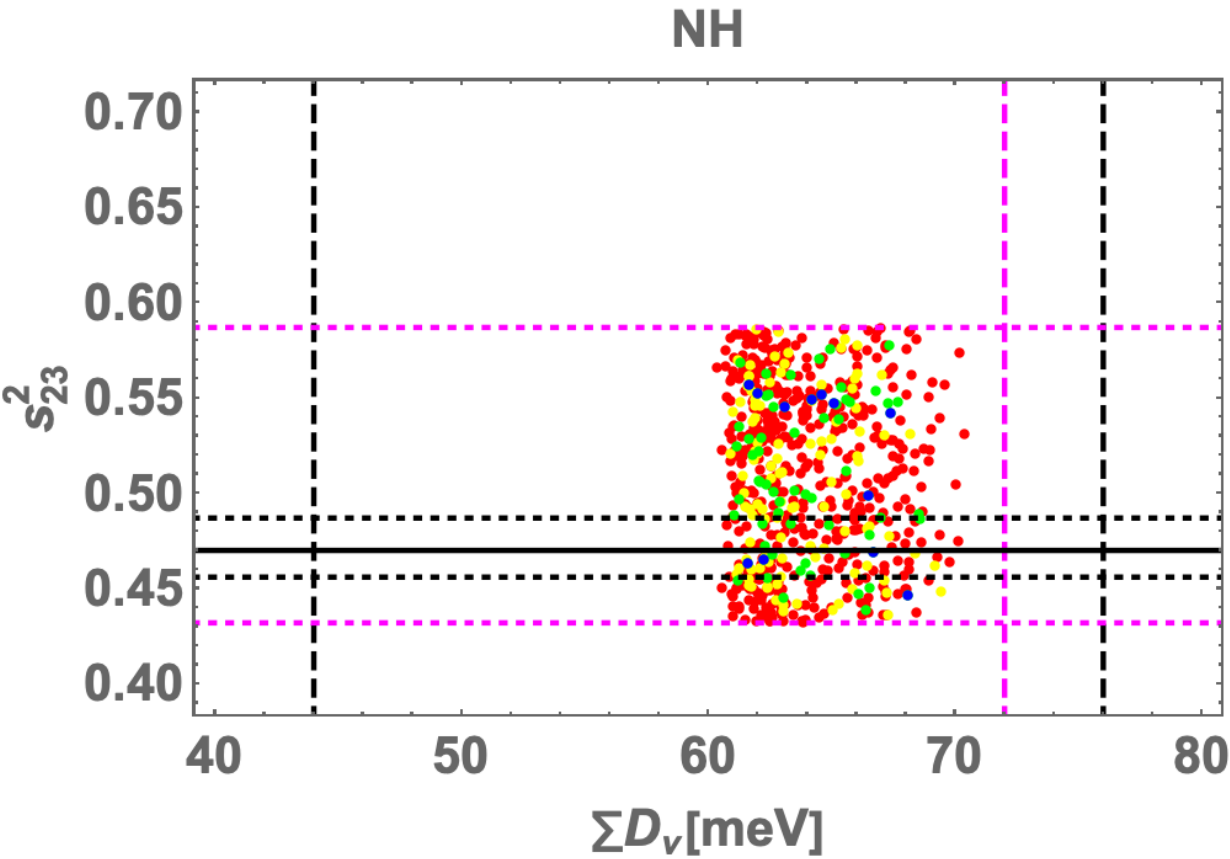}
\caption{Predicted values for neutrino observables from allowed parameter points on $\sum D_\nu$-$s^2_{12}$ (left), $\sum D_\nu$-$s^2_{13}$ (center) and $\sum D_\nu$-$s^2_{23}$ (right) plane for $\tan \beta =100$.   }
\label{fig:tb100B}
\end{center}\end{figure}

\begin{figure}[t]\begin{center}
\includegraphics[width=50mm]{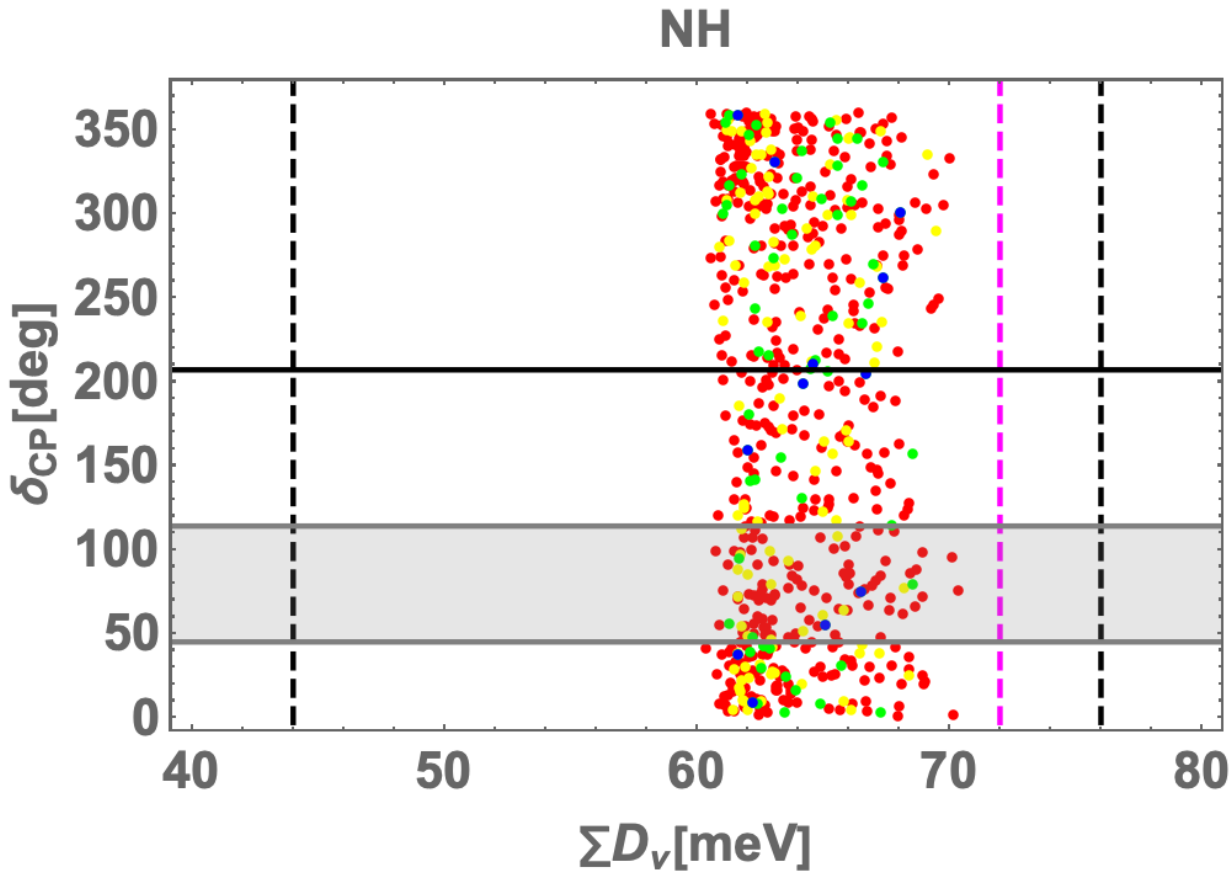} \quad
\includegraphics[width=50mm]{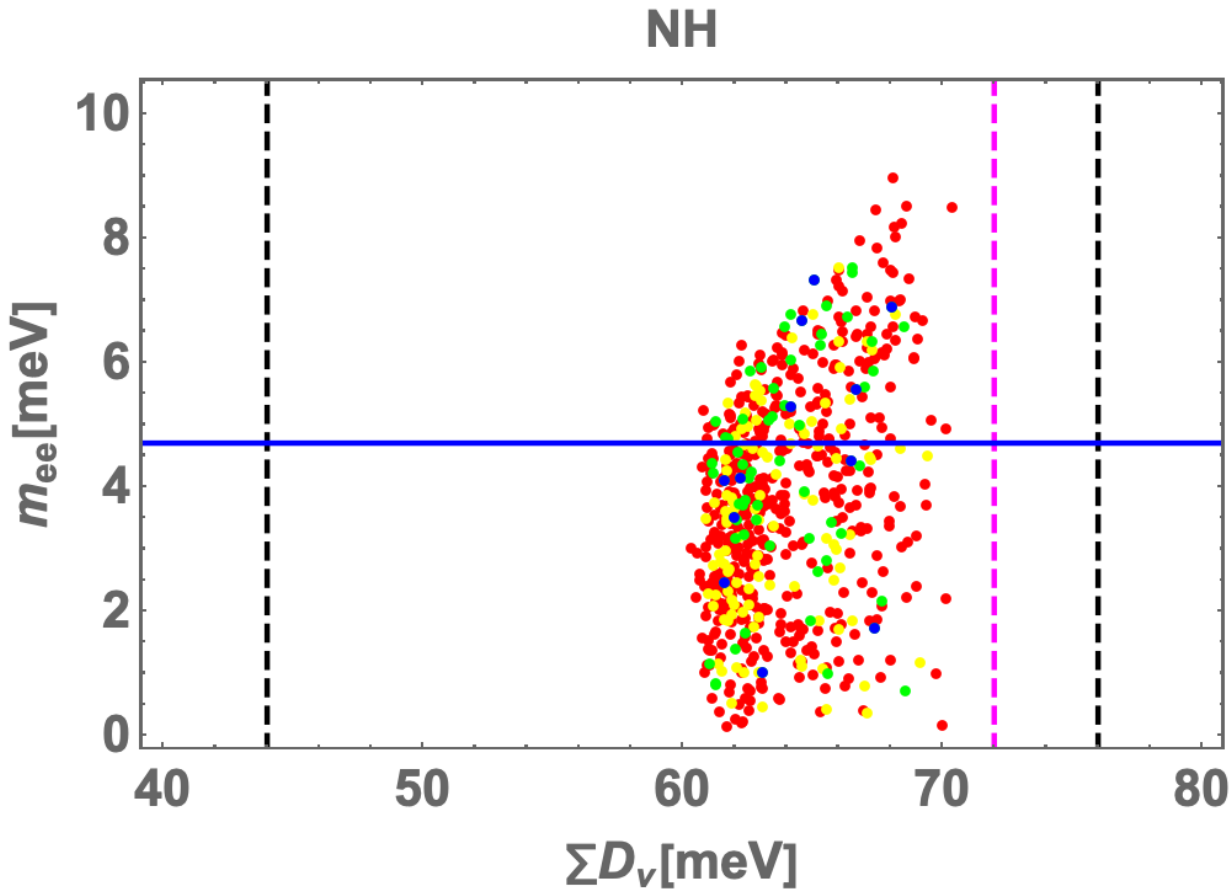} \quad
\includegraphics[width=50mm]{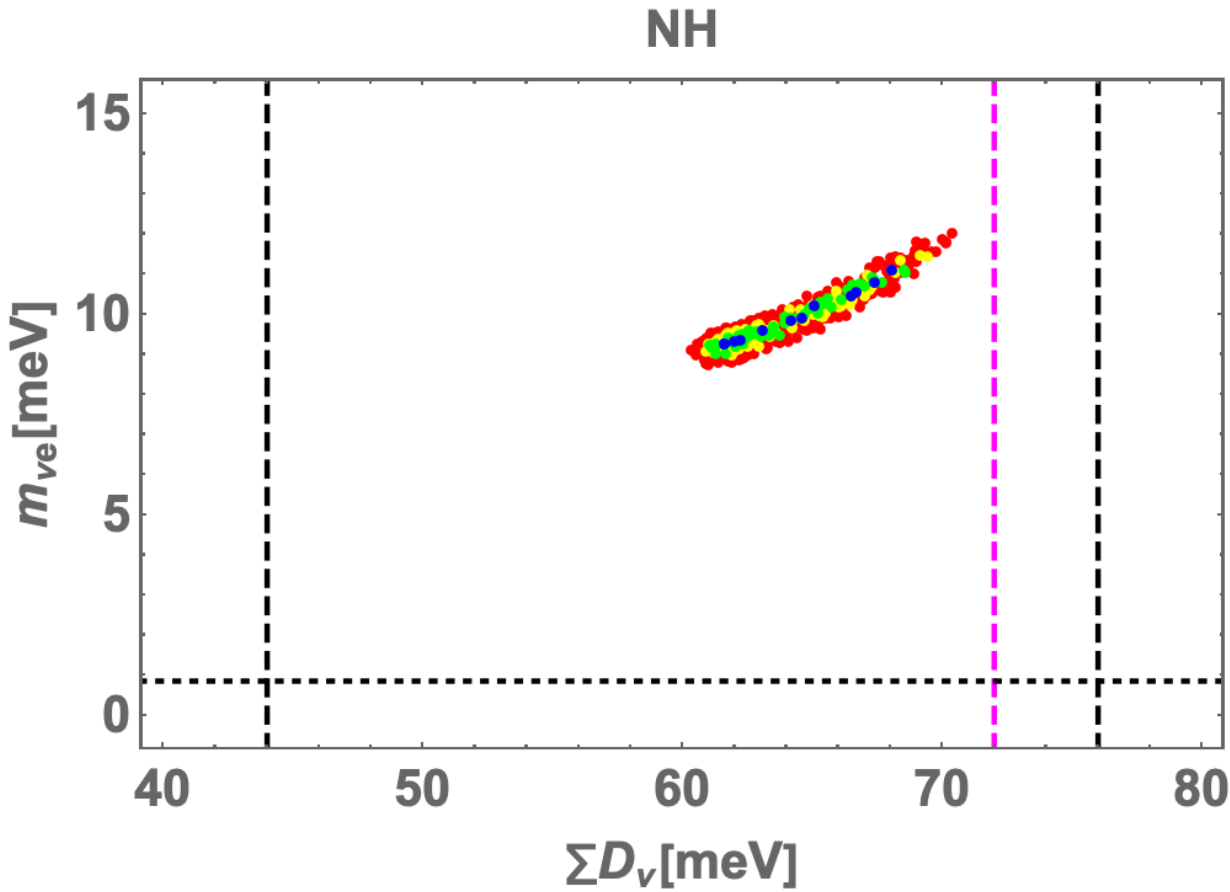}
\caption{Predicted values for neutrino observables from allowed parameter points on $\sum D_\nu$-$\delta_{\rm CP}$ (left), $\sum D_\nu$-$m_{ee}$ (center) and $\sum D_\nu$-$m_{\nu e}$ (right) plane for $\tan \beta =100$.   }
\label{fig:tb100C}
\end{center}\end{figure}

\begin{figure}[t]\begin{center}
\includegraphics[width=80mm]{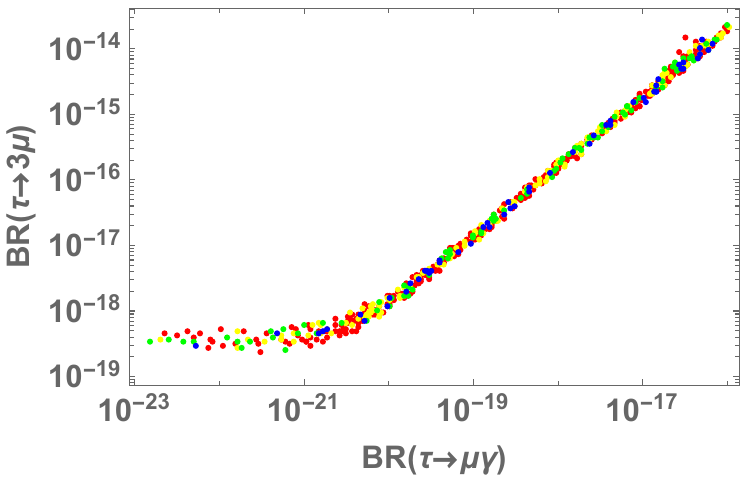}
\caption{The predicted BRs of CLFV processes $\tau \to \mu \gamma$ and $\tau \to 3 \mu$.}
\label{fig:tb100CLFV}
\end{center}\end{figure}

\noindent
(ii){\bf Results of $t_\beta =100$ case} \\
Fig.~\ref{fig:tb100A} shows the predictions regarding Dirac and Majorana CP phases where 
the left, center and right plots in Fig.~\ref{fig:tb100A} represents correlations on $\alpha_{21}$-$\delta_{\rm CP}$, $\alpha_{31}$-$\delta_{\rm CP}$ and $\alpha_{21}$-$\alpha_{31}$ plane, respectively. 
We find that the value of $\alpha_{21}$ is restricted within around 150-200 [deg].
On the other hand, the other phases $\delta_{CP}$ and $\alpha_{32}$ can be any value.
We also do not find any correlation among the phases.

Fig.~\ref{fig:tb100B} shows our predictions on $\sum D_\nu$-$s^2_{12}$ (left), $\sum D_\nu$-$s^2_{13}$ (center) and $\sum D_\nu$-$s^2_{23}$ (right) planes.
The predicted range of sum of neutrino masses is  around 60-69 meV that is below the current cosmological limit.
For mixing angles, we find points in most of the region within $3 \sigma$ level.

Fig.~\ref{fig:tb100C} shows the predicted region on $\sum D_\nu$-$\delta_{\rm CP}$ (left), $\sum D_\nu$-$m_{ee}$ (center) and $\sum D_\nu$-$m_{\nu_e}$ (right) planes.
We find the range of $m_{ee}$ is $m_{ee} \lesssim 9$ meV and a little correlation between $\sum D_\nu$ and $m_{ee}$.
The region for $m_{\nu_e}$ is within around $8$-$12$ meV where we find some correlation between $\sum D_\nu$ and $m_{\nu_e}$.

Fig.~\ref{fig:tb100CLFV} shows our prediction regarding CLFV BRs for $\tau \to \mu \gamma$ and $BR(\tau \to 3 \mu)$. We find the range of these BRs as around $BR(\tau \to \mu \gamma) \in [10^{-23}, 10^{-16}]$ and $BR(\tau \to 3 \mu) \in [10^{-19}, 10^{-14}]$.
In addition, we find clear correlation between these BRs.

\begin{figure}[t]\begin{center}
\includegraphics[width=50mm]{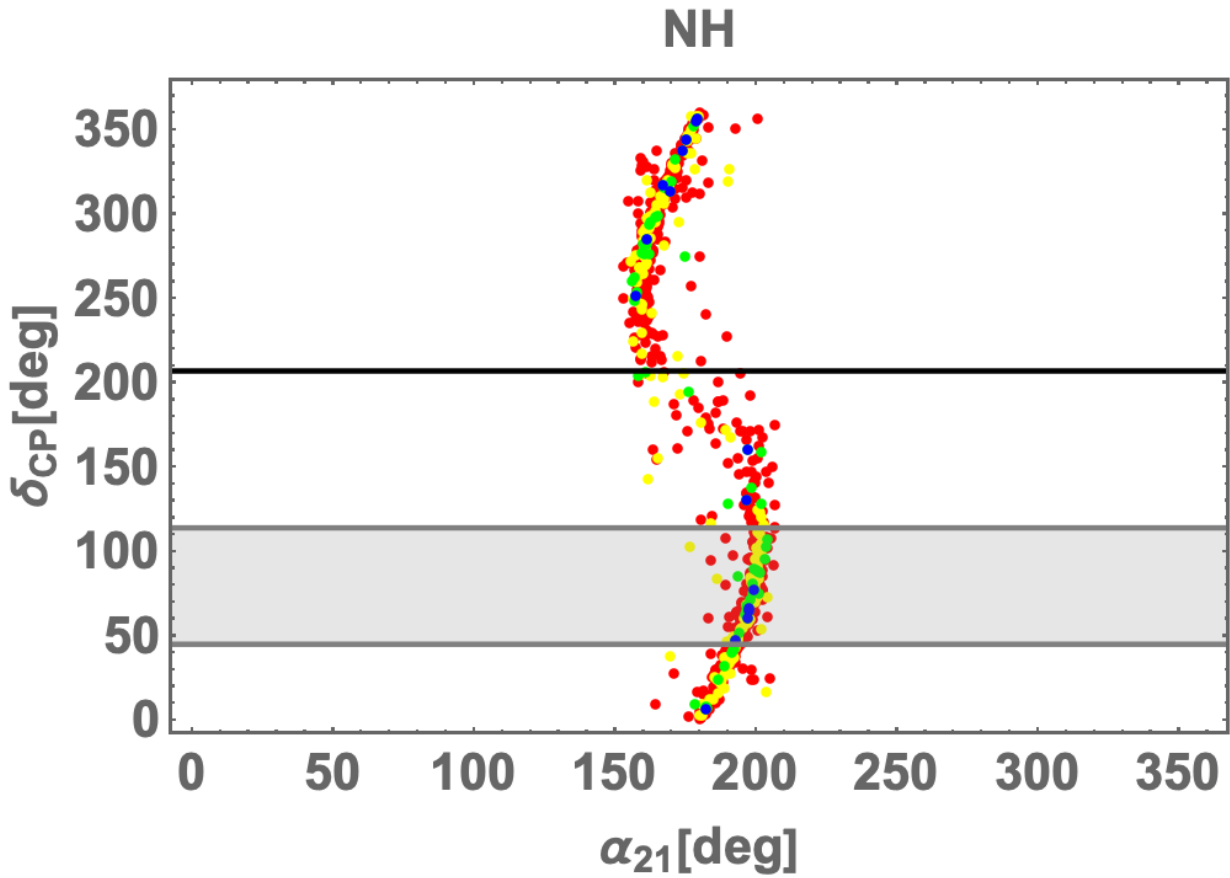} \quad
\includegraphics[width=50mm]{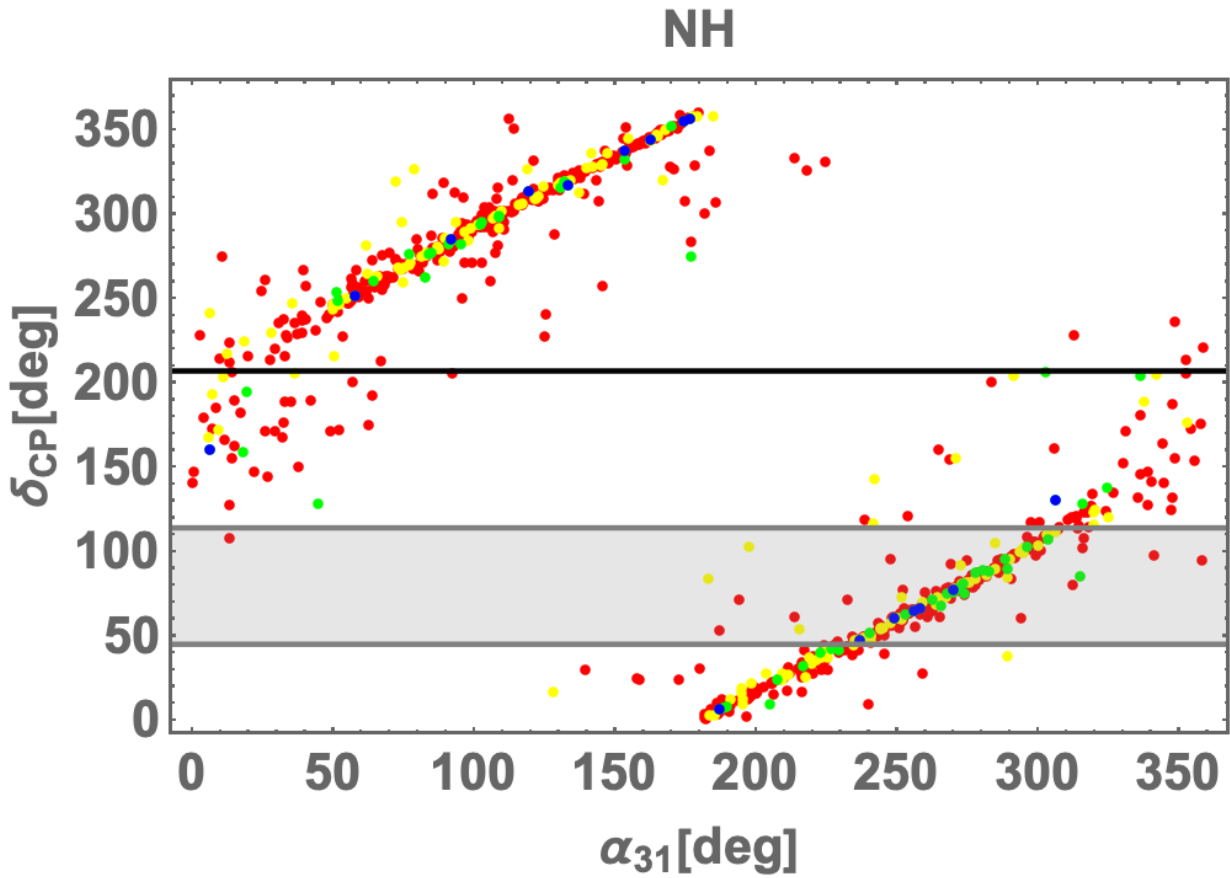} \quad
\includegraphics[width=50mm]{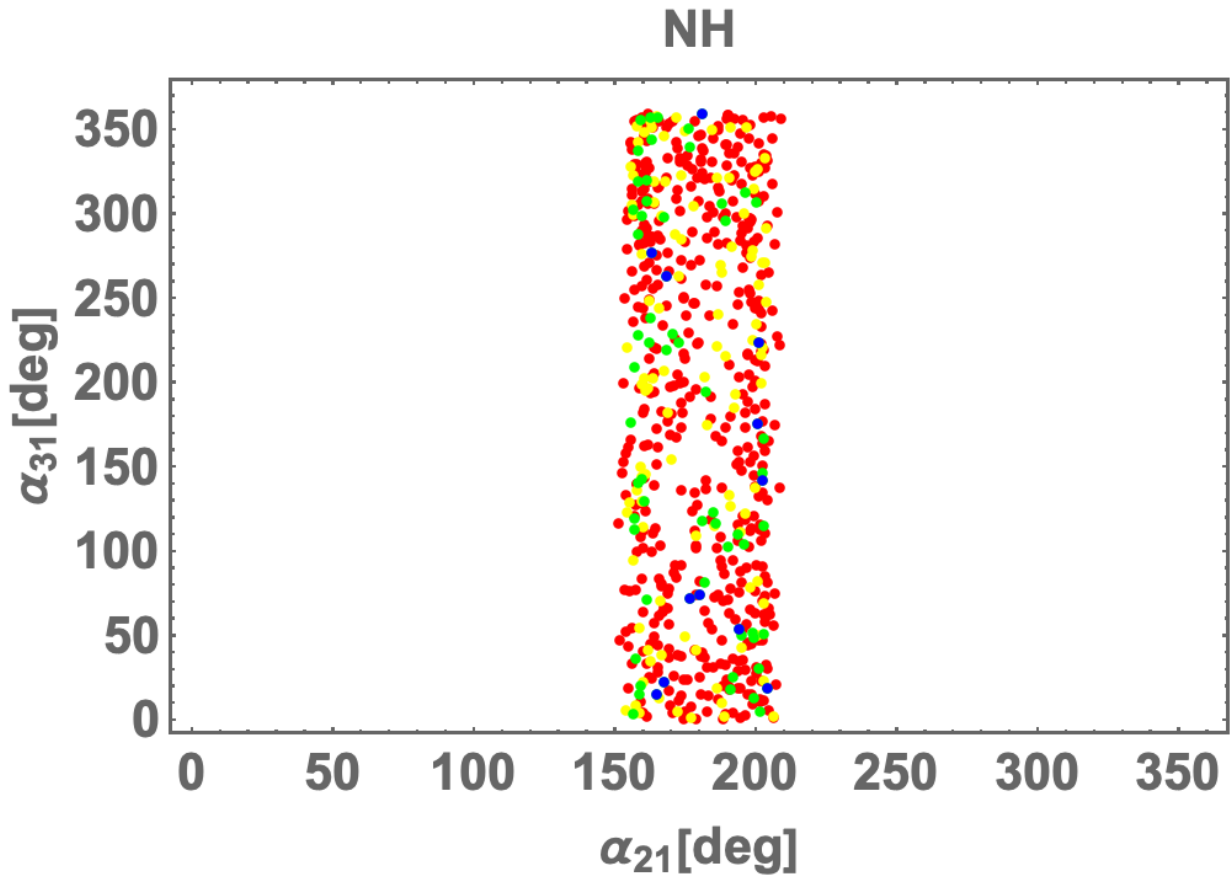}
\caption{Predicted values for neutrino observables from allowed parameter points on $\alpha_{21}$-$\delta_{\rm CP}$ (left), $\alpha_{31}$-$\delta_{\rm CP}$ (center) and $\alpha_{21}$-$\alpha_{31}$ (right) plane for $\tan \beta =10000$.    }
\label{fig:tb10000A}
\end{center}\end{figure}

\begin{figure}[t]\begin{center}
\includegraphics[width=50mm]{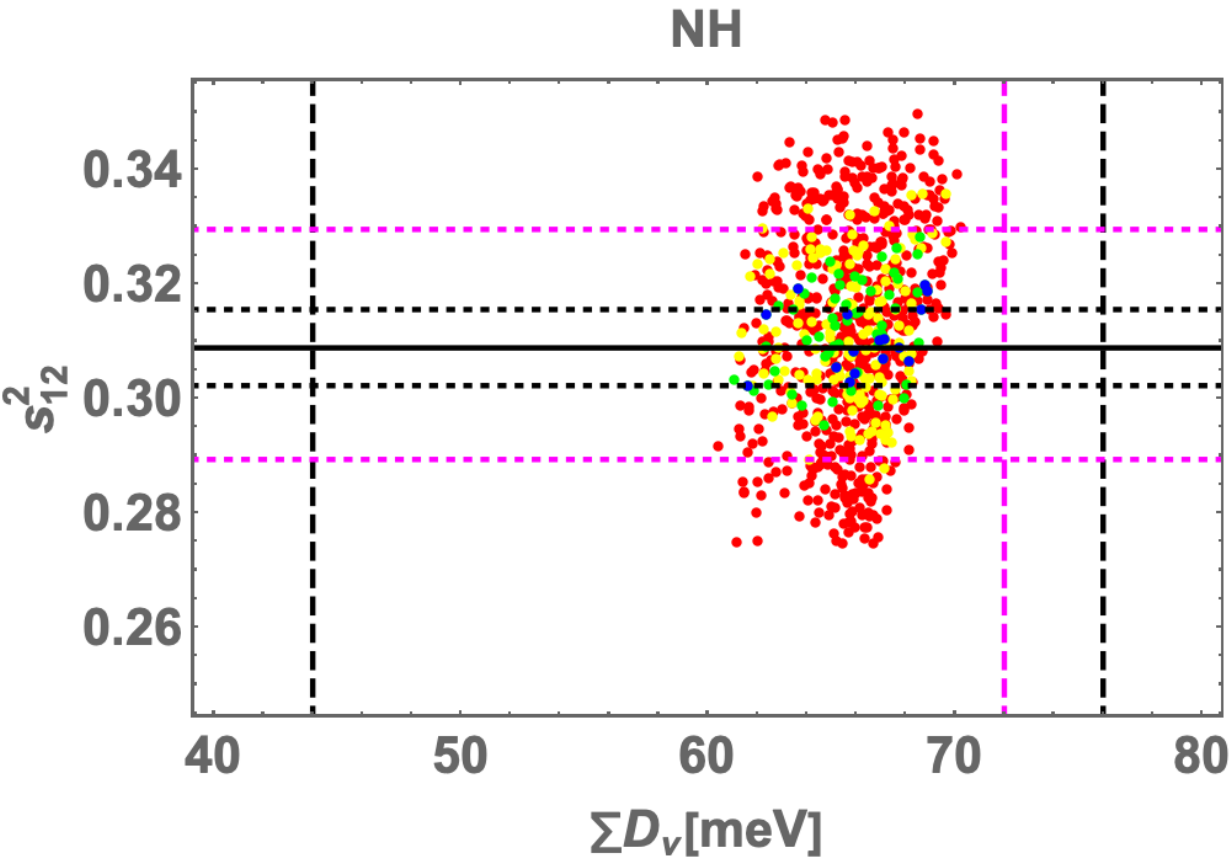} \quad
\includegraphics[width=50mm]{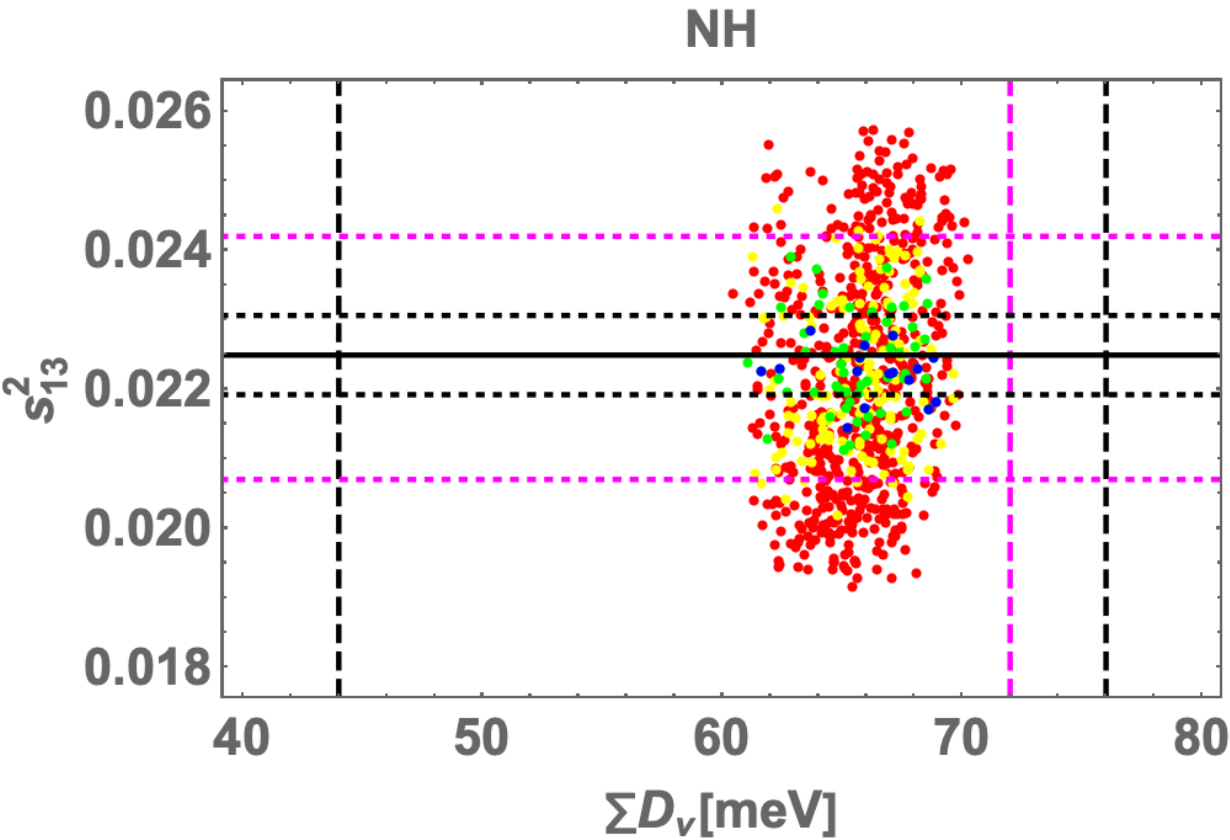} \quad
\includegraphics[width=50mm]{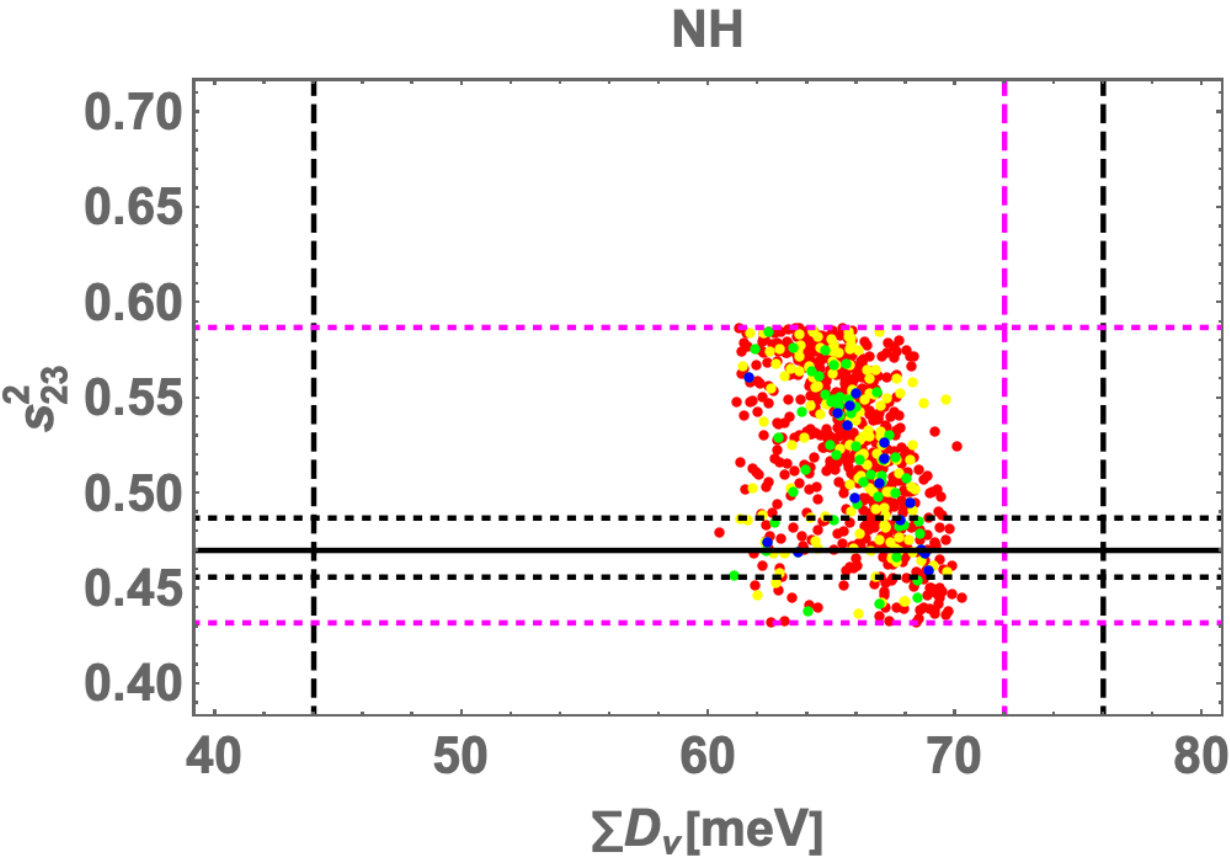}
\caption{Predicted values for neutrino observables from allowed parameter points on $\sum D_\nu$-$s^2_{12}$ (left), $\sum D_\nu$-$s^2_{13}$ (center) and $\sum D_\nu$-$s^2_{23}$ (right) plane for $\tan \beta =10000$.   }
\label{fig:tb10000B}
\end{center}\end{figure}

\begin{figure}[t]\begin{center}
\includegraphics[width=50mm]{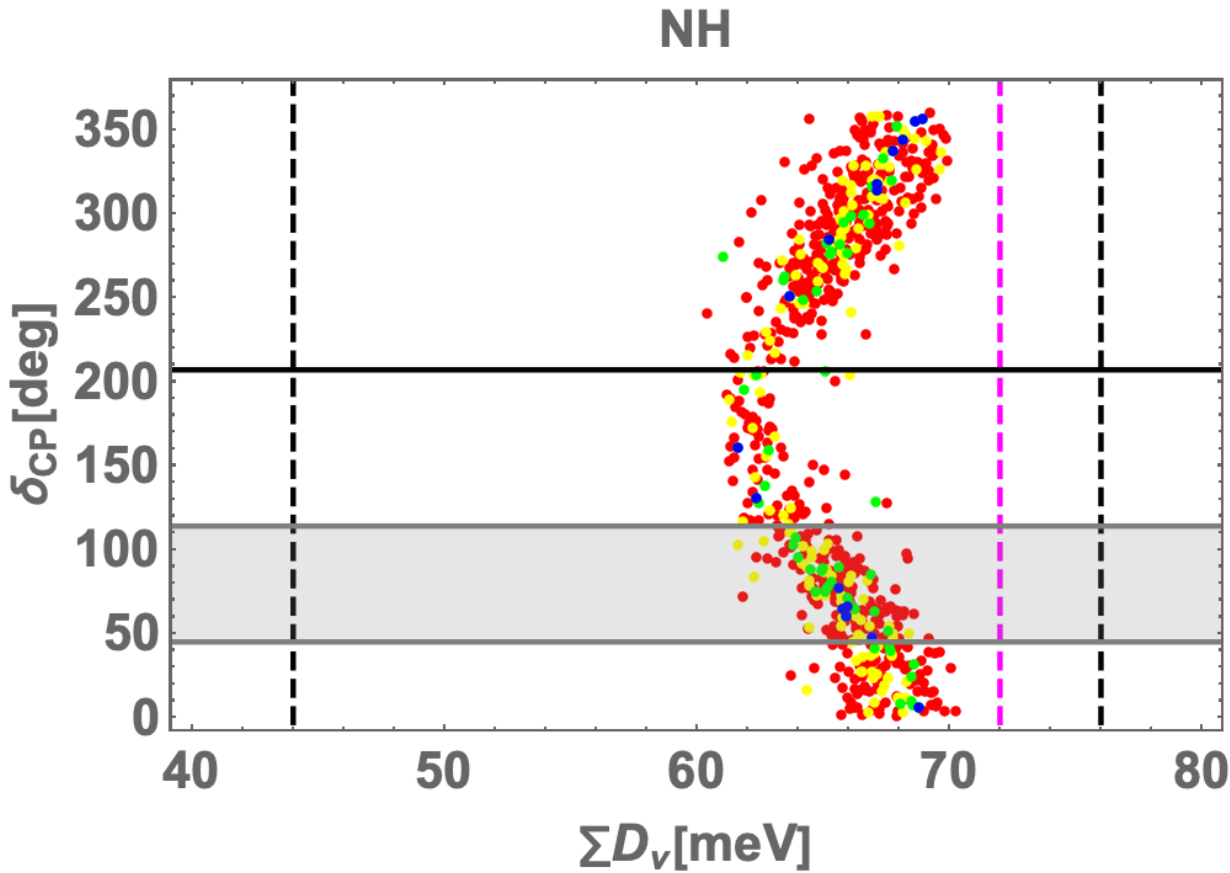} \quad
\includegraphics[width=50mm]{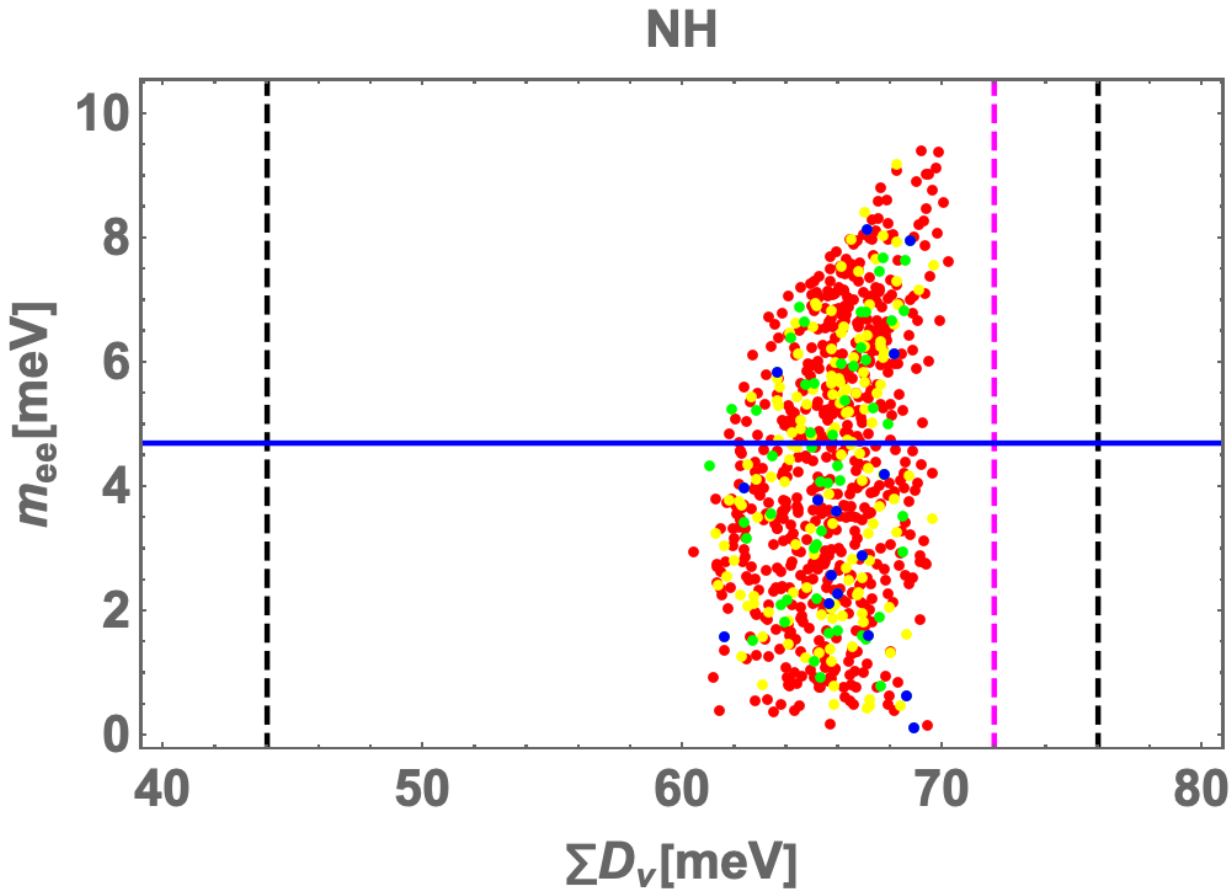} \quad
\includegraphics[width=50mm]{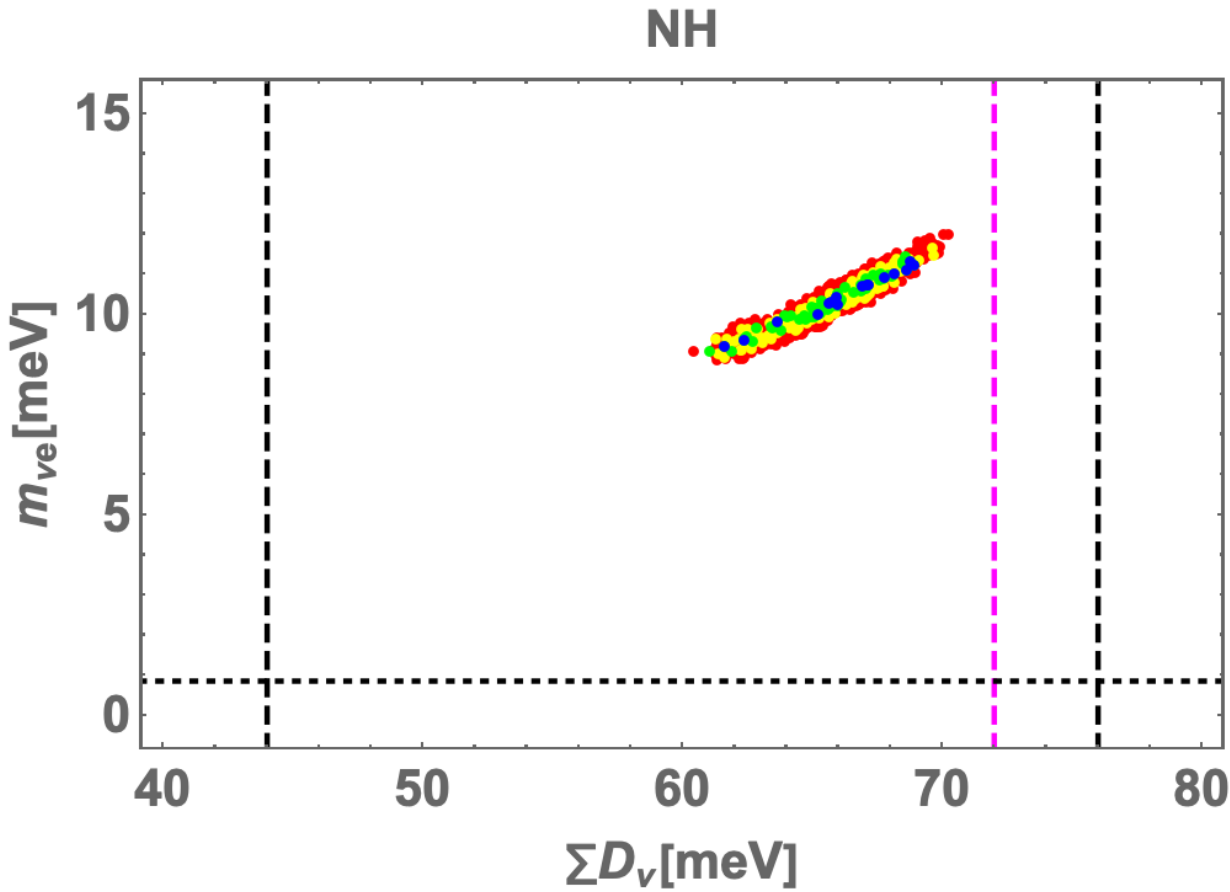}
\caption{Predicted values for neutrino observables from allowed parameter points on $\sum D_\nu$-$\delta_{\rm CP}$ (left), $\sum D_\nu$-$m_{ee}$ (center) and $\sum D_\nu$-$m_{\nu e}$ (right) plane for $\tan \beta =10000$.   }
\label{fig:tb10000C}
\end{center}\end{figure}

\begin{figure}[t]\begin{center}
\includegraphics[width=80mm]{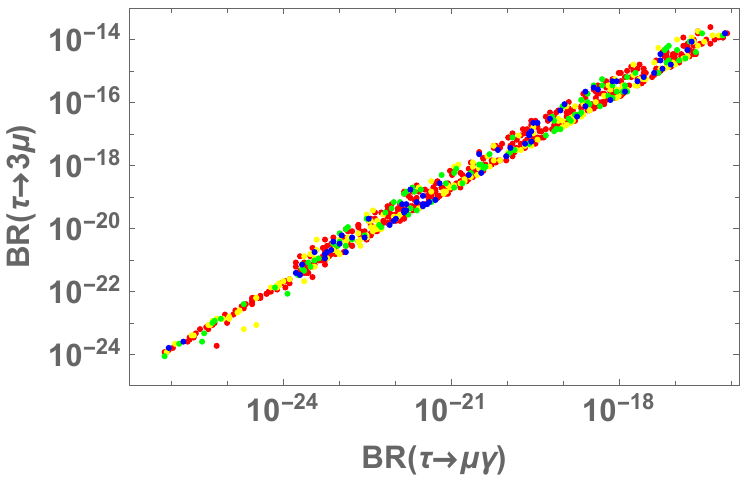}
\caption{The predicted BRs of CLFV processes $\tau \to \mu \gamma$ and $\tau \to 3 \mu$.}
\label{fig:tb10000CLFV}
\end{center}\end{figure}

\noindent
(iii){\bf Results of $t_\beta =10000$ case} \\
Fig.~\ref{fig:tb10000A} shows the predictions regarding Dirac and Majorana CP phases where 
the left, center and right plots in Fig.~\ref{fig:tb10000A} represents correlations on $\alpha_{21}$-$\delta_{\rm CP}$, $\alpha_{31}$-$\delta_{\rm CP}$ and $\alpha_{21}$-$\alpha_{31}$ plane, respectively. 
We find that the value of $\alpha_{21}$ is restricted within around 150-200 [deg] while $\delta_{CP}$ and $\alpha_{31}$ can be any values.
We also find some correlations between $\delta_{\rm CP}$ and $\alpha_{21(31)}$ as shown in the left(center) plot while there is no correlation between $\alpha_{21}$ and $\alpha_{31}$. 

Fig.~\ref{fig:tb10000B} shows our predictions on $\sum D_\nu$-$s^2_{12}$ (left), $\sum D_\nu$-$s^2_{13}$ (center) and $\sum D_\nu$-$s^2_{23}$ (right) planes.
The predicted range of sum of neutrino masses is  around 60-70 meV that is below the current cosmological limit, and we find points in most of the region within $3 \sigma$ level for mixing angles.
The results are similar to $t_\beta = 100$ case.

Fig.~\ref{fig:tb10000C} shows the predicted region on $\sum D_\nu$-$\delta_{\rm CP}$ (left), $\sum D_\nu$-$m_{ee}$ (center) and $\sum D_\nu$-$m_{\nu_e}$ (right) planes.
We find the predicted range of these values similar to the case of $t_\beta =100$; $m_{ee} \lesssim 9$ meV and $m_{\nu_e}$ is within around $8$-$12$ meV.
On the other hand, we have some correlations between $\sum D_\nu$-$\delta_{\rm CP}$ and  $\sum D_\nu$-$m_{\nu e}$ in this case in contrast to the $t_\beta = 100$ case.

Fig.~\ref{fig:tb10000CLFV} shows our prediction regarding CLFV BRs for $\tau \to \mu \gamma$ and $BR(\tau \to 3 \mu)$. We find the range of these BRs as around $BR(\tau \to \mu \gamma) \in [10^{-26}, 10^{-16}]$ and $BR(\tau \to 3 \mu) \in [10^{-24}, 10^{-14}]$.
In addition, we find clear correlation between these BRs.

\noindent
{\it Qualitative characteristic of $\tan \beta$ dependence}: 
Here we discuss the dependence of our predictions on $\tan \beta$. 
For $\tan \beta = 1$ case, $y^\Phi$ and $y^\ell$ equally contribute to $M_\ell$ and $Y$. In this case, charged lepton mass value strongly affects neutrino mass matrix and the restriction to neutrino observables are stronger than other cases as we see in Figs.~\ref{fig:tb1A}-\ref{fig:tb1C}. 
For $\tan \beta = 100$ case, the larger $\tan \beta$ value mildly suppresses $y^\Phi (y^\ell)$ contribution to $M_\ell (Y)$. In this case, neutrino observables are less restricted and we don't have predictions for CP phases as shown in Figs.~\ref{fig:tb100A}-\ref{fig:tb100C}.
For $\tan \beta = 10000$ case, $M_\ell$ becomes almost diagonal and the neutrino mass matrix is close to two-zero texture one. We then have some correlations in neutrino observables such as CP phases and masses as shown in Figs.~\ref{fig:tb10000A}-\ref{fig:tb10000C}. 

Furthermore, the model would be tested by exploring signals of heavy scalar productions.
In particular, exotic neutral scalars can decay violating lepton flavor as $H/A \to \ell^+ \ell'^-$.
These signals can be explored by the current and future experimental energy that can be realized by the LHC, FCC~\cite{FCC:2018byv,FCC:2025lpp}, CEPC~\cite{CEPCStudyGroup:2023quu,Ai:2025cpj,CEPCStudyGroup:2025kmw} and muon colliders~\cite{Accettura:2023ked,Hamada:2022mua}.
Detailed study of collider signals is beyond the scope of this paper and left as future works.

\section{Summary and discussion}

In this work, we have investigated Zee models incorporating with non-invertible symmetry $Z_M^{NI}$.
The introduction of  $Z_M^{NI}$ provides novel non-invertible selection rules that leads to specific Yukawa coupling structures, which are unattainable through traditional group-symmetric approaches.
We have systematically classified these models in terms of assignments of classes to fields under $Z_M^{NI}$ for $M=4$ to $M=7$.
By scrutinizing the resulting neutrino mass matrices, we have identified a set of 
viable models  that  are consistent with the current neutrino oscillation data.

Following this classification, we have explored a benchmark model based on $Z_7^{NI}$ symmetry.
We have found that the neutrino mass matrix exhibits either one-zero or two-zero texture, depending on the value of $\tan \beta$.
Our numerical results demonstrate that predictions for neutrino observables and charged lepton flavor violating  processes, such as $\tau\to 3\mu$, are highly sensitive to the $\tan\beta$ parameter.
%
The analysis of the benchmark model illustrates the unique phenomenological features of Zee model under $Z_M^{NI}$.
Given the distinctive signatures identified in this study, we expect these models to be rigorously tested by upcoming high-precision neutrino experiments and searches for lepton flavor violations.

\section*{Acknowledgments}
The project is supported by the Fundamental Research Funds for the Central Universities (T.~N.) 
and Zhongyuan Talent (Talent Recruitment Series) Foreign Experts Project (H.~O.).

\appendix

\section{Amplitudes for CLFVs} \label{sec:appendix-CLFV}

Here we summarize the formulas for decay amplitudes of the CLFV decays $\ell_i \to \ell_j \gamma$.
The amplitude $A_{L(R)}$ in Eq.~\eqref{eq:BRCLFV} is written as the sum of contributions associated with different scalar bosons, $\{h,H,A,H^\pm_{1,2} \}$, such that
\begin{equation}
A_{L(R)} = \sum_{\phi^0 =h,H,A} A^{\phi^0}_{L(R)} + A^{H_1^\pm}_{L(R)}+ A^{H_2^\pm}_{L(R)}. \label{eq:amp}
\end{equation}
Neutral scalar contributions in Eq.~\eqref{eq:amp} are written by
\begin{align}
\left( A^{\phi^0}_{L} \right)_{ji} = & \frac{1}{16 \pi^2 m_i} \left[ (Y^\dagger_{\phi^0})_{jk} (Y_{\phi^0})_{ki} m_i I_1 (m_i, m_j, m_k, m_{\phi^0}) + (Y_{\phi^0})_{jk} (Y^\dagger_{\phi^0})_{ki} m_j I_2 (m_i, m_j, m_k, m_{\phi^0}) \right. \nn \\
& \left. \qquad \qquad + \eta (Y^\dagger_{\phi^0})_{jk} (Y^\dagger_{\phi^0})_{ki} m_k I_3 (m_i, m_j, m_k, m_{\phi^0}) \right] \\
\left( A^{\phi^0}_{R} \right)_{ji} = & \frac{1}{16 \pi^2 m_i} \left[ (Y^\dagger_{\phi^0})_{jk} (Y_{\phi^0})_{ki} m_j I_2 (m_i, m_j, m_k, m_{\phi^0}) + (Y_{\phi^0})_{jk} (Y^\dagger_{\phi^0})_{ki} m_i I_1 (m_i, m_j, m_k, m_{\phi^0}) \right. \nn \\
& \left. \qquad \qquad + \eta (Y_{\phi^0})_{jk} (Y_{\phi^0})_{ki} m_k I_3 (m_i, m_j, m_k, m_{\phi^0}) \right], 
\end{align}
where $\eta = +1$ for $\phi^0 = h, H$ and $-1$ for $\phi^0 = A$.
\begin{align}
(A_L^{H_1^+})_{ji} &= \frac{-1}{16 \pi^2 m_i} \left[ (Y^\dagger_{H^+_{1}})_{jk} (Y_{H^+_{1}})_{ki} m_i I_4 (m_i, m_j, m_{H_1^+})  + (F^\dagger_1)_{jk} (F_1)_{ki} m_j I_5 (m_i,m_j, m_{H_1^+}) \right], \\
(A_L^{H_2^+})_{ji} &= \frac{-1}{16 \pi^2 m_i} \left[ (Y^\dagger_{H^+_{2}})_{jk} (Y_{H^+_{2}})_{ki} m_i I_4 (m_i, m_j, m_{H_2^+})  + (F^\dagger_2)_{jk} (F_2)_{ki} m_j I_5 (m_i,m_j, m_{H_2^+}) \right], \\
(A_R^{H_1^+})_{ji} &= \frac{-1}{16 \pi^2 m_i} \left[ (Y^\dagger_{H^+_{1}})_{jk} (Y_{H^+_{1}})_{ki} m_j I_5 (m_i, m_j, m_{H_1^+})  + (F^\dagger_1)_{jk} (F_1)_{ki} m_i I_4 (m_i,m_j, m_{H_1^+}) \right], \\
(A_R^{H_2^+})_{ji} &= \frac{-1}{16 \pi^2 m_i} \left[ (Y^\dagger_{H^+_{2}})_{jk} (Y_{H^+_{2}})_{ki} m_j I_5 (m_i, m_j, m_{H_2^+})  + (F^\dagger_2)_{jk} (F_2)_{ki} m_i I_4 (m_i,m_j, m_{H_2^+}) \right].
\end{align}

The loop integration factors are given by
\begin{align}
I_1(m_1,m_2,m_3,m_4) &= \int [dX] \frac{x y}{-x(1-x) m^2_1 + x z (m_1^2 - m_2^2) + (z+y) m_3^2 + x m_4^2 }, \\
I_2(m_1,m_2,m_3,m_4) &= \int [dX] \frac{x z}{-x(1-x) m^2_1 + x z (m_1^2 - m_2^2) + (z+y) m_3^2 + x m_4^2 }, \\
I_3(m_1,m_2,m_3,m_4) &= \int [dX] \frac{1-x}{-x(1-x) m^2_1 + x z (m_1^2 - m_2^2) + (z+y) m_3^2 + x m_4^2 }, \\
I_4(m_1,m_2,m_3) &= \int [dX] \frac{x y}{-x(1-x) m^2_1 + x z (m_1^2 - m_2^2) + (z+y) m_3^2}, \\
I_5(m_1,m_2,m_3) &= \int [dX] \frac{x z}{-x(1-x) m^2_1 + x z (m_1^2 - m_2^2) + (z+y) m_3^2},
\end{align}
where $\int [dX] \equiv \int_0^1 dx dy dz \delta(1-x-y-z)$.

\section{List of viable models} \label{sec:viable-models}

In this appendix, we list viable models which can in principle fit neutrino data. 
The viable models under $Z_4^{NI}$, $Z_5^{NI}$, $Z_6^{NI}$ and $Z_7^{NI}$ are summarized in Table~\ref{tab:M1}, Table~\ref{tab:M2}, Table~\ref{tab:M3}, and Tables~\ref{tab:M4}-\ref{tab:M6}, respectively.

\begin{table}[t]
\begin{tabular}{|c||c|c|c|c|c|c|c|c|}\hline\hline  
$Z_4^{NI}$ Models & $\{k_{\ell 1}, k_{\ell 2}, k_{\ell 3} \}$ & $k_S$ & $k_\Phi$ & $f$ & $y^\Phi$ & $M_\ell$ & $m_\nu$ & $m_\nu|_{t_\beta \to \infty}$ \\ \hline
(1) & $\{0, 1, 2 \}$ & $1$ & $1$ & 
$\begin{pmatrix} 0 & \times & 0 \\ \times  & 0 & \times \\ 0 & \times & 0 \end{pmatrix}$ &
$\begin{pmatrix} 0 & \times & 0 \\ \times  & 0 & \times \\ 0 & \times & 0 \end{pmatrix}$ &
$\begin{pmatrix} \times & \times & 0 \\ \times  & \times & \times \\ 0 & \times & \times \end{pmatrix}$ &
$\begin{pmatrix} \times & \times & \times \\ \times  & \times & \times \\ \times & \times & \times \end{pmatrix}$ &
$\begin{pmatrix} \times & 0 & \times \\ 0  & \times & 0 \\ \times & 0 & \times \end{pmatrix}$ \\ \hline
(2) & $\{0, 1, 2 \}$ & $2$ & $1$ & 
$\begin{pmatrix} 0 & \times & 0 \\ \times  & 0 & \times \\ 0 & \times & 0 \end{pmatrix}$ &
$\begin{pmatrix} 0 & 0 & \times \\ 0  & \times & 0 \\ \times & 0 & 0 \end{pmatrix}$ &
$\begin{pmatrix} \times & \times & 0 \\ \times  & \times & \times \\ 0 & \times & \times \end{pmatrix}$ &
$\begin{pmatrix} \times & \times & \times \\ \times  & 0 & \times \\ \times & \times & \times \end{pmatrix}$ &
$\begin{pmatrix} 0 & \times & 0 \\ \times  & 0 & \times \\ 0 & \times & 0 \end{pmatrix}$ \\ \hline
\end{tabular}
\caption{Viable assignments under $Z^{NI}_4$ and structures of Yukawa and mass matrices. The structures obtained by exchanging and/or taking permutation among $\{k_{\ell1}, k_{\ell2}, k_{\ell3}\}$ are also possible. It is the same for the tables afterwards. }\label{tab:M1}
\end{table}

\begin{table}[t]
\begin{tabular}{|c||c|c|c|c|c|c|c|c|}\hline\hline  
$Z_5^{NI}$ Models & $\{k_{\ell 1}, k_{\ell 2}, k_{\ell 3} \}$ & $k_S$ & $k_\Phi$ & $f$ & $y^\Phi$ & $M_\ell$ & $m_\nu$ & $m_\nu|_{t_\beta \to \infty}$ \\ \hline
(1) & $\{0, 1, 2 \}$ & $1$ & $1$ & 
$\begin{pmatrix} 0 & \times & 0 \\ \times  & 0 & \times \\ 0 & \times & 0 \end{pmatrix}$ &
$\begin{pmatrix} 0 & \times & 0 \\ \times  & 0 & \times \\ 0 & \times & \times \end{pmatrix}$ &
$\begin{pmatrix} \times & \times & 0 \\ \times  & \times & \times \\ 0 & \times & \times \end{pmatrix}$ &
$\begin{pmatrix} \times & \times & \times \\ \times  & \times & \times \\ \times & \times & \times \end{pmatrix}$ &
$\begin{pmatrix} \times & 0 & \times \\ 0  & \times & \times \\ \times & \times & \times \end{pmatrix}$ \\ \hline
(2) & $\{0, 1, 2 \}$ & $1$ & $2$ & 
$\begin{pmatrix} 0 & \times & 0 \\ \times  & 0 & \times \\ 0 & \times & 0 \end{pmatrix}$ &
$\begin{pmatrix} 0 & 0 & \times \\ 0  & \times & \times \\ \times & \times & 0 \end{pmatrix}$ &
$\begin{pmatrix} \times & 0 & \times \\ 0  & \times & \times \\ \times & \times & \times \end{pmatrix}$ &
$\begin{pmatrix} \times & \times & \times \\ \times  & \times & \times \\ \times & \times & \times \end{pmatrix}$ &
$\begin{pmatrix} 0 & \times & \times \\ \times  & \times & \times \\ \times & \times & \times \end{pmatrix}$ \\ \hline
(3) & $\{0, 1, 2 \}$ & $2$ & $1$ & 
$\begin{pmatrix} 0 & 0 & \times \\ 0  & 0 & \times \\ \times & \times & 0 \end{pmatrix}$ &
$\begin{pmatrix} 0 & \times & 0 \\ \times  & 0 & \times \\ 0 & \times & \times \end{pmatrix}$ &
$\begin{pmatrix} \times & \times & 0 \\ \times  & \times & \times \\ 0 & \times & \times \end{pmatrix}$ &
$\begin{pmatrix} \times & \times & \times \\ \times  & \times & \times \\ \times & \times & \times \end{pmatrix}$ &
$\begin{pmatrix} 0 & \times & \times \\ \times  & \times & \times \\ \times & \times & \times \end{pmatrix}$ \\ \hline
(4) & $\{0, 1, 2 \}$ & $2$ & $2$ & 
$\begin{pmatrix} 0 & 0 & \times \\ 0  & 0 & \times \\ \times & \times & 0 \end{pmatrix}$ &
$\begin{pmatrix} 0 & 0 & \times \\ 0  & \times & \times \\ \times & \times & \times \end{pmatrix}$ &
$\begin{pmatrix} \times & \times & 0 \\ \times  & \times & \times \\ 0 & \times & \times \end{pmatrix}$ &
$\begin{pmatrix} \times & \times & \times \\ \times  & \times & \times \\ \times & \times & \times \end{pmatrix}$ &
$\begin{pmatrix} \times & \times & 0 \\ \times  & \times & \times \\ 0 & \times & \times \end{pmatrix}$ \\ \hline
\end{tabular}
\caption{Viable assignments under $Z^{NI}_5$ and structures of Yukawa and mass matrices.}\label{tab:M2}
\end{table}


\begin{table}[t]
\begin{tabular}{|c||c|c|c|c|c|c|c|c|}\hline\hline  
$Z_6^{NI}$ Models & $\{k_{\ell 1}, k_{\ell 2}, k_{\ell 3} \}$ & $k_S$ & $k_\Phi$ & $f$ & $y^\Phi$ & $M_\ell$ & $m_\nu$ & $m_\nu|_{t_\beta \to \infty}$ \\ \hline
(1) & $\{0, 1, 2 \}$ & $1$ & $1$ & 
$\begin{pmatrix} 0 & \times & 0 \\ \times  & 0 & \times \\ 0 & \times & 0 \end{pmatrix}$ &
$\begin{pmatrix} 0 & \times & 0 \\ \times  & 0 & \times \\ 0 & \times & 0 \end{pmatrix}$ &
$\begin{pmatrix} \times & \times & 0 \\ \times  & \times & \times \\ 0 & \times & \times \end{pmatrix}$ &
$\begin{pmatrix} \times & \times & \times \\ \times  & \times & \times \\ \times & \times & \times \end{pmatrix}$ &
$\begin{pmatrix} \times & 0 & \times \\ 0  & \times & 0 \\ \times & 0 & \times \end{pmatrix}$ \\ \hline
(2) & $\{0, 1, 2 \}$ & $1$ & $3$ & 
$\begin{pmatrix} 0 & \times & 0 \\ \times  & 0 & \times \\ 0 & \times & 0 \end{pmatrix}$ &
$\begin{pmatrix} 0 & 0 & 0 \\ 0  & 0 & \times \\ 0 & \times & 0 \end{pmatrix}$ &
$\begin{pmatrix} \times & 0 & 0 \\ 0  & \times & \times \\ 0 & \times & \times \end{pmatrix}$ &
$\begin{pmatrix} 0 & \times & \times \\ \times  & \times & \times \\ \times & \times & \times \end{pmatrix}$ &
$\begin{pmatrix} 0 & 0 & \times \\ 0  & \times & 0 \\ \times & 0 & \times \end{pmatrix}$ \\ \hline
(3) & $\{0, 1, 2 \}$ & $2$ & $1$ & 
$\begin{pmatrix} 0 & 0 & \times \\ 0  & 0 & 0 \\ \times & 0 & 0 \end{pmatrix}$ &
$\begin{pmatrix} 0 & \times & 0 \\ \times  & 0 & \times \\ 0 & \times & 0 \end{pmatrix}$ &
$\begin{pmatrix} \times & \times & 0 \\ \times  & \times & \times \\ 0 & \times & \times \end{pmatrix}$ &
$\begin{pmatrix} \times & \times & \times \\ \times  & 0 & \times \\ \times & \times & \times \end{pmatrix}$ &
$\begin{pmatrix} 0 & \times & 0 \\ \times  & 0 & \times \\ 0 & \times & 0 \end{pmatrix}$ \\ \hline
(4) & $\{0, 1, 2 \}$ & $3$ & $1$ & 
$\begin{pmatrix} 0 & 0 & 0 \\ 0  & 0 & \times \\ 0 & \times & 0 \end{pmatrix}$ &
$\begin{pmatrix} 0 & \times & 0 \\ \times  & 0 & \times \\ 0 & \times & 0 \end{pmatrix}$ &
$\begin{pmatrix} \times & \times & 0 \\ \times  & \times & \times \\ 0 & \times & \times \end{pmatrix}$ &
$\begin{pmatrix} 0 & \times & \times \\ \times  & \times & \times \\ \times & \times & \times \end{pmatrix}$ &
$\begin{pmatrix} 0 & 0 & \times \\ 0  & \times & 0 \\ \times & 0 & \times \end{pmatrix}$ \\ \hline
(5) & $\{1, 2, 3 \}$ & $1$ & $1$ & 
$\begin{pmatrix} 0 & \times & 0 \\ \times  & 0 & \times \\ 0 & \times & 0 \end{pmatrix}$ &
$\begin{pmatrix} 0 & \times & 0 \\ \times  & 0 & \times \\ 0 & \times & 0 \end{pmatrix}$ &
$\begin{pmatrix} \times & \times & 0 \\ \times  & \times & \times \\ 0 & \times & \times \end{pmatrix}$ &
$\begin{pmatrix} \times & \times & \times \\ \times  & \times & \times \\ \times & \times & \times \end{pmatrix}$ &
$\begin{pmatrix} \times & 0 & \times \\ 0  & \times & 0 \\ \times & 0 & \times \end{pmatrix}$ \\ \hline
(6) & $\{1, 2, 3 \}$ & $1$ & $3$ & 
$\begin{pmatrix} 0 & \times & 0 \\ \times  & 0 & \times \\ 0 & \times & 0 \end{pmatrix}$ &
$\begin{pmatrix} 0 & \times & 0 \\ \times  & 0 & 0 \\ 0 & 0 & 0 \end{pmatrix}$ &
$\begin{pmatrix} \times & \times & 0 \\ \times  & \times & 0 \\ 0 & 0 & \times \end{pmatrix}$ &
$\begin{pmatrix} \times & \times & \times \\ \times  & \times & \times \\ \times & \times & 0 \end{pmatrix}$ &
$\begin{pmatrix} \times & 0 & \times \\ 0  & \times & 0 \\ \times & 0 & 0 \end{pmatrix}$ \\ \hline
(7) & $\{1, 2, 3 \}$ & $2$ & $1$ & 
$\begin{pmatrix} 0 & 0 & \times \\ 0  & 0 & 0 \\ \times & 0 & 0 \end{pmatrix}$ &
$\begin{pmatrix} 0 & \times & 0 \\ \times  & 0 & \times \\ 0 & \times & 0 \end{pmatrix}$ &
$\begin{pmatrix} \times & \times & 0 \\ \times  & \times & \times \\ 0 & \times & \times \end{pmatrix}$ &
$\begin{pmatrix} \times & \times & \times \\ \times  & 0 & \times \\ \times & \times & \times \end{pmatrix}$ &
$\begin{pmatrix} 0 & \times & 0 \\ \times  & 0 & \times \\ 0 & \times & 0 \end{pmatrix}$ \\ \hline
(8) & $\{1, 2, 3 \}$ & $3$ & $1$ & 
$\begin{pmatrix} 0 & \times & 0 \\ \times  & 0 & 0 \\ 0 & 0 & 0 \end{pmatrix}$ &
$\begin{pmatrix} 0 & \times & 0 \\ \times  & 0 & \times \\ 0 & \times & 0 \end{pmatrix}$ &
$\begin{pmatrix} \times & \times & 0 \\ \times  & \times & \times \\ 0 & \times & \times \end{pmatrix}$ &
$\begin{pmatrix} \times & \times & \times \\ \times  & \times & \times \\ \times & \times & 0 \end{pmatrix}$ &
$\begin{pmatrix} \times & 0 & \times \\ 0  & \times & 0 \\ \times & 0 & 0 \end{pmatrix}$ \\ \hline
\end{tabular}
\caption{Viable assignments under $Z^{NI}_6$ and structures of Yukawa and mass matrices.}\label{tab:M3}
\end{table}

\begin{table}[t]
\begin{tabular}{|c||c|c|c|c|c|c|c|c|}\hline\hline  
$Z_7^{NI}$ Models & $\{k_{\ell 1}, k_{\ell 2}, k_{\ell 3} \}$ & $k_S$ & $k_\Phi$ & $f$ & $y^\Phi$ & $M_\ell$ & $m_\nu$ & $m_\nu|_{t_\beta \to \infty}$ \\ \hline
(1) & $\{0, 1, 2 \}$ & $1$ & $3$ & 
$\begin{pmatrix} 0 & \times & 0 \\ \times  & 0 & \times \\ 0 & \times & 0 \end{pmatrix}$ &
$\begin{pmatrix} 0 & 0 & 0 \\ 0  & 0 & \times \\ 0 & \times & 0 \end{pmatrix}$ &
$\begin{pmatrix} \times & 0 & 0 \\ 0  & \times & \times \\ 0 & \times & \times \end{pmatrix}$ &
$\begin{pmatrix} 0 & \times & \times \\ \times  & \times & \times \\ \times & \times & \times \end{pmatrix}$ &
$\begin{pmatrix} 0 & 0 & \times \\ 0  & \times & \times \\ \times & \times & \times \end{pmatrix}$ \\ \hline
(2) & $\{0, 1, 2 \}$ & $2$ & $1$ & 
$\begin{pmatrix} 0 & 0 & \times \\ 0  & 0 & 0 \\ \times & 0 & 0 \end{pmatrix}$ &
$\begin{pmatrix} 0 & \times & 0 \\ \times  & 0 & \times \\ 0 & \times & 0 \end{pmatrix}$ &
$\begin{pmatrix} \times & \times & 0 \\ \times  & \times & \times \\ 0 & \times & \times \end{pmatrix}$ &
$\begin{pmatrix} \times & \times & \times \\ \times  & 0 & \times \\ \times & \times & \times \end{pmatrix}$ &
$\begin{pmatrix} 0 & \times & 0 \\ \times  & 0 & \times \\ 0 & \times & 0 \end{pmatrix}$ \\ \hline
(3) & $\{0, 1, 2 \}$ & $3$ & $1$ & 
$\begin{pmatrix} 0 & 0 & 0 \\ 0  & 0 & \times \\ 0 & \times & 0 \end{pmatrix}$ &
$\begin{pmatrix} 0 & \times & 0 \\ \times  & 0 & \times \\ 0 & \times & 0 \end{pmatrix}$ &
$\begin{pmatrix} \times & \times & 0 \\ \times  & \times & \times \\ 0 & \times & \times \end{pmatrix}$ &
$\begin{pmatrix} 0 & \times & \times \\ \times  & \times & \times \\ \times & \times & \times \end{pmatrix}$ &
$\begin{pmatrix} 0 & 0 & \times \\ 0  & \times & 0 \\ \times & 0 & 0 \end{pmatrix}$ \\ \hline
(4) & $\{0, 1, 3 \}$ & $1$ & $3$ & 
$\begin{pmatrix} 0 & \times & 0 \\ \times  & 0 & 0 \\ 0 & 0 & 0 \end{pmatrix}$ &
$\begin{pmatrix} 0 & 0 & \times \\ 0  & 0 & \times \\ \times & \times & 0 \end{pmatrix}$ &
$\begin{pmatrix} \times & 0 & \times \\ 0  & \times & \times \\ \times & \times & \times \end{pmatrix}$ &
$\begin{pmatrix} \times & \times & \times \\ \times  & 0 & \times \\ \times & \times & \times \end{pmatrix}$ &
$\begin{pmatrix} 0 & 0 & \times \\ 0  &0 & \times \\ \times & \times & 0 \end{pmatrix}$ \\ \hline
(5) & $\{0, 1, 3 \}$ & $2$ & $3$ & 
$\begin{pmatrix} 0 & 0 & 0 \\ 0  & 0 & \times \\ 0 & \times & 0 \end{pmatrix}$ &
$\begin{pmatrix} 0 & 0 & \times \\ 0  & 0 & \times \\ \times & \times & 0 \end{pmatrix}$ &
$\begin{pmatrix} \times & 0 & \times \\ 0  & \times & \times \\ \times & \times & \times \end{pmatrix}$ &
$\begin{pmatrix} 0 & \times & \times \\ \times  & \times & \times \\ \times & \times & \times \end{pmatrix}$ &
$\begin{pmatrix} 0 & \times & 0 \\ \times  & \times & 0 \\ 0 & 0 & \times \end{pmatrix}$ \\ \hline
(6) & $\{0, 1, 3 \}$ & $3$ & $2$ & 
$\begin{pmatrix} 0 & 0 & \times \\ 0  & 0 & \times \\ \times & \times & 0 \end{pmatrix}$ &
$\begin{pmatrix} 0 & 0 & 0 \\ 0  & \times & \times \\ 0 & \times & 0 \end{pmatrix}$ &
$\begin{pmatrix} \times & 0 & 0 \\ 0  & \times & \times \\ 0 & \times & \times \end{pmatrix}$ &
$\begin{pmatrix} 0 & \times & \times \\ \times  & \times & \times \\ \times & \times & \times \end{pmatrix}$ &
$\begin{pmatrix} 0 & \times & 0 \\ \times  & \times & \times \\ 0 & \times & \times \end{pmatrix}$ \\ \hline
(7) & $\{0, 1, 3 \}$ & $3$ & $3$ & 
$\begin{pmatrix} 0 & 0 & \times \\ 0  & 0 & \times \\ \times & \times & 0 \end{pmatrix}$ &
$\begin{pmatrix} 0 & 0 & \times \\ 0  & 0 & \times \\ \times & \times & 0 \end{pmatrix}$ &
$\begin{pmatrix} \times & 0 & \times \\ 0  & \times & \times \\ \times & \times & \times \end{pmatrix}$ &
$\begin{pmatrix} \times & \times & \times \\ \times  & \times & \times \\ \times & \times & \times \end{pmatrix}$ &
$\begin{pmatrix} \times & \times & 0 \\ \times  & \times & 0 \\ 0 & 0 & \times \end{pmatrix}$ \\ \hline
(8) & $\{0, 2, 3 \}$ & $1$ & $2$ & 
$\begin{pmatrix} 0 & 0 & 0 \\ 0  & 0 & \times \\ 0 & \times & 0 \end{pmatrix}$ &
$\begin{pmatrix} 0 & \times & 0 \\ \times  & 0 & \times \\ 0 & \times & 0 \end{pmatrix}$ &
$\begin{pmatrix} \times & \times & 0 \\ \times  & \times & \times \\ 0 & \times & \times \end{pmatrix}$ &
$\begin{pmatrix} 0 & \times & \times \\ \times  & \times & \times \\ \times & \times & \times \end{pmatrix}$ &
$\begin{pmatrix} 0 & 0 & \times \\ 0  & \times & 0 \\ \times & 0 & \times \end{pmatrix}$ \\ \hline
(9) & $\{0, 2, 3 \}$ & $2$ & $1$ & 
$\begin{pmatrix} 0 & \times & 0 \\ \times  & 0 & \times \\ 0 & \times & 0 \end{pmatrix}$ &
$\begin{pmatrix} 0 & 0 & 0 \\ 0  & 0 & \times \\ 0 & \times & \times \end{pmatrix}$ &
$\begin{pmatrix} \times & 0 & 0 \\ 0  & \times & \times \\ 0 & \times & \times \end{pmatrix}$ &
$\begin{pmatrix} 0 & \times & \times \\ \times  & \times & \times \\ \times & \times & \times \end{pmatrix}$ &
$\begin{pmatrix} 0 & 0 & \times \\ 0  & \times & \times \\ \times & \times & \times \end{pmatrix}$ \\ \hline
\end{tabular}
\caption{Viable assignments under $Z^{NI}_7$ and structures of Yukawa and mass matrices.}\label{tab:M4}
\end{table}

\begin{table}[t]
\begin{tabular}{|c||c|c|c|c|c|c|c|c|}\hline\hline  
$Z_7^{NI}$ Models & $\{k_{\ell 1}, k_{\ell 2}, k_{\ell 3} \}$ & $k_S$ & $k_\Phi$ & $f$ & $y^\Phi$ & $M_\ell$ & $m_\nu$ & $m_\nu|_{t_\beta \to \infty}$ \\ \hline
(10) & $\{0, 2, 3 \}$ & $2$ & $2$ & 
$\begin{pmatrix} 0 & \times & 0 \\ \times  & 0 & \times \\ 0 & \times & 0 \end{pmatrix}$ &
$\begin{pmatrix} 0 & \times & 0 \\ \times  & 0 & \times \\ 0 & \times & 0 \end{pmatrix}$ &
$\begin{pmatrix} \times & \times & 0 \\ \times  & \times & \times \\ 0 & \times & \times \end{pmatrix}$ &
$\begin{pmatrix} \times & \times & \times \\ \times  & \times & \times \\ \times & \times & \times \end{pmatrix}$ &
$\begin{pmatrix} \times & 0 & \times \\ 0  & \times & 0 \\ \times & 0 & \times \end{pmatrix}$ \\ \hline
(11) & $\{0, 2, 3 \}$ & $3$ & $2$ & 
$\begin{pmatrix} 0 & 0 & \times \\ 0  & 0 & 0 \\ \times & 0 & 0 \end{pmatrix}$ &
$\begin{pmatrix} 0 & \times & 0 \\ \times  & 0 & \times \\ 0 & \times & 0 \end{pmatrix}$ &
$\begin{pmatrix} \times & \times & 0 \\ \times  & \times & \times \\ 0 & \times & \times \end{pmatrix}$ &
$\begin{pmatrix} \times & \times & \times \\ \times  & 0 & \times \\ \times & \times & \times \end{pmatrix}$ &
$\begin{pmatrix} 0 & \times & 0 \\ \times  & 0 & \times \\ 0 & \times & 0 \end{pmatrix}$ \\ \hline
(12) & $\{1, 2, 3 \}$ & $1$ & $1$ & 
$\begin{pmatrix} 0 & \times & 0 \\ \times  & 0 & \times \\ 0 & \times & 0 \end{pmatrix}$ &
$\begin{pmatrix} 0 & \times & 0 \\ \times  & 0 & \times \\ 0 & \times & \times \end{pmatrix}$ &
$\begin{pmatrix} \times & \times & 0 \\ \times  & \times & \times \\ 0 & \times & \times \end{pmatrix}$ &
$\begin{pmatrix} \times & \times & \times \\ \times  & \times & \times \\ \times & \times & \times \end{pmatrix}$ &
$\begin{pmatrix} \times & 0 & \times \\  0  & \times & \times \\ \times & \times & \times \end{pmatrix}$ \\ \hline
(13) & $\{1, 2, 3 \}$ & $1$ & $2$ & 
$\begin{pmatrix} 0 & \times & 0 \\ \times  & 0 & \times \\ 0 & \times & 0 \end{pmatrix}$ &
$\begin{pmatrix} \times & 0 & \times \\ 0  & 0 & \times \\ \times & \times & \times \end{pmatrix}$ &
$\begin{pmatrix} \times & 0 & \times \\ 0  & \times & \times \\ \times & \times & \times \end{pmatrix}$ &
$\begin{pmatrix} \times & \times & \times \\ \times  & \times & \times \\ \times & \times & \times \end{pmatrix}$ &
$\begin{pmatrix} 0 & \times & \times \\  \times  & \times & \times \\ \times & \times & \times \end{pmatrix}$ \\ \hline
(14) & $\{1, 2, 3 \}$ & $1$ & $3$ & 
$\begin{pmatrix} 0 & \times & 0 \\ \times  & 0 & \times \\ 0 & \times & 0 \end{pmatrix}$ &
$\begin{pmatrix} 0 & \times & \times \\ \times  & \times & 0 \\ \times & 0 & 0 \end{pmatrix}$ &
$\begin{pmatrix} \times & \times & \times \\ \times  & \times & 0 \\ \times & 0 & \times \end{pmatrix}$ &
$\begin{pmatrix} \times & \times & \times \\ \times  & \times & \times \\ \times & \times & \times \end{pmatrix}$ &
$\begin{pmatrix} \times & \times & \times \\  \times  & \times & \times \\ \times & \times & 0 \end{pmatrix}$ \\ \hline
(15) & $\{1, 2, 3 \}$ & $2$ & $1$ & 
$\begin{pmatrix} 0 & 0 & \times \\ 0  & 0 & \times \\ \times & \times & 0 \end{pmatrix}$ &
$\begin{pmatrix} 0 & \times & 0 \\ \times  & 0 & \times \\ 0 & \times & \times \end{pmatrix}$ &
$\begin{pmatrix} \times & \times & 0 \\ \times  & \times & \times \\ 0 & \times & \times \end{pmatrix}$ &
$\begin{pmatrix} \times & \times & \times \\ \times  & \times & \times \\ \times & \times & \times \end{pmatrix}$ &
$\begin{pmatrix} 0 & \times & \times \\  \times  & \times & \times \\ \times & \times & \times \end{pmatrix}$ \\ \hline
(16) & $\{1, 2, 3 \}$ & $2$ & $2$ & 
$\begin{pmatrix} 0 & 0 & \times \\ 0  & 0 & \times \\ \times & \times & 0 \end{pmatrix}$ &
$\begin{pmatrix} \times & 0 & \times \\ 0  & 0 & \times \\ \times & \times & \times \end{pmatrix}$ &
$\begin{pmatrix} \times & 0 & \times \\ 0  & \times & \times \\ \times & \times & \times \end{pmatrix}$ &
$\begin{pmatrix} \times & \times & \times \\ \times  & \times & \times \\ \times & \times & \times \end{pmatrix}$ &
$\begin{pmatrix} \times & \times & \times \\  \times  & \times & 0 \\ \times & 0 & \times \end{pmatrix}$ \\ \hline
(17) & $\{1, 2, 3 \}$ & $2$ & $3$ & 
$\begin{pmatrix} 0 & 0 & \times \\ 0  & 0 & \times \\ \times & \times & 0 \end{pmatrix}$ &
$\begin{pmatrix} 0 & \times & \times \\ \times  & \times & 0 \\ \times & 0 & 0 \end{pmatrix}$ &
$\begin{pmatrix} \times & \times & \times \\ \times  & \times & 0 \\ \times & 0 & \times \end{pmatrix}$ &
$\begin{pmatrix} \times & \times & \times \\ \times  & \times & \times \\ \times & \times & \times \end{pmatrix}$ &
$\begin{pmatrix} \times & \times & \times \\  \times  & 0 & \times \\ \times & \times & \times \end{pmatrix}$ \\ \hline
(18) & $\{1, 2, 3 \}$ & $3$ & $1$ & 
$\begin{pmatrix} 0 & \times & \times \\ \times  & 0 & 0 \\ \times & 0 & 0 \end{pmatrix}$ &
$\begin{pmatrix} 0 & \times & 0 \\ \times & 0 & \times \\ 0 & \times & \times \end{pmatrix}$ &
$\begin{pmatrix} \times & \times & 0 \\ \times  & \times & \times \\ 0 & \times & \times \end{pmatrix}$ &
$\begin{pmatrix} \times & \times & \times \\ \times  & \times & \times \\ \times & \times & \times \end{pmatrix}$ &
$\begin{pmatrix} \times & \times & \times \\  \times  & \times & \times \\ \times & \times & 0 \end{pmatrix}$ \\ \hline
\end{tabular}
\caption{Viable assignments under $Z^{NI}_7$ and structures of Yukawa and mass matrices.}\label{tab:M5}
\end{table}


\begin{table}[t]
\begin{tabular}{|c||c|c|c|c|c|c|c|c|}\hline\hline  
$Z_7^{NI}$ Models & $\{k_{\ell 1}, k_{\ell 2}, k_{\ell 3} \}$ & $k_S$ & $k_\Phi$ & $f$ & $y^\Phi$ & $M_\ell$ & $m_\nu$ & $m_\nu|_{t_\beta \to \infty}$ \\ \hline
(19) & $\{1, 2, 3 \}$ & $3$ & $2$ & 
$\begin{pmatrix} 0 & \times & \times \\ \times  & 0 & 0 \\ \times & 0 & 0 \end{pmatrix}$ &
$\begin{pmatrix} \times & 0 & \times \\ 0  & 0 & \times \\ \times & \times & 0 \end{pmatrix}$ &
$\begin{pmatrix} \times & 0 & \times \\ 0  & \times & \times \\ \times & \times & \times \end{pmatrix}$ &
$\begin{pmatrix} \times & \times & \times \\ \times  & \times & \times \\ \times & \times & \times \end{pmatrix}$ &
$\begin{pmatrix} \times & \times & \times \\ \times  & 0 & \times \\ \times & \times & \times \end{pmatrix}$ \\ \hline
(20) & $\{1, 2, 3 \}$ & $3$ & $3$ & 
$\begin{pmatrix} 0 & \times & \times \\ \times  & 0 & 0 \\ \times & 0 & 0 \end{pmatrix}$ &
$\begin{pmatrix} 0 & \times & \times \\ \times  & \times & 0 \\ \times & 0 & 0 \end{pmatrix}$ &
$\begin{pmatrix} \times & \times & \times \\ \times  & \times & 0 \\ \times & 0 & \times \end{pmatrix}$ &
$\begin{pmatrix} \times & \times & \times \\ \times  & \times & \times \\ \times & \times & \times \end{pmatrix}$ &
$\begin{pmatrix} \times & \times & 0 \\ \times  & \times & \times \\ 0 & \times & \times \end{pmatrix}$ \\ \hline
\end{tabular}
\caption{Viable assignments under $Z^{NI}_7$ and structures of Yukawa and mass matrices.}\label{tab:M6}
\end{table}

\bibliography{Zee.bib}

@article{Zee:1980ai,
      author         = "Zee, A.",
      title          = "{A Theory of Lepton Number Violation, Neutrino Majorana
                        Mass, and Oscillation}",
      journal        = "Phys. Lett.",
      volume         = "93B",
      year           = "1980",
      pages          = "389",
      doi            = "10.1016/0370-2693(80)90349-4,
                        10.1016/0370-2693(80)90193-8",
      note           = "[Erratum: Phys. Lett.95B,461(1980)]",
      reportNumber   = "UPR-0150T",
      SLACcitation   = "%%CITATION = PHLTA,93B,389;%%"
}

@article{BaBar:2009hkt,
    author = "Aubert, Bernard and others",
    collaboration = "BaBar",
    title = "{Searches for Lepton Flavor Violation in the Decays $\tau^\pm \to e^\pm \gamma$ and $\tau^\pm \to \mu^\pm \gamma$}",
    eprint = "0908.2381",
    archivePrefix = "arXiv",
    primaryClass = "hep-ex",
    reportNumber = "SLAC-PUB-13753, BABAR-PUB-09-026",
    doi = "10.1103/PhysRevLett.104.021802",
    journal = "Phys. Rev. Lett.",
    volume = "104",
    pages = "021802",
    year = "2010"
}

@article{KamLAND-Zen:2024eml,
    author = "Abe, S. and others",
    collaboration = "KamLAND-Zen",
    title = "{Search for Majorana Neutrinos with the Complete KamLAND-Zen Dataset}",
    eprint = "2406.11438",
    archivePrefix = "arXiv",
    primaryClass = "hep-ex",
    month = "6",
    year = "2024"
}

@article{Esteban:2024eli,
    author = "Esteban, Ivan and Gonzalez-Garcia, M. C. and Maltoni, Michele and Martinez-Soler, Ivan and Pinheiro, Jo{\~a}o Paulo and Schwetz, Thomas",
    title = "{NuFit-6.0: updated global analysis of three-flavor neutrino oscillations}",
    eprint = "2410.05380",
    archivePrefix = "arXiv",
    primaryClass = "hep-ph",
    reportNumber = "IFT-UAM/CSIC-24-140, YITP-SB-2024-24, IPPP/24/64, IPPP/24/64, IFT-UAM/CSIC-24-140, YITP-SB-2024-24",
    doi = "10.1007/JHEP12(2024)216",
    journal = "JHEP",
    volume = "12",
    pages = "216",
    year = "2024"
}

@article{DESI:2024mwx,
    author = "Adame, A. G. and others",
    collaboration = "DESI",
    title = "{DESI 2024 VI: cosmological constraints from the measurements of baryon acoustic oscillations}",
    eprint = "2404.03002",
    archivePrefix = "arXiv",
    primaryClass = "astro-ph.CO",
    reportNumber = "FERMILAB-PUB-24-0154-PPD",
    doi = "10.1088/1475-7516/2025/02/021",
    journal = "JCAP",
    volume = "02",
    pages = "021",
    year = "2025"
}

@article{Planck:2018vyg,
    author = "Aghanim, N. and others",
    collaboration = "Planck",
    title = "{Planck 2018 results. VI. Cosmological parameters}",
    eprint = "1807.06209",
    archivePrefix = "arXiv",
    primaryClass = "astro-ph.CO",
    doi = "10.1051/0004-6361/201833910",
    journal = "Astron. Astrophys.",
    volume = "641",
    pages = "A6",
    year = "2020",
    note = "[Erratum: Astron.Astrophys. 652, C4 (2021)]"
}

@inbook{Feruglio:2017spp,
    author = "Feruglio, Ferruccio",
    editor = "Levy, Aharon and Forte, Stefano and Ridolfi, Giovanni",
    title = "{Are neutrino masses modular forms?}",
    booktitle = "{From My Vast Repertoire ...}: {Guido Altarelli's Legacy}",
    eprint = "1706.08749",
    archivePrefix = "arXiv",
    primaryClass = "hep-ph",
    reportNumber = "DFPD-2017-TH-09",
    doi = "10.1142/9789813238053_0012",
    pages = "227--266",
    year = "2019"
}

@article{Gomes:2023ahz,
    author = "Gomes, Pedro R. S.",
    title = "{An introduction to higher-form symmetries}",
    eprint = "2303.01817",
    archivePrefix = "arXiv",
    primaryClass = "hep-th",
    doi = "10.21468/SciPostPhysLectNotes.74",
    journal = "SciPost Phys. Lect. Notes",
    volume = "74",
    pages = "1",
    year = "2023"
}

@article{Schafer-Nameki:2023jdn,
    author = "Schafer-Nameki, Sakura",
    title = "{ICTP lectures on (non-)invertible generalized symmetries}",
    eprint = "2305.18296",
    archivePrefix = "arXiv",
    primaryClass = "hep-th",
    doi = "10.1016/j.physrep.2024.01.007",
    journal = "Phys. Rept.",
    volume = "1063",
    pages = "1--55",
    year = "2024"
}

@article{Bhardwaj:2023kri,
    author = "Bhardwaj, Lakshya and Bottini, Lea E. and Fraser-Taliente, Ludovic and Gladden, Liam and Gould, Dewi S. W. and Platschorre, Arthur and Tillim, Hannah",
    title = "{Lectures on generalized symmetries}",
    eprint = "2307.07547",
    archivePrefix = "arXiv",
    primaryClass = "hep-th",
    doi = "10.1016/j.physrep.2023.11.002",
    journal = "Phys. Rept.",
    volume = "1051",
    pages = "1--87",
    year = "2024"
}

@article{Shao:2023gho,
    author = "Shao, Shu-Heng",
    title = "{What's Done Cannot Be Undone: TASI Lectures on Non-Invertible Symmetries}",
    eprint = "2308.00747",
    archivePrefix = "arXiv",
    primaryClass = "hep-th",
    reportNumber = "YITP-SB-2023-19",
    month = "8",
    year = "2023"
}

@article{Kobayashi:2024cvp,
    author = "Kobayashi, Tatsuo and Otsuka, Hajime and Tanimoto, Morimitsu",
    title = "{Yukawa textures from non-invertible symmetries}",
    eprint = "2409.05270",
    archivePrefix = "arXiv",
    primaryClass = "hep-ph",
    reportNumber = "EPHOU-24-013, KYUSHU-HET-295",
    doi = "10.1007/JHEP12(2024)117",
    journal = "JHEP",
    volume = "12",
    pages = "117",
    year = "2024"
}

@article{Delgado:2024pcv,
    author = "Delgado, Antonio and Koren, Seth",
    title = "{Non-invertible Peccei-Quinn symmetry, natural 2HDM alignment, and the visible axion}",
    eprint = "2412.05362",
    archivePrefix = "arXiv",
    primaryClass = "hep-ph",
    doi = "10.1007/JHEP02(2025)178",
    journal = "JHEP",
    volume = "02",
    pages = "178",
    year = "2025"
}

@article{Funakoshi:2024uvy,
    author = "Funakoshi, Shuta and Kobayashi, Tatsuo and Otsuka, Hajime",
    title = "{Quantum aspects of non-invertible flavor symmetries in intersecting/magnetized D-brane models}",
    eprint = "2412.12524",
    archivePrefix = "arXiv",
    primaryClass = "hep-th",
    reportNumber = "EPHOU-24-017, KYUSHU-HET-304",
    doi = "10.1007/JHEP04(2025)183",
    journal = "JHEP",
    volume = "04",
    pages = "183",
    year = "2025"
}

@article{Kobayashi:2025znw,
    author = "Kobayashi, Tatsuo and Nishioka, Yume and Otsuka, Hajime and Tanimoto, Morimitsu",
    title = "{More about quark Yukawa textures from selection rules without group actions}",
    eprint = "2503.09966",
    archivePrefix = "arXiv",
    primaryClass = "hep-ph",
    reportNumber = "EPHOU-25-003, KYUSHU-HET-312",
    doi = "10.1007/JHEP05(2025)177",
    journal = "JHEP",
    volume = "05",
    pages = "177",
    year = "2025"
}

@article{Liang:2025dkm,
    author = "Liang, Qiuyue and Yanagida, Tsutomu T.",
    title = "{Non-invertible symmetry as an axion-less solution to the strong CP problem}",
    eprint = "2505.05142",
    archivePrefix = "arXiv",
    primaryClass = "hep-ph",
    doi = "10.1016/j.physletb.2025.139706",
    journal = "Phys. Lett. B",
    volume = "868",
    pages = "139706",
    year = "2025"
}

@article{Kobayashi:2025ldi,
    author = "Kobayashi, Tatsuo and Otsuka, Hajime and Tanimoto, Morimitsu and Uchida, Haruki",
    title = "{Lepton mass textures from non-invertible multiplication rules}",
    eprint = "2505.07262",
    archivePrefix = "arXiv",
    primaryClass = "hep-ph",
    reportNumber = "EPHOU-25-007, KYUSHU-HET-321",
    month = "5",
    year = "2025"
}

@article{Kobayashi:2025cwx,
    author = "Kobayashi, Tatsuo and Okada, Hiroshi and Otsuka, Hajime",
    title = "{Radiative neutrino mass models from non-invertible selection rules}",
    eprint = "2505.14878",
    archivePrefix = "arXiv",
    primaryClass = "hep-ph",
    reportNumber = "EPHOU-25-008, KYUSHU-HET-324",
    month = "5",
    year = "2025"
}

@article{Kobayashi:2025lar,
    author = "Kobayashi, Tatsuo and Mita, Hironobu and Otsuka, Hajime and Sakuma, Riku",
    title = "{Matter symmetries in supersymmetric standard models from non-invertible selection rules}",
    eprint = "2506.10241",
    archivePrefix = "arXiv",
    primaryClass = "hep-ph",
    reportNumber = "EPHOU-25-009, KYUSHU-HET-325",
    month = "6",
    year = "2025"
}

@article{Nomura:2025sod,
    author = "Nomura, Takaaki and Okada, Hiroshi",
    title = "{Radiative lepton seesaw model in a non-invertible fusion rule and gauged $B-L$ symmetry}",
    eprint = "2506.16706",
    archivePrefix = "arXiv",
    primaryClass = "hep-ph",
    month = "6",
    year = "2025"
}

@article{Dong:2025jra,
    author = "Dong, Jun and Jeric, Tim and Kobayashi, Tatsuo and Nishida, Ryusei and Otsuka, Hajime",
    title = "{On discrete gauging and non-invertible selection rules}",
    eprint = "2507.02375",
    archivePrefix = "arXiv",
    primaryClass = "hep-th",
    month = "7",
    year = "2025"
}

@article{Suzuki:2025oov,
    author = "Suzuki, Motoo and Xu, Ling-Xiao",
    title = "{Phenomenological implications of a class of non-invertible selection rules}",
    eprint = "2503.19964",
    archivePrefix = "arXiv",
    primaryClass = "hep-ph",
    month = "3",
    year = "2025"
}

@article{Heckman:2024obe,
    author = "Heckman, Jonathan J. and McNamara, Jacob and Montero, Miguel and Sharon, Adar and Vafa, Cumrun and Valenzuela, Irene",
    title = "{Fate of stringy noninvertible symmetries}",
    eprint = "2402.00118",
    archivePrefix = "arXiv",
    primaryClass = "hep-th",
    reportNumber = "CERN-TH-2024-019",
    doi = "10.1103/PhysRevD.110.106001",
    journal = "Phys. Rev. D",
    volume = "110",
    number = "10",
    pages = "106001",
    year = "2024"
}

@article{Kaidi:2024wio,
    author = "Kaidi, Justin and Tachikawa, Yuji and Zhang, Hao Y.",
    title = "{On a class of selection rules without group actions in field theory and string theory}",
    eprint = "2402.00105",
    archivePrefix = "arXiv",
    primaryClass = "hep-th",
    doi = "10.21468/SciPostPhys.17.6.169",
    journal = "SciPost Phys.",
    volume = "17",
    number = "6",
    pages = "169",
    year = "2024"
}

@article{Kobayashi:2024yqq,
    author = "Kobayashi, Tatsuo and Otsuka, Hajime",
    title = "{Non-invertible flavor symmetries in magnetized extra dimensions}",
    eprint = "2408.13984",
    archivePrefix = "arXiv",
    primaryClass = "hep-th",
    reportNumber = "EPHOU-24-010, KYUSHU-HET-292",
    doi = "10.1007/JHEP11(2024)120",
    journal = "JHEP",
    volume = "11",
    pages = "120",
    year = "2024"
}

@article{Cordova:2022fhg,
    author = "Cordova, Clay and Hong, Sungwoo and Koren, Seth and Ohmori, Kantaro",
    title = "{Neutrino Masses from Generalized Symmetry Breaking}",
    eprint = "2211.07639",
    archivePrefix = "arXiv",
    primaryClass = "hep-ph",
    doi = "10.1103/PhysRevX.14.031033",
    journal = "Phys. Rev. X",
    volume = "14",
    number = "3",
    pages = "031033",
    year = "2024"
}

@article{KATRIN:2024cdt,
    author = "Aker, Max and others",
    collaboration = "KATRIN",
    title = "{Direct neutrino-mass measurement based on 259 days of KATRIN data}",
    eprint = "2406.13516",
    archivePrefix = "arXiv",
    primaryClass = "nucl-ex",
    doi = "10.1126/science.adq9592",
    journal = "Science",
    volume = "388",
    number = "6743",
    pages = "adq9592",
    year = "2025"
}

@article{LEGEND:2021bnm,
    author = "Abgrall, N. and others",
    collaboration = "LEGEND",
    title = "{The Large Enriched Germanium Experiment for Neutrinoless $\beta\beta$ Decay}: {LEGEND-1000 Preconceptual Design Report}",
    eprint = "2107.11462",
    archivePrefix = "arXiv",
    primaryClass = "physics.ins-det",
    month = "7",
    year = "2021"
}

@article{nEXO:2021ujk,
    author = "Adhikari, G. and others",
    collaboration = "nEXO",
    title = "{nEXO: neutrinoless double beta decay search beyond 10$^{28}$ year half-life sensitivity}",
    eprint = "2106.16243",
    archivePrefix = "arXiv",
    primaryClass = "nucl-ex",
    doi = "10.1088/1361-6471/ac3631",
    journal = "J. Phys. G",
    volume = "49",
    number = "1",
    pages = "015104",
    year = "2022"
}

@article{MEGII:2025gzr,
    author = "Afanaciev, K. and others",
    collaboration = "MEG II",
    title = "{New limit on the ${\upmu ^+ \rightarrow e^+ \upgamma }$ decay with the MEG II experiment}",
    eprint = "2504.15711",
    archivePrefix = "arXiv",
    primaryClass = "hep-ex",
    doi = "10.1140/epjc/s10052-025-14906-3",
    journal = "Eur. Phys. J. C",
    volume = "85",
    number = "10",
    pages = "1177",
    year = "2025",
    note = "[Erratum: Eur.Phys.J.C 85, 1317 (2025)]"
}

@article{MEGII:2023ltw,
    author = "Afanaciev, K. and others",
    collaboration = "MEG II",
    title = "{A search for $\upmu ^+ \rightarrow \textrm{e}^+ \upgamma $ with the first dataset of the MEG~II experiment}",
    eprint = "2310.12614",
    archivePrefix = "arXiv",
    primaryClass = "hep-ex",
    doi = "10.1140/epjc/s10052-024-12416-2",
    journal = "Eur. Phys. J. C",
    volume = "84",
    number = "3",
    pages = "216",
    year = "2024",
    note = "[Erratum: Eur.Phys.J.C 84, 1042 (2024)]"
}

@article{Belle:2021ysv,
    author = "Abdesselam, A. and others",
    collaboration = "Belle",
    title = "{Search for lepton-flavor-violating tau-lepton decays to $\ell\gamma$ at Belle}",
    eprint = "2103.12994",
    archivePrefix = "arXiv",
    primaryClass = "hep-ex",
    doi = "10.1007/JHEP10(2021)019",
    journal = "JHEP",
    volume = "10",
    pages = "19",
    year = "2021"
}

@article{Nomura:2024nwh,
    author = "Nomura, Takaaki and Okada, Hiroshi",
    title = "{Zee model in a non-holomorphic modular A4 symmetry}",
    eprint = "2412.18095",
    archivePrefix = "arXiv",
    primaryClass = "hep-ph",
    doi = "10.1016/j.physletb.2025.139618",
    journal = "Phys. Lett. B",
    volume = "867",
    pages = "139618",
    year = "2025"
}

@article{Nomura:2021pld,
    author = "Nomura, Takaaki and Okada, Hiroshi and Qi, Yong-hui",
    title = "{Zee model in a modular $A_4$ symmetry}",
    eprint = "2111.10944",
    archivePrefix = "arXiv",
    primaryClass = "hep-ph",
    reportNumber = "APCTP Pre2021-028, CTP-SCU/2021034",
    doi = "10.1140/epjc/s10052-025-13864-0",
    journal = "Eur. Phys. J. C",
    volume = "85",
    number = "2",
    pages = "134",
    year = "2025"
}

@article{Nomura:2019dhw,
    author = "Nomura, Takaaki and Yagyu, Kei",
    title = "{Zee Model with Flavor Dependent Global $U(1)$ Symmetry}",
    eprint = "1905.11568",
    archivePrefix = "arXiv",
    primaryClass = "hep-ph",
    reportNumber = "KIAS-P19030, OU-HET 1013",
    doi = "10.1007/JHEP10(2019)105",
    journal = "JHEP",
    volume = "10",
    pages = "105",
    year = "2019"
}

@article{Herrero-Garcia:2017xdu,
    author = "Herrero-Garc{\'\i}a, Juan and Ohlsson, Tommy and Riad, Stella and Wir{\'e}n, Jens",
    title = "{Full parameter scan of the Zee model: exploring Higgs lepton flavor violation}",
    eprint = "1701.05345",
    archivePrefix = "arXiv",
    primaryClass = "hep-ph",
    reportNumber = "ADP-17-1-T1007",
    doi = "10.1007/JHEP04(2017)130",
    journal = "JHEP",
    volume = "04",
    pages = "130",
    year = "2017"
}

@article{Nomura:2025yoa,
    author = "Nomura, Takaaki and Popov, Oleg",
    title = "{No-group Scotogenic Model}",
    eprint = "2507.10299",
    archivePrefix = "arXiv",
    primaryClass = "hep-ph",
    month = "7",
    year = "2025"
}

@article{Chen:2025awz,
    author = "Chen, Jingqian and Geng, Chao-Qiang and Okada, Hiroshi and Wu, Jia-Jun",
    title = "{A radiative lepton model in a non-invertible fusion rule}",
    eprint = "2507.11951",
    archivePrefix = "arXiv",
    primaryClass = "hep-ph",
    doi = "10.1016/j.nuclphysb.2026.117391",
    journal = "Nucl. Phys. B",
    volume = "1025",
    pages = "117391",
    year = "2026"
}

@article{Okada:2025kfm,
    author = "Okada, Hiroshi and Shigekami, Yoshihiro",
    title = "{Three-loop induced neutrino mass model in a non-invertible symmetry}",
    eprint = "2507.16198",
    archivePrefix = "arXiv",
    primaryClass = "hep-ph",
    month = "7",
    year = "2025"
}

@article{Kobayashi:2025thd,
    author = "Kobayashi, Tatsuo and Otsuka, Hajime and Yanagida, Tsutomu T.",
    title = "{Noninvertible symmetry as a solution to the strong CP problem in a GUT-inspired standard model}",
    eprint = "2508.12287",
    archivePrefix = "arXiv",
    primaryClass = "hep-ph",
    reportNumber = "EPHOU-25-014, KYUSHU-HET-334",
    doi = "10.1103/vbd5-5cp3",
    journal = "Phys. Rev. D",
    volume = "113",
    number = "5",
    pages = "055016",
    year = "2026"
}

@article{Suzuki:2025bxg,
    author = "Suzuki, Motoo and Xu, Ling-Xiao and Zhang, Hao Y.",
    title = "{Spurion Analysis for Non-Invertible Selection Rules from Near-Group Fusions}",
    eprint = "2508.14970",
    archivePrefix = "arXiv",
    primaryClass = "hep-ph",
    month = "8",
    year = "2025"
}

@article{Jangid:2025krp,
    author = "Jangid, Shilpa and Okada, Hiroshi",
    title = "{A natural realization of inverse seesaw model in a non-invertible selection rule}",
    eprint = "2508.16174",
    archivePrefix = "arXiv",
    primaryClass = "hep-ph",
    month = "8",
    year = "2025"
}

@article{Jangid:2025thp,
    author = "Jangid, Shilpa and Okada, Hiroshi",
    title = "{A radiative seesaw model in a non-invertible selection rule with the assistance of a non-holomorphic modular $A_4$ symmetry}",
    eprint = "2510.17292",
    archivePrefix = "arXiv",
    primaryClass = "hep-ph",
    month = "10",
    year = "2025"
}

@article{Nomura:2025tvz,
    author = "Nomura, Takaaki and Okada, Hiroshi and Shigekami, Yoshihiro",
    title = "{Radiative lepton model in a non-invertible fusion rule}",
    eprint = "2510.17156",
    archivePrefix = "arXiv",
    primaryClass = "hep-ph",
    month = "10",
    year = "2025"
}

@article{Suzuki:2025kxz,
    author = "Suzuki, Motoo and Xu, Ling-Xiao",
    title = "{Spurion analysis of {\ensuremath{\mathbb{Z}}}$_{M}$/{\ensuremath{\mathbb{Z}}}$_{2}$ non-invertible selection rules: low-order versus all-order zeros}",
    eprint = "2510.18972",
    archivePrefix = "arXiv",
    primaryClass = "hep-ph",
    doi = "10.1007/JHEP02(2026)227",
    journal = "JHEP",
    volume = "02",
    pages = "227",
    year = "2026"
}

@article{Okada:2025adm,
    author = "Okada, Hiroshi and Shoji, Yutaro",
    title = "{A novel realization of linear seesaw model in a non-invertible selection rule with the assistance of $\mathbb Z_3$ symmetry}",
    eprint = "2512.20891",
    archivePrefix = "arXiv",
    primaryClass = "hep-ph",
    month = "12",
    year = "2025"
}

@article{Nakai:2025thw,
    author = "Nakai, Yuichiro and Otsuka, Hajime and Shigekami, Yoshihiro and Zhang, Zhihao",
    title = "{The Minimal Supersymmetric Standard Model with Non-Invertible Selection Rules}",
    eprint = "2512.21509",
    archivePrefix = "arXiv",
    primaryClass = "hep-ph",
    reportNumber = "KYUSHU-HET-347",
    month = "12",
    year = "2025"
}

@article{Okada:2026gxl,
    author = "Okada, Hiroshi and Shigekami, Yoshihiro",
    title = "{Two-loop rainbow neutrino masses in a non-invertible symmetry}",
    eprint = "2601.15749",
    archivePrefix = "arXiv",
    primaryClass = "hep-ph",
    month = "1",
    year = "2026"
}

@article{Kang:2026osw,
    author = "Kang, Sin Kyu and Kumar, Ranjeet and Okada, Hiroshi",
    title = "{Radiative Dirac Neutrino Masses from Modular $S_3$ Symmetry in an Axion Model}",
    eprint = "2601.22740",
    archivePrefix = "arXiv",
    primaryClass = "hep-ph",
    month = "1",
    year = "2026"
}

@article{Kashav:2026jjg,
    author = "Kashav, Monal",
    title = "{Dominant One-Loop Seesaw Contribution Induced by Non-Invertible Fusion Algebra}",
    eprint = "2602.14644",
    archivePrefix = "arXiv",
    primaryClass = "hep-ph",
    month = "2",
    year = "2026"
}

@article{Okada:2026iob,
    author = "Okada, Hiroshi and Wu, Jia-Jun",
    title = "{Dynamical Determination of the Cut-off Scale in Loop-Induced Neutrino Mass Models with Non-Invertible Symmetry}",
    eprint = "2603.17587",
    archivePrefix = "arXiv",
    primaryClass = "hep-ph",
    month = "3",
    year = "2026"
}

@article{Okada:2026bpp,
    author = "Okada, Hiroshi and Otsuka, Hajime",
    title = "{Dynamical CP Violation from Non-Invertible Selection Rules}",
    eprint = "2604.04423",
    archivePrefix = "arXiv",
    primaryClass = "hep-ph",
    reportNumber = "KYUSHU-HET-353",
    month = "4",
    year = "2026"
}

@article{Xu:2026nwh,
    author = "Xu, Ling-Xiao",
    title = "{A General Prescription for Spurion Analysis of Non-Invertible Selection Rules}",
    eprint = "2604.09345",
    archivePrefix = "arXiv",
    primaryClass = "hep-ph",
    month = "4",
    year = "2026"
}

@article{Okada:2026pek,
    author = "Okada, Hiroshi and Singh, Labh",
    title = "{Dirac one-loop seesaw in a non-invertible fusion rule}",
    eprint = "2604.11308",
    archivePrefix = "arXiv",
    primaryClass = "hep-ph",
    month = "4",
    year = "2026"
}

@article{Nomura:2026hcu,
    author = "Nomura, Takaaki and Okada, Hiroshi",
    title = "{A theoretical account of tiny multi-Higgs vacuum expectation values from non-invertible symmetry}",
    eprint = "2604.27612",
    archivePrefix = "arXiv",
    primaryClass = "hep-ph",
    month = "4",
    year = "2026"
}

@article{Cai:2017jrq,
    author = "Cai, Yi and Herrero-Garc{\'\i}a, Juan and Schmidt, Michael A. and Vicente, Avelino and Volkas, Raymond R.",
    title = "{From the trees to the forest: a review of radiative neutrino mass models}",
    eprint = "1706.08524",
    archivePrefix = "arXiv",
    primaryClass = "hep-ph",
    reportNumber = "ADP-17-29-T1035",
    doi = "10.3389/fphy.2017.00063",
    journal = "Front. in Phys.",
    volume = "5",
    pages = "63",
    year = "2017"
}

@article{Koide:2001xy,
    author = "Koide, Yoshio",
    title = "{Can the Zee model explain the observed neutrino data?}",
    eprint = "hep-ph/0104226",
    archivePrefix = "arXiv",
    reportNumber = "US-01-02",
    doi = "10.1103/PhysRevD.64.077301",
    journal = "Phys. Rev. D",
    volume = "64",
    pages = "077301",
    year = "2001"
}

@article{Frampton:2001eu,
    author = "Frampton, Paul H. and Oh, Myoung C. and Yoshikawa, Tadashi",
    title = "{Zee model confronts SNO data}",
    eprint = "hep-ph/0110300",
    archivePrefix = "arXiv",
    reportNumber = "IFP-801-UNC",
    doi = "10.1103/PhysRevD.65.073014",
    journal = "Phys. Rev. D",
    volume = "65",
    pages = "073014",
    year = "2002"
}

@article{He:2003ih,
    author = "He, Xiao-Gang",
    title = "{Is the Zee model neutrino mass matrix ruled out?}",
    eprint = "hep-ph/0307172",
    archivePrefix = "arXiv",
    doi = "10.1140/epjc/s2004-01669-8",
    journal = "Eur. Phys. J. C",
    volume = "34",
    pages = "371--376",
    year = "2004"
}

@article{Fukuyama:2010ff,
    author = "Fukuyama, Takeshi and Sugiyama, Hiroaki and Tsumura, Koji",
    title = "{Phenomenology in the Zee Model with the $A_4$ Symmetry}",
    eprint = "1012.4886",
    archivePrefix = "arXiv",
    primaryClass = "hep-ph",
    reportNumber = "MISC-2010-21",
    doi = "10.1103/PhysRevD.83.056016",
    journal = "Phys. Rev. D",
    volume = "83",
    pages = "056016",
    year = "2011"
}

@article{Kanemura:2015maa,
    author = "Kanemura, Shinya and Shindou, Tetsuo and Sugiyama, Hiroaki",
    title = "{R-Parity Conserving Supersymmetric Extension of the Zee Model}",
    eprint = "1508.05616",
    archivePrefix = "arXiv",
    primaryClass = "hep-ph",
    reportNumber = "UT-HET-104, KU-PH-018",
    doi = "10.1103/PhysRevD.92.115001",
    journal = "Phys. Rev. D",
    volume = "92",
    number = "11",
    pages = "115001",
    year = "2015"
}

@article{Matsui:2021khj,
    author = "Matsui, Toshinori and Nomura, Takaaki and Yagyu, Kei",
    title = "{Flavor dependent U(1) symmetric Zee model with a vector-like lepton}",
    eprint = "2102.09247",
    archivePrefix = "arXiv",
    primaryClass = "hep-ph",
    reportNumber = "OU-HET-1080, CTP-SCU/2021004",
    doi = "10.1016/j.nuclphysb.2021.115523",
    journal = "Nucl. Phys. B",
    volume = "971",
    pages = "115523",
    year = "2021"
}

@article{Okada:2019uoy,
    author = "Okada, Hiroshi and Tanimoto, Morimitsu",
    title = "{Towards unification of quark and lepton flavors in $A_4$ modular invariance}",
    eprint = "1905.13421",
    archivePrefix = "arXiv",
    primaryClass = "hep-ph",
    reportNumber = "APCTP Pre2019-011",
    doi = "10.1140/epjc/s10052-021-08845-y",
    journal = "Eur. Phys. J. C",
    volume = "81",
    number = "1",
    pages = "52",
    year = "2021"
}

@article{FCC:2025lpp,
    author = "Benedikt, M. and others",
    collaboration = "FCC",
    title = "{Future Circular Collider Feasibility Study Report: Volume 1, Physics, Experiments, Detectors}",
    eprint = "2505.00272",
    archivePrefix = "arXiv",
    primaryClass = "hep-ex",
    reportNumber = "CERN-FCC-PHYS-2025-0002",
    doi = "10.1140/epjc/s10052-025-15077-x",
    journal = "Eur. Phys. J. C",
    volume = "85",
    number = "12",
    pages = "1468",
    year = "2025"
}

@article{FCC:2018byv,
    author = "Abada, A. and others",
    collaboration = "FCC",
    title = "{FCC Physics Opportunities}: {Future Circular Collider Conceptual Design Report Volume 1}",
    reportNumber = "CERN-ACC-2018-0056",
    doi = "10.1140/epjc/s10052-019-6904-3",
    journal = "Eur. Phys. J. C",
    volume = "79",
    number = "6",
    pages = "474",
    year = "2019"
}

@article{CEPCStudyGroup:2025kmw,
    author = "Adhya, Souvik Priyam and others",
    collaboration = "CEPC Study Group",
    title = "{CEPC Technical Design Report - Reference Detector}",
    eprint = "2510.05260",
    archivePrefix = "arXiv",
    primaryClass = "hep-ex",
    reportNumber = "IHEP-CEPC-DR-2025-01, IHEP-EP-2025-01",
    month = "10",
    year = "2025"
}

@article{Hamada:2022mua,
    author = "Hamada, Yu and Kitano, Ryuichiro and Matsudo, Ryutaro and Takaura, Hiromasa and Yoshida, Mitsuhiro",
    title = "{$\mu$TRISTAN}",
    eprint = "2201.06664",
    archivePrefix = "arXiv",
    primaryClass = "hep-ph",
    reportNumber = "KEK-TH-2385",
    doi = "10.1093/ptep/ptac059",
    journal = "PTEP",
    volume = "2022",
    number = "5",
    pages = "053B02",
    year = "2022"
}

@article{Ai:2025cpj,
    author = "Ai, Xiaocong and others",
    title = "{New physics search at the CEPC: a general perspective}",
    eprint = "2505.24810",
    archivePrefix = "arXiv",
    primaryClass = "hep-ex",
    doi = "10.1088/1674-1137/ae1194",
    journal = "Chin. Phys. C",
    volume = "49",
    pages = "123108",
    year = "2025"
}

@article{Accettura:2023ked,
    author = "Accettura, Carlotta and others",
    title = "{Towards a muon collider}",
    eprint = "2303.08533",
    archivePrefix = "arXiv",
    primaryClass = "physics.acc-ph",
    reportNumber = "FERMILAB-PUB-23-123-AD-PPD-T",
    doi = "10.1140/epjc/s10052-023-11889-x",
    journal = "Eur. Phys. J. C",
    volume = "83",
    number = "9",
    pages = "864",
    year = "2023",
    note = "[Erratum: Eur.Phys.J.C 84, 36 (2024)]"
}

@article{CEPCStudyGroup:2023quu,
    author = "Abdallah, Waleed and others",
    collaboration = "CEPC Study Group",
    title = "{CEPC Technical Design Report: Accelerator}",
    eprint = "2312.14363",
    archivePrefix = "arXiv",
    primaryClass = "physics.acc-ph",
    reportNumber = "IHEP-CEPC-DR-2023-01, IHEP-AC-2023-01",
    doi = "10.1007/s41605-024-00463-y",
    journal = "Radiat. Detect. Technol. Methods",
    volume = "8",
    number = "1",
    pages = "1--1105",
    year = "2024",
    note = "[Erratum: Radiat.Detect.Technol.Methods 9, 184--192 (2025)]"
}

@article{Kuno:1999jp,
    author = "Kuno, Yoshitaka and Okada, Yasuhiro",
    title = "{Muon decay and physics beyond the standard model}",
    eprint = "hep-ph/9909265",
    archivePrefix = "arXiv",
    reportNumber = "KEK-PREPRINT-99-69, KEK-TH-639",
    doi = "10.1103/RevModPhys.73.151",
    journal = "Rev. Mod. Phys.",
    volume = "73",
    pages = "151--202",
    year = "2001"
}

@article{Kitano:2002mt,
    author = "Kitano, Ryuichiro and Koike, Masafumi and Okada, Yasuhiro",
    title = "{Detailed calculation of lepton flavor violating muon electron conversion rate for various nuclei}",
    eprint = "hep-ph/0203110",
    archivePrefix = "arXiv",
    reportNumber = "KEK-TH-808",
    doi = "10.1103/PhysRevD.76.059902",
    journal = "Phys. Rev. D",
    volume = "66",
    pages = "096002",
    year = "2002",
    note = "[Erratum: Phys.Rev.D 76, 059902 (2007)]"
}

@article{Davidson:2018kud,
    author = "Davidson, Sacha and Kuno, Yoshitaka and Yamanaka, Masato",
    title = "{Selecting $\mu \to e$ conversion targets to distinguish lepton flavour-changing operators}",
    eprint = "1810.01884",
    archivePrefix = "arXiv",
    primaryClass = "hep-ph",
    doi = "10.1016/j.physletb.2019.01.042",
    journal = "Phys. Lett. B",
    volume = "790",
    pages = "380--388",
    year = "2019"
}

@article{COMET:2009qeh,
    author = "Cui, Y. G. and others",
    collaboration = "COMET",
    title = "{Conceptual design report for experimental search for lepton flavor violating mu- - e- conversion at sensitivity of 10**(-16) with a slow-extracted bunched proton beam (COMET)}",
    reportNumber = "KEK-2009-10",
    month = "6",
    year = "2009"
}

@article{SINDRUM:1987nra,
    author = "Bellgardt, U. and others",
    collaboration = "SINDRUM",
    title = "{Search for the Decay $\mu^+ \to e^+ e^+ e^-$}",
    reportNumber = "SIN-PR-87-09",
    doi = "10.1016/0550-3213(88)90462-2",
    journal = "Nucl. Phys. B",
    volume = "299",
    pages = "1--6",
    year = "1988"
}

@article{Hayasaka:2010np,
    author = "Hayasaka, K. and others",
    title = "{Search for Lepton Flavor Violating Tau Decays into Three Leptons with 719 Million Produced Tau+Tau- Pairs}",
    eprint = "1001.3221",
    archivePrefix = "arXiv",
    primaryClass = "hep-ex",
    doi = "10.1016/j.physletb.2010.03.037",
    journal = "Phys. Lett. B",
    volume = "687",
    pages = "139--143",
    year = "2010"
}

@article{Abazajian:2019eic,
    author = "Abazajian, Kevork and others",
    title = "{CMB-S4 Science Case, Reference Design, and Project Plan}",
    eprint = "1907.04473",
    archivePrefix = "arXiv",
    primaryClass = "astro-ph.IM",
    reportNumber = "FERMILAB-PUB-19-431-AE-SCD",
    month = "7",
    year = "2019"
}

@article{Matsumura:2013aja,
    author = "Matsumura, T. and others",
    title = "{Mission design of LiteBIRD}",
    eprint = "1311.2847",
    archivePrefix = "arXiv",
    primaryClass = "astro-ph.IM",
    doi = "10.1007/s10909-013-0996-1",
    journal = "J. Low Temp. Phys.",
    volume = "176",
    pages = "733",
    year = "2014"
}

@article{SajjadAthar:2021prg,
    author = "Sajjad Athar, Mohammad and others",
    title = "{Status and perspectives of neutrino physics}",
    eprint = "2111.07586",
    archivePrefix = "arXiv",
    primaryClass = "hep-ph",
    reportNumber = "FERMILAB-PUB-21-621-ND",
    doi = "10.1016/j.ppnp.2022.103947",
    journal = "Prog. Part. Nucl. Phys.",
    volume = "124",
    pages = "103947",
    year = "2022"
}

\end{document}